\documentclass[useAMS,usenatbib]{mnras}

\usepackage{epsfig}
\usepackage{aas_macros}
\usepackage{natbib}
\usepackage{times}
\usepackage{graphicx, bm, amssymb, xcolor}
\usepackage{verbatim}
\usepackage[all]{hypcap}
\usepackage{xspace}
\usepackage{multirow}
\usepackage{pgf,tikz}
\usepackage{mathrsfs}
\usetikzlibrary{arrows}
\usepackage{amsmath}

\usepackage{environ}
\makeatletter
\newsavebox{\measure@tikzpicture}
\NewEnviron{scaletikzpicturetowidth}[1]{%
  \def\tikz@width{#1}%
  \begin{lrbox}{\measure@tikzpicture}%
  \BODY
  \end{lrbox}%
  \pgfmathparse{#1/\wd\measure@tikzpicture}%
  \BODY
}
\makeatother

\newcommand{\be}{\begin{equation}}
\newcommand{\ee}{\end{equation}}
\newcommand{\bea}{\begin{eqnarray}}
\newcommand{\eea}{\end{eqnarray}}

\title{A general theory for the lifetimes of giant molecular clouds under the influence of galactic dynamics}
\author{Sarah~M.~R.~Jeffreson\thanks{s.jeffreson@uni-heidelberg.de} and J.~M.~Diederik~Kruijssen\\
Astronomisches Rechen-Institut, Zentrum f\"{u}r Astronomie der Universit\"{a}t Heidelberg, M\"{o}nchhofstra\ss e 12-14, 69120 Heidelberg, Germany}

\hypersetup{draft}
\begin{document}

\date{Accepted 2018 March 1. Received 2018 March 1; in original form 2017 November 28.}

\pagerange{\pageref{firstpage}--\pageref{lastpage}} \pubyear{2017}

\maketitle

\label{firstpage}

\begin{abstract}
We propose a simple analytic theory for environmentally-dependent molecular cloud lifetimes, based on the large-scale (galactic) dynamics of the interstellar medium. Within this theory, the cloud lifetime is set by the time-scales for gravitational collapse, galactic shear, spiral arm interactions, epicyclic perturbations and cloud-cloud collisions. It is dependent on five observable quantities, accessible through measurements of the galactic rotation curve, the gas and stellar surface densities, and the gas and stellar velocity dispersions of the host galaxy. We determine how the relative importance of each dynamical mechanism varies throughout the space of observable galactic properties, and conclude that gravitational collapse and galactic shear play the greatest role in setting the cloud lifetime for the considered range of galaxy properties, while cloud-cloud collisions exert a much lesser influence. All five environmental mechanisms are nevertheless required to obtain a complete picture of cloud evolution. We apply our theory to the galaxies M31, M51, M83, and the Milky Way, and find a strong dependence of the cloud lifetime upon galactocentric radius in each case, with a typical cloud lifetime between $10$ and $50$~Myr. Our theory is ideally-suited for systematic observational tests with the Atacama Large Millimetre/submillimetre array.
\end{abstract}


\section{Introduction}
\label{Sec::Introduction}
As the sites of the majority of galactic star formation, giant molecular clouds (GMCs) and their life-cycles are of critical importance in understanding and predicting the galactic star formation efficiency (SFE). In particular, the molecular cloud lifetime sets a time-scale for star formation that is degenerate with the SFE in the observational relation between the galactic star formation rate (SFR) and the gas mass~\citep{Kennicutt1998}. The SFE for a given unit of gas quantifies its ability to form stars, and so offers crucial insight into the conditions most conducive to star formation in the interstellar medium (ISM). To constrain this quantity from observations of the SFR, a reliable theory of the molecular cloud lifetime is required.

In contrast to past observational data supporting the notion of `long' cloud lifetimes of order $\gtrapprox 100$ Myr~\citep{Scoville1979,Scoville2004,Koda2009}, recent observations of the molecular ISM at high spatial resolution by~\cite{Engargiola2003},~\cite{Blitz2006},~\cite{Kawamura2009},~\cite{Murray2011},~\cite{Miura2012} and~\cite{Meidt2015} have favoured much shorter lifetimes between $10$ and $55$ Myr. These shorter lifetimes are consistent with the characteristic time-scale of collapse for overdense clumps within the cloud substructure~\citep{Elmegreen2000,Hartmann2001}. While theories of long-lived GMCs support the view of clouds as distinct, gravitationally-bound, virialised entities, as distinguished by a tight correlation between virial mass and CO luminosity~\citep[e.g.][]{Solomon1987}, short-lived GMCs appear to be dynamic and continually evolving, with highly complex life-cycles~\citep{Dobbs2013b}. Observations such as those by~\cite{Colombo2014} support this view, demonstrating a large scatter in the relationship between virial mass and CO luminosity, and thus a significant fraction of GMCs that may be gravitationally unbound. In fact,~\cite{Dobbs2011} point out that over 50\% of the clouds observed by~\cite{Heyer2009} are strictly unbound, with virial parameters $\alpha_{\rm vir}>2$. These observations are in line with numerical simulations of molecular cloud evolution~\citep{Dobbs2011,Dobbs2013b}, which produce largely unbound GMCs with star formation occurring in localised bound regions.

The diversity in the observed dynamical states of GMCs presents a challenge to theories of cloud formation and evolution that rely on theoretical assumptions about what constitutes a molecular cloud. Cloud evolution shaped by frequent inter-cloud collisions is proposed by~\cite{Tan2000} to be in agreement with the scaling relations between the gas surface density and the star formation rate surface density observed by~\cite{Kennicutt1998}, however the theory accounts only for those clouds that are gravitationally bound~\citep{Gammie1991} and supported against collapse by hydrostatic and magnetised turbulent pressure, such that they live long enough for collisions to actually occur. Additionally, the anti-correlation between star formation efficiency and the shear parameter $\beta$ expected in this model is not observed in spiral and dwarf galaxies~\citep{Leroy2008}. Although cloud-cloud collisions may contribute to the evolution of certain clouds at certain values of $\beta$, it does not account for molecular clouds in all observable regions of the ISM.

Molecular clouds that persist in a state of gravitational free-fall throughout their lives without reaching virial equilibrium, via global collapse~\citep{Elmegreen1993,Ballesteros-Paredes1999a,Ballesteros-Paredes1999b,Hartmann2001,Vazquez-Semadeni2003,Vazquez-Semadeni2006,Heitsch2005,Heitsch2006} or via hierarchical collapse~\citep{Elmegreen2007,Zamora-Aviles2012,Zamora-Aviles2014,Ibanez-Mejia2016}, are gravitationally bound by definition. Therefore, theories of cloud evolution that are dominated by gravitational collapse, as first proposed by~\cite{Goldreich1974} and~\cite{Liszt1974}, do not account for clouds that are virialised or unbound. If all clouds were bound and collapsing, we would expect a clear correlation between the Toomre $Q$ stability parameter and the number of GMCs. Given that the majority of star formation occurs in molecular clouds, this would lead to a correlation between $Q$ and the SFE per unit gas and per unit time, which is not observed~\citep{Leroy2008}. Clouds that are not bound or collapsing may also play an important role in star formation. Like theories of cloud-cloud collisions, theories of cloud evolution dominated by gravitational collapse do not account for the wide variety of observable GMC properties.

The great variety and complexity of physics that may influence molecular clouds is further emphasised by the large number of physical processes that can successfully account for their large non-thermal line-widths~\citep{Fukui2001,Engargiola2003,Rosolowsky2005}. These include, but are not limited to, bulk radial motions due to a persistent state of gravitational free-fall~\citep{Ballesteros-Paredes1999,Hartmann2001,Heitsch2008}, energy input due to cloud-cloud collisions~\citep{Tan2000,Tasker2009,Tasker2011}, external driving by supernovae~\citep[e.g.][]{Kim2015b,Kim2015a}, and accretion of material via converging flows associated with cloud formation~\citep{Klessen2010}. Given that the virial state of GMCs scales with their velocity dispersions squared, it is to be expected that the variety of processes affecting GMC velocity dispersions also affect their boundedness, and hence their lifecycles. It is therefore undesirable to limit the scope of cloud evolution theories to clouds with size-scales, mass-scales and structures imposed by the assumptions of gravitational boundedness and virialisation. The spread of $\approx 50$ Myr among `short' observed cloud lifetimes points towards a diverse range of astrophysical objects that can be observationally identified as GMCs. If a theory is to successfully account for the entire population, it must be correspondingly flexible.

A solution to this problem is suggested by the work of~\cite{Leroy2017b}, which shows that the SFE per unit time (i.e.~the inverse of the gas depletion time) scales nearly linearly with the ratio of the squared velocity dispersion to the surface density, an indicator of cloud boundedness. As the majority of star formation occurs in GMCs, this environmental dependence of the SFR indicates that the evolution of GMCs is also strongly environmentally dependent, and a theory for molecular cloud lifetimes should be able to capture this environmental influence on cloud evolution. Using only the observable properties of the ISM, the time-scales of large-scale dynamical processes can be derived and combined to predict the environmentally-dependent cloud lifetime, independent of the precise properties of GMCs, and of the theoretical distinction between processes of cloud formation and evolution.

In this work, we take a systematic approach to predicting the cloud lifetime, combining the time-scales for those dynamical processes with the greatest potential to influence molecular cloud evolution. These include gravitational collapse and cloud-cloud collisions, as well as epicyclic perturbations in the plane of the host galaxy, galactic shear, and spiral arm crossings, where applicable. We propose a theory of molecular cloud lifetimes that quantifies the complexities of GMC evolution as naively as possible, providing a platform upon which increasing levels of detail can be built in future work. Taking a simple but expansive approach allows us to dispense with arbitrary theoretical definitions of what exactly constitutes a molecular cloud, such that we do not need to make assumptions about its state of gravitational-boundedness, state of virialisation, scale or structure. Rather than focusing on one particular evolutionary mechanism, we account for the coexistence of different mechanisms and how they may augment each other or compete against each other. Using this theory, we can provide systematic predictions of cloud lifetimes throughout the parameter space of observed galactic properties.

Our theory of GMC formation and evolution, being dependent upon large-scale galactic dynamics, will be observationally testable by applying the new statistical technique of the `uncertainty principle for star formation'~\citep{Kruijssen2014} to the recently available wealth of high-resolution observations of the ISM, both at low and high redshifts~\citep[e.g.][]{Hodge2012,Leroy2016}. With this method, we gain access to cloud lifetimes, feedback time-scales and cloud separation lengths for a statistically representative sample of galaxies~\citep{Kruijssen2018b,Haydon2017,Hygate2017}, enabling the evolutionary lifecycle of GMCs to be probed beyond the limited environment of the most nearby galaxies (Kruijssen et al.~in prep., Hygate et al.~in prep., Chevance et al.~in prep., Schruba et al.~in prep.). This will provide a much more complete and systematic perspective than previously available, making a global theory of cloud evolution pertinent, viable and testable.

The structure of this paper is as follows. In Section~\ref{Sec::time-scales}, we derive the time-scales of galactic-scale dynamics which may determine the rate of GMC formation and evolution. In Section~\ref{Sec::Comptime-scales}, we describe how our model combines these time-scales to provide a comprehensive picture of cloud evolution over all of parameter space. In Section~\ref{Sec::Coexistence}, we examine the competition between time-scales and their different regions of dominance in parameter space, the interpretation of which we discuss in Section~\ref{Sec::Discussion}. In Section~\ref{Sec::Galaxies}, we apply our theory to predict cloud lifetimes in real galaxies, and finally we summarise our results and conclusions in Section~\ref{Sec::Conclusions}.

\section{Time-scales of cloud evolution} \label{Sec::time-scales}
Here we derive the time-scales for six different large-scale dynamical processes that affect the ISM and its constituent molecular clouds. Each time-scale is dependent only on the physical, observable properties of the ISM, presented in Table~\ref{Tab::Params}. In this work, we systematically develop a theory of cloud evolution and the cloud lifetime based exclusively on the observable large-scale dynamics of the ISM. As such, we refer to these time-scales throughout the remainder of this paper as the `time-scales of cloud evolution'. The first five of these time-scales have a compressive effect on molecular clouds, while the sixth has a dispersive effect. We implicitly assume that the compressive processes lead to star formation and subsequently to cloud dispersal on a time-scale appropriate to stellar feedback, assumed to be shorter than the time-scales considered here.

\subsection{Gravitational collapse of the ISM ($\tau_{\text{ff,g}}$)} \label{Sec::tau_ffg}
On scales shorter than the Toomre length $\lambda_T$ within the galactic disc, approximately spherical regions of the ISM are susceptible to collapse on the local free-fall time-scale. We follow the derivation of~\cite{KrumholzMcKee2005}, beginning with the generalisation of gas pressure within the galactic plane to include the contribution of the stellar population, such that

\begin{equation}
\label{Eqn::pressure}
 P_g \approx \frac{\pi}{2} \phi_P G\Sigma_g^2.
\end{equation}
The quantity $\phi_P$ is given according to ~\cite{Elmegreen1989} as

\begin{equation}
\label{Eqn::phiP_defn}
 \phi_P = 1+\frac{\Sigma_s}{\Sigma_g}\frac{\sigma_g}{\sigma_s},
\end{equation}
where $\Sigma_s$ and $\Sigma_g$ are the stellar and gas surface densities respectively, while $\sigma_s$ and $\sigma_g$ are the stellar and gas velocity dispersions. The case of pure gas then corresponds to $\phi_P=1$. Given that the midplane ISM volume density $\rho_g$ scales with the midplane ISM pressure $P_g$ we can scale the Jeans equations by the constant factor $\phi_P$ from Equation~(\ref{Eqn::pressure}), to obtain the scale height $h_g$ as

\begin{equation}
 h_g = \frac{\sigma_g}{\sqrt{2\pi G\phi_P \rho_g}},
\end{equation}
which leads to the definition of $\rho_g$~\citep[Equation 36 in][]{KrumholzMcKee2005},

\begin{equation}
\label{Eqn::rho_g}
 \rho_g = \frac{\phi_P \kappa^2}{2\pi Q^2 G},
\end{equation}
where $Q = \kappa \sigma_g/\pi G \Sigma_g$ is the Toomre stability parameter~\citep{Toomre1964}. Using this midplane ISM density we can straight-forwardly calculate the free-fall time-scale as

\begin{equation}
\label{Eqn::tau_ffg}
 \tau_{\text{ff,g}} = \sqrt{\frac{3\pi^2}{32\phi_P (1+\beta)}} \frac{Q}{\Omega},
\end{equation}
in terms of the Toomre $Q$, the orbital speed $\Omega$ and the shear parameter $\beta$, which is defined by the rotation curve of the galaxy $v_c(R)$ as

\begin{equation}
 \beta = \frac{d\ln{v_c(R)}}{d\ln{R}}.
\end{equation}
Lower values of $\beta$ indicate a higher degree of differential rotation, and for a given value of the angular speed $\Omega$, this corresponds to a higher degree of galactic shearing. Note that $\tau_{\text{ff,g}}$ is dependent only upon the properties $\beta$, $Q$, $\Omega$ and $\phi_P$ of the ISM, via Equation~(\ref{Eqn::tau_ffg}). It increases linearly as $Q$ and decreases inversely as $\Omega^{-1}$, with weaker dependences on $\beta$ and $\phi_P$ as $(1+\beta)^{-1/2}$ and $\phi_P^{-1/2}$, respectively. The time-scale $\tau_{\rm ff,g}$ is therefore dependent only on the rotation curve, the surface density profiles $\Sigma_g$ and $\Sigma_s$, and the velocity dispersion profiles $\sigma_g$ and $\sigma_s$ of the host galaxy, via the definition of the Toomre $Q$ parameter and the definition of $\phi_P$ (Equation~(\ref{Eqn::phiP_defn})). Importantly, $\tau_{\rm ff,g}$ does not assume any properties of the cloud itself, and can therefore be used to quantify the rate of collapse across the varied spectrum of objects that are observationally classified as GMCs. In the next section, we will see that the introduction of a characteristic cloud mass scale restricts the applicability of the free-fall time-scale to gravitationally-bound clouds of a particular size and structure. These assumptions are not appropriate to all GMCs, so we will argue that $\tau_{\rm ff,g}$ is the gravitational free-fall time-scale we require to encode gravitational free-fall in our theory.

\subsection{Gravitational collapse of a Toomre-mass cloud ($\tau_{\text{ff,cl}}$)} \label{Sec::tau_ffcl}
\begin{figure}
\label{Fig::grav_collapse}
\centering
\begin{tikzpicture}[line cap=round,line join=round,>=triangle 45,x=0.5cm,y=0.5cm]
\clip(5.5,3.) rectangle (20.5,8.5);
\draw(8.,6.) circle (1.cm);
\draw [->] (10.,6.) -- (8.898937624871147,6.);
\draw [->] (8.,8.) -- (8.,6.898937624871145);
\draw [->] (6.,6.) -- (7.101062375128855,6.);
\draw [->] (8.,4.) -- (8.,5.101062375128855);
\draw(12.984519656856023,6.005936157027841) circle (1.cm);
\draw(12.011762030192955,6.827375930654432) circle (0.24288898727169347cm);
\draw [->] (12.42248191700625,6.567973896877613) -- (12.011762030192955,6.827375930654432);
\draw [->] (11.66826513024226,6.483879030703737) -- (12.011762030192955,6.827375930654432);
\draw [->] (11.648687045207954,7.1501092506410995) -- (12.011762030192955,6.827375930654432);
\draw [->] (12.306847913084859,7.2132574698207685) -- (12.011762030192955,6.827375930654432);
\draw(13.287155362928976,7.130011636727386) circle (0.24288898727169347cm);
\draw [->] (13.58224124582088,7.515893175893723) -- (13.287155362928976,7.130011636727386);
\draw [->] (12.924080377943975,7.452744956714054) -- (13.287155362928976,7.130011636727386);
\draw [->] (12.94365846297828,6.786514736776692) -- (13.287155362928976,7.130011636727386);
\draw [->] (13.697875249742271,6.870609602950568) -- (13.287155362928976,7.130011636727386);
\draw(14.,6.) circle (0.24288898727169347cm);
\draw(12.573799770042728,5.768150959399091) circle (0.24288898727169347cm);
\draw(13.179071182188636,4.730542824291819) circle (0.24288898727169347cm);
\draw [->] (12.230302870092032,5.424654059448397) -- (12.573799770042728,5.768150959399091);
\draw [->] (13.656503100049305,5.656503100049306) -- (14.,6.);
\draw [->] (12.878807954534077,4.387045924341124) -- (13.222304854484772,4.730542824291819);
\draw [->] (12.210724785057726,6.090884279385759) -- (12.573799770042728,5.768150959399091);
\draw [->] (13.636925015014999,6.322733319986668) -- (14.,6.);
\draw [->] (12.837613033351703,5.0316593081304175) -- (13.200688018336704,4.70892598814375);
\draw [->] (12.868885652934631,6.154032498565428) -- (12.573799770042728,5.768150959399091);
\draw [->] (14.295085882891904,6.385881539166337) -- (14.,6.);
\draw [->] (13.495773901228608,5.094807527310087) -- (13.200688018336704,4.70892598814375);
\draw [->] (12.984519656856023,5.508748925622273) -- (12.573799770042728,5.768150959399091);
\draw [->] (13.589791069001931,4.471140790515) -- (13.179071182188636,4.730542824291819);
\draw [->] (14.357680916867942,5.671298612324219) -- (14.,6.);
\draw(18.,6.) circle (1.cm);
\draw(17.696989937134884,6.870609602950568) circle (0.24288898727169425cm);
\draw(17.805074117875225,5.076412202660909) circle (0.24288898727169425cm);
\draw [->] (17.992075820026788,7.256491142116906) -- (17.696989937134884,6.870609602950568);
\draw [->] (17.333914952149883,7.193342922937237) -- (17.696989937134884,6.870609602950568);
\draw [->] (18.05467085400283,6.541908215274788) -- (17.696989937134916,6.870609602950608);
\draw [->] (17.35349303718422,6.527112702999919) -- (17.696989937134916,6.870609602950608);
\draw [->] (17.420382296742158,5.3775286864995095) -- (17.783457281727156,5.054795366512841);
\draw [->] (18.078543164619056,5.440676905679179) -- (17.783457281727156,5.054795366512841);
\draw [->] (18.141138198595097,4.7260939788370635) -- (17.783457281727156,5.054795366512842);
\draw [->] (17.45068101249401,4.744168666366018) -- (17.80499718727121,5.03500063744333);
\draw (6.8,3.9) node[anchor=north west] {$\alpha_{\rm vir}\lessapprox 1$};
\draw (11.9,3.9) node[anchor=north west] {$\alpha_{\rm vir}\lessapprox 1$};
\draw (16.8,3.9) node[anchor=north west] {$\alpha_{\rm vir} \geq 2$};
\end{tikzpicture}
\caption{Illustration of possible modes of gravitational collapse. Global gravitational collapse towards the centre of mass necessarily means that the molecular cloud is gravitationally-bound, with $\alpha_{\rm vir}\lessapprox 1$ (left). A molecular cloud can also be gravitationally bound if it is collapsing hierarchically (centre), such that smaller regions within a larger diffuse envelope are collapsing. A cloud which is collapsing hierarchically can also be gravitationally unbound (right) if its overall internal kinetic energy is larger than its potential energy.}
\end{figure}
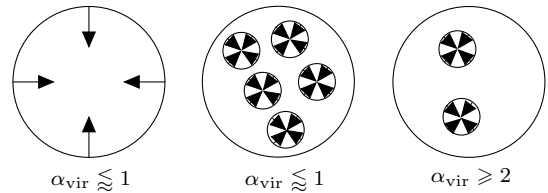

Given that the free-fall time of a roughly spherical region of gas scales as $\tau \propto \rho^{-1/2}$, we can relate the collapse time-scale $\tau_{\text{ff,cl}}$ for a cloud of mean density $\rho_{\text{cl}}$ to the collapse time-scale $\tau_{\text{ff,g}}$ in the midplane, as

\begin{equation}
\label{Eqn::phi_rho_defn}
 \frac{\tau_{\text{ff,cl}}}{\tau_{\text{ff,g}}} = \Big(\frac{\rho_{\text{cl}}}{\rho_g}\Big)^{-1/2} \equiv \phi_{\rho}^{-1/2},
\end{equation}
where $\rho_g$ is the mean midplane density of the ISM gas and the second equality defines the ratio of cloud to midplane densities $\phi_{\rho} = \rho_{\text{cl}}/\rho_g$. Given that the cloud is formed via gravitational collapse of the ISM, it must have a higher overall density than the surrounding gas, such that $\phi_\rho>1$. To calculate this ratio we again follow~\cite{KrumholzMcKee2005} to obtain the cloud density $\rho_{\text{cl}}$ in terms of its pressure $P_{\text{cl}}$ and mass $M_{\text{cl}}$ as

\begin{equation}
 \rho_{\text{cl}} = \Big(\frac{375}{4\pi}\Big)^{1/4} \Big(\frac{P_{\text{cl}}^3}{\alpha_{\text{vir}}^3 G^3 M_{\text{cl}}^2}\Big)^{1/4},
\end{equation}
where $\alpha_{\text{vir}}$ is the standard virial parameter~\citep[e.g.][]{MacLaren1988, BertoldiMcKee1992}, given by

\begin{equation}{}
\label{Eqn::alpha_vir}
 \alpha_{\text{vir}} = \frac{5 \sigma_{\text{cl}}^2}{G\sqrt{\pi M_{\text{cl}} \Sigma_{\text{cl}}}},
\end{equation}{}
and $P_{\text{cl}}$ is given by their Equation 47 as,

\begin{equation}
 P_{\text{cl}} = \frac{3\pi}{20} \alpha_{\text{vir}} G\Sigma_{\text{cl}}^2.
\end{equation}
Rather than taking the typical cloud mass scale $M_{\rm cl}$ to be the Jeans mass $M_J$, we instead opt to use the Toomre mass $M_T$~\citep{Toomre1964}, as in~\cite{Reina-Campos2017}. This ensures that we are looking at the largest collapsing scales, given by the Toomre length. In practice, this change is consistent with~\cite{KrumholzMcKee2005} due to their assumption of $Q \approx 1$, which gives an approximate equality between $M_T$ and $M_J$, as mentioned in their Section 3. The critical Toomre mass corresponding to the maximum unstable length-scale is given by

\begin{equation}
 M_T = \frac{4 \pi^5 G^2 \Sigma_g^3}{\kappa^4}.
\end{equation}
Following through with the derivation of $\phi_\rho$ we obtain

\begin{equation}
\label{Eqn::phi_rho_tauffcl}
 \phi_{\rho} = \Big(\frac{375}{32\pi^4}\Big)^{1/4} \Big(\frac{\phi_{\bar{P}}^3}{\phi_P \alpha_{\text{vir}}^3}\Big)^{1/4} Q^2,
\end{equation}
where $\phi_{\bar{P}}$ is defined as in Equation 46 of~\cite{KrumholzMcKee2005}, i.e.

\begin{equation}
 P_{\text{cl}}=\phi_{\bar{P}} P_{\rm mp} = \frac{\pi}{2} \phi_{\bar{P}} \phi_{\rm P} G \Sigma_g^2,
\end{equation}
c.f.~\cite{Elmegreen1989,Blitz2004}, with $P_{\rm mp}$ the mid-plane gas pressure in the galaxy. Since $\phi_\rho \geq 1$ by definition and $\phi_P \geq 1$ according to Equation~(\ref{Eqn::phiP_defn}), the range of $\alpha_{\text{vir}}$ and $\phi_{\overline{P}}$ values possible for a cloud of mass $M_T$ is limited by Equation~(\ref{Eqn::phi_rho_tauffcl}) to

\begin{equation}
\frac{\phi_{\bar{P}}}{\alpha_{\rm vir}} \gtrapprox \frac{2}{Q^{8/3}}.
\end{equation}
The value of the time-scale, given by substituting Equation~(\ref{Eqn::phi_rho_tauffcl}) into Equation~(\ref{Eqn::phi_rho_defn}), is

\begin{equation}
\label{Eqn::tau_ffcl}
 \tau_{\text{ff,cl}}=\frac{\pi^{3/2}}{2} \Big(\frac{3\alpha_{\text{vir}}}{10 \phi_P \phi_{\bar{P}}}\Big)^{3/8} \frac{1}{\Omega} \frac{1}{\sqrt{1+\beta}}.
\end{equation}
We note that in contrast to the time-scale $\tau_{\text{ff,g}}$ in Section~\ref{Sec::tau_ffg}, $\tau_{\text{ff,cl}}$ depends upon a set of pre-defined properties for each molecular cloud. Namely, it depends on the virial parameter $\alpha_{\text{vir}}$ that quantifies the gravitational boundedness of the cloud, and the ratio of cloud pressure to that of the ambient gas, $\phi_{\overline{P}}$. The values of these properties must be determined for each cloud, in order to compute its free-fall time-scale. Additionally, the introduction of a characteristic collapsing mass scale $M_T$ assumes that the cloud is in a state of {\it global collapse} (i.e.~collapsing radially towards its centre of mass) as depicted in the left panel of Figure~\ref{Fig::grav_collapse}. Global gravitational collapse necessarily implies gravitational boundedness ($\alpha_{\rm vir}<1$). Combined with the assumption of a roughly spherical cloud, the mass scale $M_T$ also sets a characteristic length-scale $\lambda_T$ for the cloud, corresponding to the Toomre scale. The use of the time-scale $\tau_{\rm ff,cl}$ to quantify the influence of gravitational collapse on GMCs therefore implies a restriction to gravitationally-bound, globally-collapsing clouds of diameter $\lambda_T$.

In the middle and right panels of Figure~\ref{Fig::grav_collapse}, we depict two situations in which a time-scale for gravitational collapse is required to describe the evolution of molecular clouds that do not fit the restrictions described above. In the middle panel we demonstrate the possibility of a gravitationally-bound cloud which is not collapsing globally but hierarchically, as theorised by~\cite{Elmegreen2007} and observed in simulations by~\cite{Ibanez-Mejia2016}. In the right-hand panel, we demonstrate the possibility of hierarchical collapse within a molecular cloud that is not gravitationally-bound at all, but that may be in fact be gravitationally unbound ($\alpha_{\rm vir} \geq 2$).

In this work, we aim to examine the processes affecting the evolution of all molecular clouds, rather than restricting ourselves to processes affecting the subset of gravitationally-bound, globally collapsing clouds of radius $\lambda_T$. As such, we opt hereafter to work with the free-fall time-scale of the ISM $\tau_{\text{ff,g}}$, which also provides an upper bound on $\tau_{\text{ff,cl}}$ via the definition of $\phi_\rho > 1$. While this allows our theory to be as general as possible, it is conceivable that it leads to an overestimation of the molecular cloud lifetime in atomic gas-dominated environments, in which only the gas density peaks are molecular. We return to this caveat in Section~\ref{Sec::Galaxies}. For the remainder of this paper, we will refer to $\tau_{\rm ff,g}$ alone as the `gravitational free-fall time-scale'. We do not need to assume values of the cloud properties $\alpha_{\text{vir}}$ and $\phi_{\bar{P}}$, leaving only the properties $\beta$, $Q$, $\Omega$ and $\phi_P$ of the ISM. These quantities can be determined observationally, and do not require us to impose theoretical constraints on the nature of molecular clouds.

\subsection{Cloud-cloud collisions ($\tau_{\text{cc}}$)}
\label{Sec::tau_cc}
\cite{Tan2000} derives a time-scale for collisions between gravitationally-bound molecular clouds, as

\begin{equation}
\label{Eqn::tau_cc}
 \tau_{\text{cc}} \approx \frac{2\pi Q}{9.4 f_G \Omega (1+0.3 \beta)(1-\beta)}.
\end{equation}
The parameter $f_G$ represents the `probability of collision' associated with any single encounter between GMCs. \cite{Tan2000} hypothesises that such collisions trigger star formation by inducing compressions in parts of the interacting clouds. We will use $\tau_{\rm cc}$ to quantify the effect of cloud-cloud collisions on cloud evolution and the cloud lifetime, via compressions~\citep{Tan2000}, shocks, or simply mixing between regions of gas with different surface densities.

In principle, the time-scale $\tau_{\rm cc}$ increases monotonically with $\beta$. However, the range of $\beta$ is limited to $0<\beta<0.12$, due to the underlying assumptions. Equation (\ref{Eqn::tau_cc}) is based on the work of~\cite{Gammie1991}, who derive gas velocity dispersions for $0 < \beta < 0.12$, by considering the gravitational scattering between clouds due to galactic differential rotation, using an expansion about the equilibrium perturbative solution at $\beta=0$ to first order in $\beta$. The perturbative solution is then numerically extended over a range corresponding to $-0.06<\beta<0.12$. By including higher orders, we could obtain a real (physical) value of the velocity dispersion up to $\beta=0.36$, but this would involve an extrapolation of the numerical fit outside the range for which it has been verified in~\cite{Gammie1991}. Throughout this paper, we therefore use the value of $\tau_{\rm cc}$ at $\beta=0.12$ for higher values of $\beta$, which is justified because it acts as a lower limit on the collision time-scale at $\beta>0.12$. Given that the collisions are caused by differential rotation, the assumption that the collision time-scale will monotonically increase with $\beta$ is also justified intuitively. Thus the time-scale $\tau_{\rm cc}$ in our model increases with $\beta$ over the range $0\leq\beta<0.12$ and remains constant with $\beta$ over the range $0.12\leq\beta$. Like the time-scale $\tau_{\rm ff,g}$ for gravitational collapse, $\tau_{\rm cc}$ decreases inversely as $\Omega^{-1}$, since a higher angular speed gives a shorter time between collisions for clouds at different galactocentric radii, when differential rotation is present ($\beta \neq 1$).

The $Q$-dependence of $\tau_{\rm cc}$ is linear, because the mean free time between cloud collisions scales inversely with the surface number density of gravitationally-bound clouds~\citep{Tan2000}, which in turn is inversely proportional to $Q$. However, we must note that, in using $\tau_{\rm cc}$ for values of gravitational stability $Q \gg 1$, we assume that these relations hold for very low number densities of bound clouds. In reality, for very small numbers of bound clouds in an ISM of finite scale height $h_g$, the time between collisions will approach infinity much faster than $Q$ does. This assumption is therefore likely to overestimate the number of cloud collisions per unit time, and thus (again) to provide a lower limit to the cloud lifetime due to cloud-cloud collisions. Despite this, we will see later that cloud-cloud collisions are rarely significant in comparison to the other dynamical mechanisms of cloud evolution, from $Q \sim 0.5$ up to $Q \sim 15$. We will therefore use $\tau_{\rm cc}$ as the time-scale for cloud-cloud collisions for all $Q$, and will comment on the interpretation of the resulting cloud lifetimes, where appropriate.

\subsection{Pattern speed perturbations ($\tau_{\Omega_{\rm P}}$)}
\label{Sec::tau_Omega}
The passage of spiral arms through the ISM at pattern speed $\Omega_{\rm P}$ has a wide variety of possible effects on GMCs~\citep[for more detailed discussions, see][]{Meidt2013,Dobbs2013c}. The sudden change in mass surface density, pressure and gravitational potential associated with a spiral arm encounter induces a shock in the ISM~\citep{Roberts1969}, although whether this shock triggers gravitational collapse remains uncertain~\citep[e.g.][]{Elmegreen1986,Eden2012,Eden2015}. Aside from the shock itself, the  cloud is subjected to varying tidal and Coriolis forces by a non-axisymmetric gravitational potential~\citep{Elmegreen1979}, with a generally compressive effect. The conservation of angular momentum then implies that the degree of azimuthal shearing is also lower in the spiral arm than in the galactic disc~\citep{Elmegreen1994}. Both of these effects promote gravitational collapse, possibly explaining the high density of high-mass GMCs in spiral arms, an interpretation that is supported by the formation of massive, bound fragments in the MHD simulations of~\cite{Kim2002} and~\cite{Kim2006}. \cite{Meidt2013} show that in the case of M51, the gravitational potential associated with the grand design spiral structure creates torques that drive large radial excursions of gas along the spiral arms at a much greater amplitude than within the galactic disc.

In addition to these dynamical effects, it is observed in simulations by~\cite{Dobbs2008} that the rate of cloud-cloud collisions is increased within spiral arms, due to increased levels of radial motion and an overall increase in cloud density. This may lead to an increased rate of gravitational collapse and star formation as in~\cite{Tan2000}, or an increased rate of cloud agglomeration to produce more massive clouds. The non-axisymmetric structure of the galaxy itself may also influence the rate at which clouds are subjected to external sources of stellar feedback, as the majority of star formation in M51 is observed to preferentially occur along its spiral arms~\citep{Meidt2013}.

The higher density of massive GMCs observed in simulations by~\citep{Dobbs2011b} also raises the question of whether clouds are swept up in spiral arms, or whether they preferentially form in spiral arms~\citep[e.g.][]{Casoli1982,Dobbs2008b}. Once a cloud enters a spiral arm, the time for which it remains there is not generally known.

We take the simplest possible approach to quantifying the effect of spiral arm interactions on GMCs. We do not consider the precise mechanisms by which clouds are affected by spiral arms, but instead consider only the time-scale on which encounters occur. This time-scale is given by

\begin{equation}
\label{Eqn::tau_OmegaP}
 \tau_{\Omega_{\rm P}} = \frac{2\pi}{m\Omega|1-\Omega_{\rm P}/\Omega|},
\end{equation}
where $m$ is the number of spiral arms in the galaxy. In common with the time-scales derived previously, $\tau_{\Omega_{\rm P}}$ depends inversely on $\Omega$. However, it has no dependence on $\beta$ or $Q$, and is instead a function of two new parameters, $m$ and $\Omega_{\rm P}/\Omega$, associated with the presence of spiral arms. The time-scale decreases inversely with increasing numbers of spiral arms $m$, and increases asymptotically towards the radius of co-rotation as $\Omega_{\rm P}/\Omega$ approaches unity.

Our use of $\tau_{\Omega_{\rm P}}$ to quantify the effect of spiral arms on GMCs implicitly assumes that the effect of each encounter is life-changing for the cloud, i.e.~that in the absence of all other effects, the average cloud lifetime is determined by the average time before which it encounters a spiral arm. Given the large differences in cloud properties observed between arm and interarm regions~\citep{Elmegreen1983,Meidt2013,Dobbs2008b,Dobbs2011b}, along with the large differences in the density of clouds between these regions, this assumption appears to be justified.

Inclusion of this spiral arm crossing time-scale must be considered on a case-by-case basis when considering molecular clouds in real galaxies. For example, it would not be appropriate for clouds which spend their whole lives inside a spiral arm, in which case it would be better to consider just the environmental conditions within the spiral arm. The case of very weak spiral arms might equally be treated by setting $m=0$ (see Section~\ref{Sec::MilkyWay}).

\subsection{Epicyclic perturbations ($\tau_{\kappa}$)}
\label{Sec::tau_kappa}
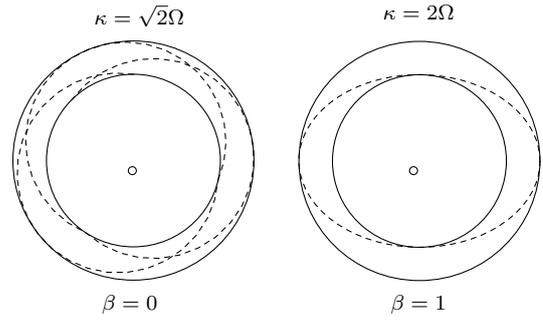
\begin{figure}
\centering
\begin{tikzpicture}[line cap=round,line join=round,>=triangle 45,x=1.0cm,y=1.0cm]
\label{Fig::Epicycles_tikz}
\clip(2.7,0.3) rectangle (10.2,4.8);
\draw [rotate around={-0.27678905217776173:(8.377465178429551,2.5283477629493247)},dash pattern=on 2pt off 2pt] (8.377465178429551,2.5283477629493247) ellipse (1.5855380957040242cm and 1.1441228062183206cm);
\draw(8.377465178429551,2.5283477629493243) circle (1.585536483446527cm);
\draw(8.377465178429551,2.5283477629493243) circle (1.1441228062183235cm);
\draw(4.614342059329806,2.534025620973247) circle (1.585536483446527cm);
\draw(4.614342059329806,2.534025620973247) circle (1.144122806218323cm);
\draw [shift={(4.407478079309705,2.3706380522536468)},dash pattern=on 2pt off 2pt]  plot[domain=1.4097619386792613:6.210442279794634,variable=\t]({1.*1.324635209988166*cos(\t r)+0.*1.324635209988166*sin(\t r)},{0.*1.324635209988166*cos(\t r)+1.*1.324635209988166*sin(\t r)});
\draw [shift={(4.513913560289052,2.786180598928477)},dash pattern=on 2pt off 2pt]  plot[domain=-0.39877721374393627:4.298416814454838,variable=\t]({1.*1.318121001392105*cos(\t r)+0.*1.318121001392105*sin(\t r)},{0.*1.318121001392105*cos(\t r)+1.*1.318121001392105*sin(\t r)});
\draw [shift={(4.875207415795406,2.5610110712264436)},dash pattern=on 2pt off 2pt]  plot[domain=-2.3081281715299036:2.514286677109874,variable=\t]({1.*1.3260252448453875*cos(\t r)+0.*1.3260252448453875*sin(\t r)},{0.*1.3260252448453875*cos(\t r)+1.*1.3260252448453875*sin(\t r)});
\draw (4.,4.7) node[anchor=north west] {$\kappa = \sqrt{2} \Omega$};
\draw (7.8,4.7) node[anchor=north west] {$\kappa = 2 \Omega$};
\draw (4.1,0.85) node[anchor=north west] {$\beta = 0$};
\draw (7.9,0.85) node[anchor=north west] {$\beta = 1$};
\begin{scriptsize}
\draw (8.3,2.4) circle (1.5pt);
\draw (4.6,2.4) circle (1.5pt);
\end{scriptsize}
\end{tikzpicture}
\caption{Epicyclic orbits in the plane of the galactic disc for the case of a flat rotation curve (left, $\kappa=\sqrt{2}\Omega$) and a solid body rotation curve (right, $\kappa=2\Omega$). Note that in the case of a higher shear parameter $\beta$ (right), the number of epicycles performed per orbit of the galactic centre is greater, and thus for fixed $\Omega$, the number of epicycles performed per unit time is also greater. This effect outweighs the slight decrease in epicyclic amplitude as $\beta$ increases at fixed $\Omega$.}
\end{figure}

Within the epicyclic approximation, the ambient ISM and its constituent molecular clouds are subjected to harmonic radial perturbations relative to the guiding centres of circular orbits about the galactic centre. The epicyclic frequency $\kappa$ of these oscillations is set by the angular speed and the shear parameter $\beta$ as

\begin{equation}
\kappa = \Omega \sqrt{2(1+\beta)},
\end{equation}
which has extreme values of $\kappa = \sqrt{2} \Omega$ for a flat rotation curve (left panel of Figure~\ref{Fig::Epicycles_tikz}) and $\kappa = 2 \Omega$ in the solid-body regime (right panel of Figure~\ref{Fig::Epicycles_tikz}). On the galactic scale, the primary effect of epicycles is to introduce eccentricity to an otherwise circular orbit around the galactic centre. On cloud scales, epicycles cause a molecular cloud to perform small elliptical circuits about a guiding centre that moves with the angular velocity $\Omega$ of the bulk ISM at some galactocentric radius $R_g$. These circuits subject the cloud to radial variations in galactic environment, such as variations in tidal force and pressure, and also to tangential variations relative to the guiding centre (i.e.~relative to a cloud that is moving on a non-epicyclic, circular orbit at angular frequency $\Omega$). Such tangential variations may include an increased number of interactions with other objects (that may be undergoing similar excursions), as well as stretching and compression due to acceleration relative to the guiding centre.

As with the other cloud evolutionary mechanisms, we aim to quantify all the possible effects of epicycles using a single time-scale $\tau_\kappa$. We consider the time-scale

\begin{equation}
\label{Eqn::tau_kappa_primary}
\tau_\kappa = \frac{\pi}{\kappa} \frac{\pi R_g/N}{\pi \sqrt{\frac{X^2+Y^2}{2}}},
\end{equation}
where $N$ is the number of epicycles around the guiding centre per revolution around the galactic centre, given by

\begin{equation}
\label{Eqn::N}
N = \frac{\kappa}{\Omega},
\end{equation}
and $(X,Y)$ are the amplitudes of epicyclic oscillations from the guiding centre in the radial and tangential directions, respectively. The first term in Equation~\ref{Eqn::tau_kappa_primary}, $\pi/\kappa$, gives the time required for the cloud to move from its orbital apocentre to its orbital pericentre, or vice versa.\footnote{Relative to the guiding centre, epicyclic motion is symmetric between the first and second halves of the circuit, and thus its time-scale is calculated for half an epicycle.} This is scaled by a second term, which quantifies the effect of epicycles relative to an object at the same galactocentric radius on a circular orbit, moving with the bulk ISM at an angular velocity $\Omega$. Relative to such an object, the cloud with epicyclic motion moves a distance $\pi \sqrt{(X^2+Y^2)/2}$ during the time $\pi/\kappa$, while the object itself moves a distance $\pi R_g/N$ relative to the galactic centre. Any large-scale galactic variations experienced by the cloud as it orbits the galactic centre, such as encounters with spiral arms or with other clouds, will be experienced by any object at the radius $R_g$, regardless of whether epicycles are performed. Thus we normalise the relative epicyclic motion, $\pi \sqrt{(X^2+Y^2)/2}$, by the guiding centre motion, $\pi R_g/N$. Due to the expression $X^2 + Y^2$, the influence of epicycles becomes greater if the amplitude of epicyclic motion is greater, while the terms $\kappa$ and $N$ indicate that the influence of epicycles becomes greater when they occur at a higher frequency relative to the bulk motion of the ISM.

Our theory aims to be as general as possible, so does not assume a precise cause for epicyclic perturbations. We do not set initial conditions for epicyclic motion and so do not explicitly calculate the quantity $X/R_g$. Instead, we quantify the magnitude of this ratio by taking the average value of a uniform distribution between $X/R_g=0$ and the upper bound $X/R_g=1/2$, set by the conservation of angular momentum within the epicyclic approximation (see Appendix~\ref{Sec::AppendixA}). This gives a value of

\begin{equation}
\label{Eqn::X_over_R}
\frac{X}{R_g} \approx \frac{1}{4}.
\end{equation}
Bringing together Equations~(\ref{Eqn::tau_kappa_primary})-(\ref{Eqn::X_over_R}), and substituting $\gamma = \sqrt{2/(1+\beta)}$ (see Equation~(\ref{Eqn::gamma})), the expression for $\tau_\kappa$ can be reduced to

\begin{equation}
\label{Eqn::tau_kappa}
\begin{split}
\tau_\kappa &= \frac{\pi}{\kappa} \frac{R_g}{X} \sqrt{\frac{2}{1+\gamma^2}} \frac{\Omega}{\kappa} \\
&= \frac{4\pi}{\kappa} \sqrt{\frac{2(1+\beta)}{3+\beta}} \frac{1}{\sqrt{2(1+\beta})} \\
&= \frac{4\pi}{\Omega\sqrt{2(1+\beta)}} \frac{1}{\sqrt{3+\beta}}. \\
\end{split}
\end{equation}
That is, the time-scale on which epicyclic oscillations make a significant contribution to cloud evolution increases weakly as the rotation curve flattens (towards low $\beta$). This is due to the reduction in epicyclic frequency $\kappa$ with decreasing $\beta$ at fixed $\Omega$, which increases the number of epicyclic oscillations performed per unit time, and per unit distance travelled around the galactic centre by a cloud on a circular, non-epicyclic orbit. This effect balances the increasing circumference of each epicycle with decreasing $\beta$. Like the time-scales previously derived, $\tau_\kappa$ is inversely dependent on $\Omega$, due to the linear correspondence between $\kappa$ and $\Omega$ for fixed $\beta$. It is also independent of $Q$, having nothing to do with gravitational stability.

\subsection{Shear within the galaxy ($\tau_{\beta}$)}
\label{Sec::tau_beta}
The shear time-scale $\tau_\beta$ is the time-scale on which a cloud is pulled apart by differential rotation, thus limiting the available growth time of gravitational instabilities~\citep{Goldreich1965,Elmegreen1987}. It is given by the inverse Oort constant, such that

\begin{equation}
 \tau_{\beta} = \frac{2}{\Omega(1-\beta)}.
 \label{Eqn::tau_beta}
\end{equation}
If the shear time-scale is comparable to or shorter than the gravitational free-fall time-scale $\tau_{\rm ff,g}$, the separation of radially correlated gas reduces the rate at which overdensities form, as well as slowing their subsequent collapse, thus increasing the cloud lifetime. If the shear time-scale is shorter than the gravitational free-fall time-scale $\tau_{\text{ff,g}}$, shear may disperse clouds completely, or break them up into smaller entities~\citep{Dobbs2011,Dobbs2013b}. This scenario also increases the cloud lifetime according to the statistical method for measuring cloud lifetimes from~\cite{Kruijssen2014} and~\cite{Kruijssen2018b}, which adds up the total time a Lagrangian mass element spends as a cloud, prior to star formation. That is, if a cloud is dispersed by shear and later reforms, the durations of both phases are added to calculate the observed lifetime. Galactic shear therefore works against the formation of collapsing overdensities in the ISM to generally elongate the lifetimes of GMCs. Shearing is strongest at high degrees of differential rotation (low $\beta$) and at high values of the angular speed $\Omega$, such that $\tau_\beta$ displays the same $\Omega^{-1}$ dependence as all other time-scales derived in this section.

\section{Comparison of time-scales} \label{Sec::Comptime-scales}
\label{Sec::ParamSpace}
\begin{table}
\begin{center}
\label{Tab::Params}
  \caption{Dynamic ranges for the parameters in Section~\ref{Sec::time-scales}, listed in the first column. The second column gives the parameter ranges explored in this work (corresponding to the black vertical bars in Figure~\protect\ref{Fig::DynamicRanges}). The choice of variable range for each of the other parameters is justified in Section~\ref{Sec::VariableRanges}. The third column gives the fiducial values of each parameter corresponding to the open circles in Figure~\ref{Fig::DynamicRanges}, and the literature references for these values are given in the fourth column. References: (1)~\protect\cite{KrumholzMcKee2005}, (2)~\protect\cite{Tan2000}, (3)~\protect\cite{Gerhard2011}.}
  \begin{tabular}{@{}l c c c c@{}}
  \hline
   Variable & Dynamic range & Fiducial value & References \\
  \hline
   $\beta$ & $[0,1)$ & $0.5$ & - \\
   $\Omega_{\rm P}/\Omega$ & $[0.01,8]$ & $2$ & (3) \\
   $m$ & $1,2,4$ & $1,2,4$ & - \\
   $Q$ & $[0.5,15]$ & $1.3$ &  (1) \\
   $\phi_P$ & $[1,9]$ & $3$ & (1) \\
   $f_G$ & $0.5$ & $0.5$ & (2) \\
  \hline
\end{tabular}
\end{center}
\end{table}

\subsection{Variable ranges} \label{Sec::VariableRanges}
We now turn to a comparison of the time-scales derived in Section~\ref{Sec::time-scales}. A crucial insight in this regard is that all time-scales carry an inverse dependence on the orbital speed $\Omega$, such that this quantity is irrelevant for the relative importance of the different time-scales, and acts as an overall normalisation of the cloud lifetime. Throughout large parts of this paper, we will therefore express the time-scales in the dimensionless form $\tau/\Omega^{-1}$. As discussed in Sections~\ref{Sec::tau_ffg} and~\ref{Sec::tau_ffcl}, we also omit the parameters $\alpha_{\rm vir}$ and $\phi_{\bar{P}}$ that are associated with the free-fall time-scale $\tau_{\rm ff,cl}$ for globally-collapsing, gravitationally-bound clouds of size equal to the Toomre scale. We opt instead to work with the more general time-scale $\tau_{\rm ff,g}$ for the gravitational collapse of the ISM, to quantify the influence of gravity on GMCs that do not necessarily adhere to these assumptions. The parameter space over which $\tau/\Omega^{-1}$ varies is then spanned by the six variables $\beta$, $Q$, $\Omega_{\rm P}/\Omega$, $m$, $\phi_P$ and $f_G$. Out of these, we fix $f_G$ to its fiducial value of $f_G=0.5$, and both $m$ and $\Omega_{\rm P}/\Omega$ are relevant only in the presence of spiral arms. The fundamental parameter space is therefore spanned by $\beta$ and $Q$, with a secondary dependence on $\phi_P$. This parameter space is extended with $m$ and $\Omega_{\rm P}/\Omega$ when considering spiral galaxies.

\begin{figure}
 \includegraphics[width=\linewidth]{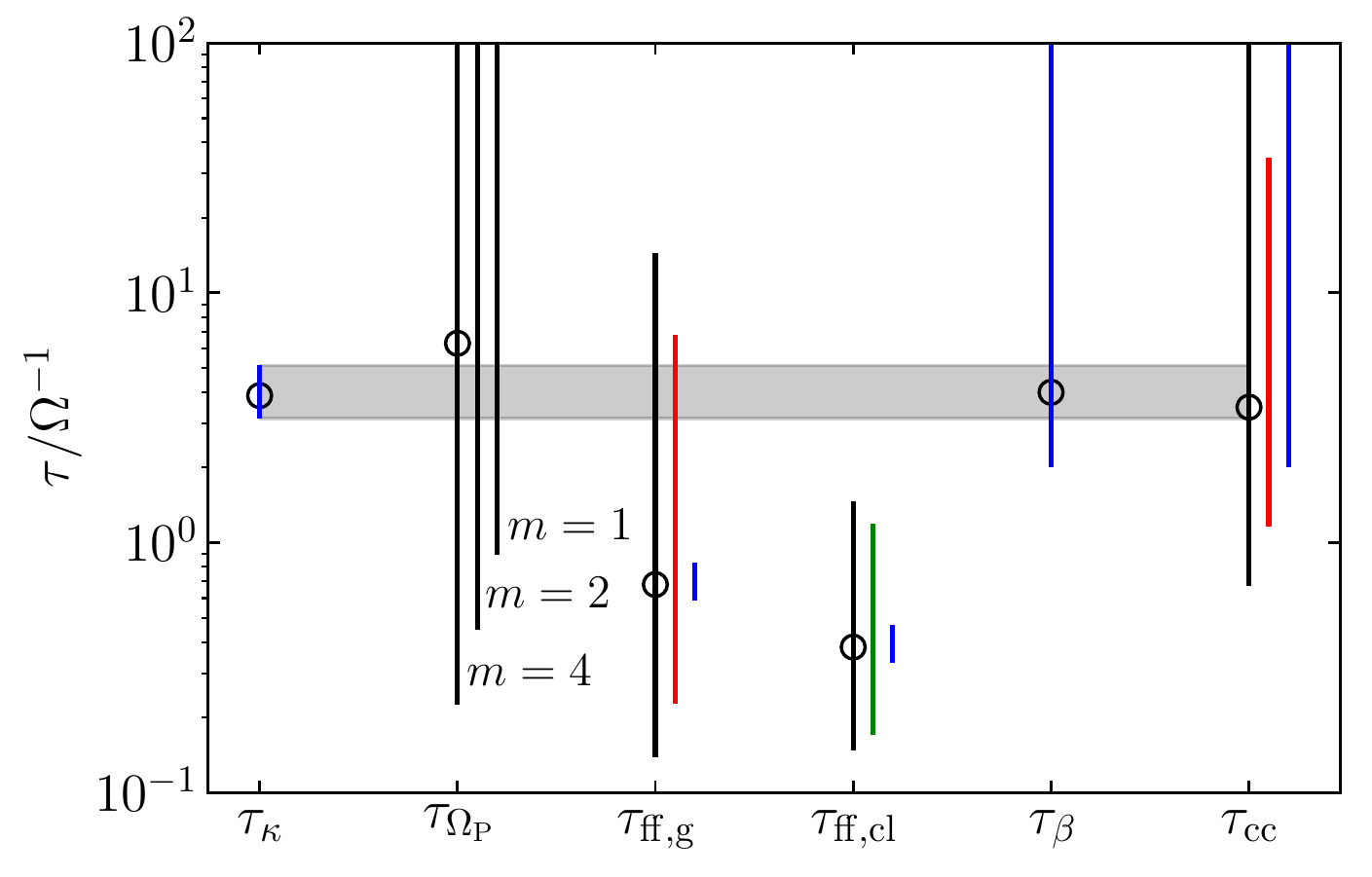}
 \caption{Dynamic ranges of each time-scale for the ranges of physical quantities given in Table~\ref{Tab::Params}, normalised by the orbital period $\Omega^{-1}$. Black bars represent the total dynamic ranges of each time-scale, red bars represent the dynamic ranges due to varying Toomre $Q$ alone, and blue bars represent the dynamic ranges due to varying the shear parameter $\beta$ alone. The green bar represents the dynamic range of $\tau_{\text{ff,cl}}$ due to simultaneously varying all three of the parameters $\alpha_{\text{vir}}$, $\phi_P$ and $\phi_{\bar{P}}$. The black open circles indicate the fiducial values from the literature, listed in the third column of Table~\ref{Tab::Params}. The grey shaded region demonstrates that there exists a region of overlap for the dynamic ranges of all time-scales other than $\tau_{\text{ff,cl}}$, which we omit in favour of $\tau_{\text{ff,g}}$ (see Sections~\ref{Sec::tau_ffg} and~\ref{Sec::tau_ffcl}). The non-zero width of this grey shaded region is a necessary but not sufficient condition that, for at least one choice of quantities in Table~\ref{Tab::Params}, any one of these five relevant time-scales might be shorter than all others.}
 \label{Fig::DynamicRanges}
\end{figure}

The importance of the fundamental plane $(\beta,Q,\Omega)$ in determining cloud evolution and the cloud lifetime is demonstrated visually in Figure~\ref{Fig::DynamicRanges}. The black vertical bars represent the total dynamic ranges of the five cloud evolutionary time-scales used in this work, plus $\tau_{\rm ff,cl}$, due to simultaneously varying all the quantities in Table~\ref{Tab::Params} (other than $f_G$, which is fixed). The blue bars represent the dynamic ranges for each time-scale due to variation in $\beta$ alone, while the red bars represent the dynamic ranges for each time-scale due to variation in $Q$ alone. We see that the time-scales $\tau_\kappa$, $\tau_\beta$, and $\tau_{\rm cc}$ are completely determined by the fundamental plane variables $(\beta,Q,\Omega)$. The time-scale $\tau_{\rm ff,g}$ is additionally controlled by $\phi_P$, but to a much weaker extent than $Q$, as demonstrated by comparing the the sizes of its black and red bars in Figure~\ref{Fig::DynamicRanges}. Thus the time-scales in galaxies without spiral arms vary primarily within the fundamental plane $(\beta,Q,\Omega)$, with secondary variations in $\phi_P$. The only time-scale that is not accounted for by $(\beta,Q,\Omega)$ is the spiral arm crossing time-scale $\tau_{\Omega_{\rm P}}$, which depends instead on the quantities $m$ and $\Omega_{\rm P}/\Omega$ in the extended parameter space $(\beta,Q,\Omega,m,\Omega_{\rm P}/\Omega)$.

One of the goals of our theory is to cover the entire parameter space of galactic properties which may feasibly be observed. In order to perform our first comparison of cloud evolutionary time-scales, we therefore choose upper and lower limits for the dynamic ranges in Table~\ref{Tab::Params} that correspond to the highest and lowest observed values of these parameters. Natural upper and lower limits on the shear parameter $\beta$ are defined by solid body rotation and a flat rotation curve, respectively. The dynamic range for $\Omega_{\rm P}/\Omega$ is calculated using the pattern speed for the Milky Way, $\Omega_{\rm P} = 0.026 \pm 0.002 {\rm Myr}^{-1}$~\citep{Gerhard2011}, and a reasonable range of angular speeds $\Omega$ between $\log{(\Omega/{\rm Myr}^{-1})} = -2.5$ and $\log{(\Omega/{\rm Myr}^{-1})} = 0.5$. We choose an upper limit for the Toomre $Q$ parameter of the ISM according to the work of~\cite{Leroy2008}, which shows that values of $Q>10$ may be readily obtained in both spiral and dwarf galaxies, and may be as large as $Q \approx 15$, a range which is matched in observations of the central $\sim 500$~pc of the Milky Way by~\cite{Kruijssen2014b}. Our lower limit for the Toomre $Q$ parameter is extended to $Q<1$ according to the work of~\cite{Genzel2014}, who find Toomre $Q$ values as low as $Q \approx 0.2$ for a sample of 19 main-sequence star-forming galaxies. We will see later in the paper that the strong influence of gravitational free-fall at low $Q$ makes it uninteresting to look at values of $Q<0.5$, so we adopt a range of $0.5<Q<15$. \cite{KrumholzMcKee2005} estimate a value of the stellar contribution to the midplane gas pressure, $\phi_P \sim 3$, by considering a variety of cases from normal disc galaxies to starbursts. Since the lowest possible value of $\phi_P$ is unity (no stellar contribution to midplane gas pressure), we adopt a conservative range of $1<\phi_P<9$.

We emphasise that our theory is able to deal with values of $Q$, $\beta$, $\phi_P$, $m$ and $\Omega_{\rm P}/\Omega$ which fall outside the extended parameter space defined here. The analytic time-scales derived in Section~\ref{Sec::time-scales} do not impose any restrictions on the values of these quantities, and so we are free to substitute any physically-reasonable, observed values that we choose. The boundaries defined here represent a physically and observationally motivated window through which we can view our parameter space, to draw conclusions about the trends in cloud evolutionary mechanisms and the cloud lifetime for the vast majority of real-Universe galactic environments. Our theory is also sufficiently flexible that we can extend the window at any time, to account for future observations.

\subsection{Preliminary comparison of time-scales}
In Figure~\ref{Fig::DynamicRanges}, we make a preliminary comparison between the dynamic ranges of all six time-scales, represented by vertical black bars, over all values of the quantities in Table~\ref{Tab::Params}. In the initial comparative discussion carried out here, we assume that the value of a given time-scale can be adjusted without altering the values of the other time-scales, or equivalently that no two time-scales are dependent upon the same physical quantities. This assumption, although incorrect, gives a preliminary indication of whether the coexistence of cloud evolutionary mechanisms is an important consideration in theories of the cloud lifetime. Where the dynamic ranges of two time-scales in Figure~\ref{Fig::DynamicRanges} overlap, it is {\it possible} that there exist values of the quantities in Table~\ref{Tab::Params} for which these time-scales are equal. Conversely, if the ranges of two time-scales do not overlap in Figure~\ref{Fig::DynamicRanges}, then there exist {\it no values} of the quantities in Table~\ref{Tab::Params} for which equality between these time-scales is obtained. This would indicate that a subset of the mechanisms does not govern the cloud lifecycle anywhere in parameter space.

For the fiducial values of the physical quantities in Table~\ref{Tab::Params}, represented by open circles in Figure~\ref{Fig::DynamicRanges}, the two time-scales for gravitational collapse $\tau_{\text{ff,g}}$ and $\tau_{\text{ff,cl}}$ are up to an order of magnitude lower than all other time-scales. For these values it is clear that gravitational collapse dominates the formation and evolution of molecular clouds. The dynamic range of $\tau_{\rm ff,cl}$ in particular does not overlap with the dynamic range of the shear time-scale $\tau_\beta$ or the time-scale of epicyclic perturbations $\tau_\kappa$, indicating that no choice of values from Table~\ref{Tab::Params} allows significant competition between $\tau_{\rm ff,cl}$ and either of these time-scales. However, in Sections~\ref{Sec::tau_ffg} and~\ref{Sec::tau_ffcl} we have justified the use of $\tau_{\rm ff,g}$ rather than $\tau_{\rm ff,cl}$ to quantify the effect of gravitational collapse on GMCs. In this case, Figure~\ref{Fig::DynamicRanges} shows a region of overlap between the dynamic range for gravitational free-fall $\tau_{\rm ff,g}$ and the dynamic ranges of {\it all} other time-scales. This is indicated by the grey shaded region in Figure~\ref{Fig::DynamicRanges}. It is therefore possible that all time-scales of cloud evolution have comparable values in some region of our extended parameter space. Coexistence between these mechanisms is potentially an important factor in determining the cloud lifetime, which requires further investigation. In the rest of this paper, we will examine the coexistence of different cloud evolutionary mechanisms in much greater depth, taking into account that the quantities in Table~\ref{Tab::Params} may affect several different time-scales simultaneously. We will also examine the nature of coexistence, and in particular the potential for time-scales of comparable magnitude to augment, or compete against, each other.

\subsection{Variation of time-scales with $\beta$}
\label{Sec::beta_Q_variation}
\begin{figure*}
 \includegraphics[width=\linewidth]{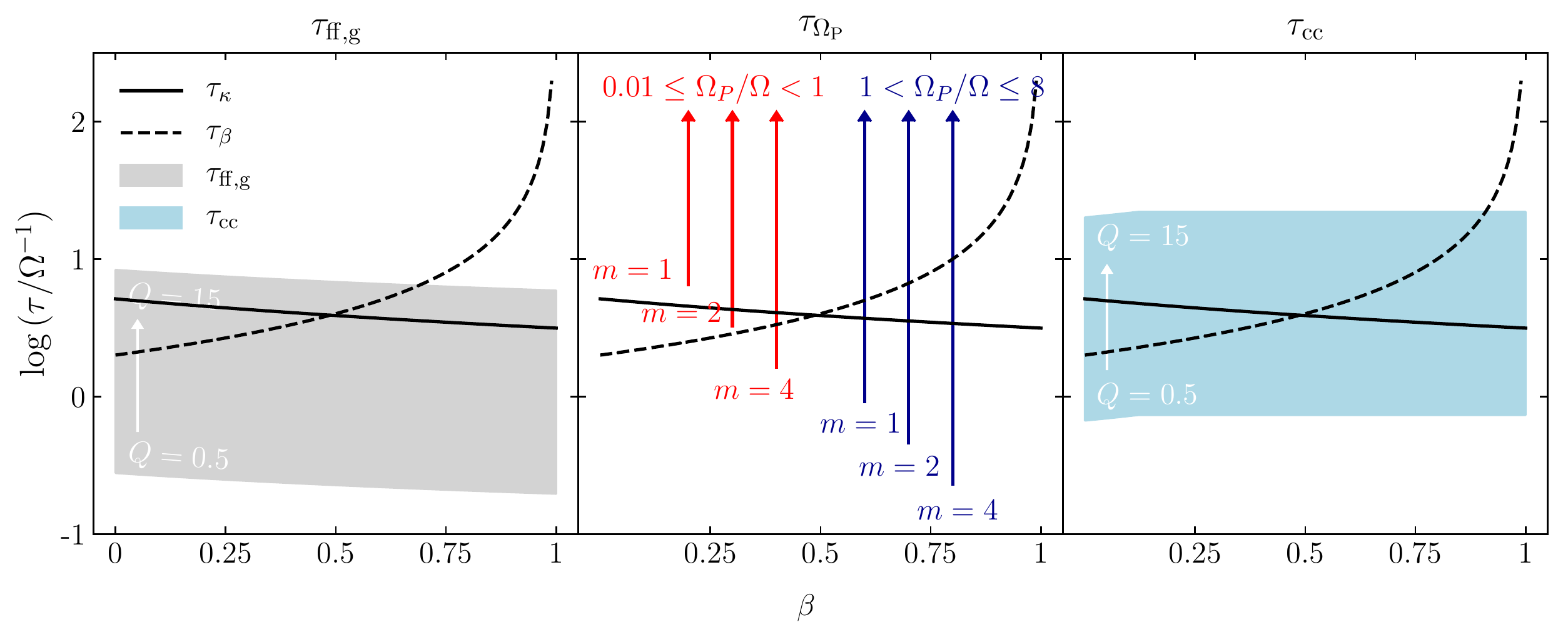}
 \caption{Variation of each time-scale, normalised by the orbital period $\Omega^{-1}$, with the shear parameter $\beta$. The two time-scales $\tau_{\kappa}$ (radial perturbations) and $\tau_{\beta}$ (shear) are dependent only on $\beta$ and are therefore represented by lines in each panel. Left: the grey region highlights the range of values for the free-fall time-scale $\tau_{\text{ff,g}}$ over the range of $Q$ values given in Table~\ref{Tab::Params}, with $\phi_P=3$. For high values of $Q$ we see that $\tau_{\beta}$ and $\tau_{\kappa}$ become the dominant time-scales relative to $\tau_{\rm ff,g}$. Centre: the red arrows indicate the range of values for the arm-crossing time-scale $\tau_{\Omega_{\rm P}}$ within the radius of co-rotation of the Milky Way~\protect\citep[$\approx 8$ kpc][]{Gerhard2011}, while the blue arrows indicate the range of $\tau_{\Omega_{\rm P}}$ outside the radius of co-rotation. The variable $m$ indicates the number of spiral arms in the galaxy. At the radius of co-rotation this time-scale becomes infinite, because the spiral arms do not move relative to the ISM. Note that $\tau_{\Omega_{\rm P}}$ is independent of $\beta$, so the horizontal placement of the  arrows is arbitrary. Right: the blue region highlights the range of values for the cloud-cloud collision time-scale $\tau_{\text{cc}}$ over the extended range of $Q$ values. This region appears independent of $\beta$ for $\beta>0.12$ because Equation~(\ref{Eqn::tau_cc}) is only valid within the range $0<\beta<0.12$ and thus we take a lower bound on $\tau_{\text{cc}}$ for $\beta>0.12$ (see Section~\ref{Sec::tau_cc}). Although we see an overlap between the blue and grey regions on the left and right, $\tau_{\text{cc}}$ is in fact always larger than $\tau_{\text{ff,g}}$, as they both depend linearly on $Q$.}
 \label{Fig::betaPlots}
\end{figure*}

As a first step towards a full analysis of all five cloud evolutionary time-scales throughout our entire parameter space, we present here a comparison of these time-scales as a function of $\beta$, the most prevalent physical quantity in our models. Each panel in Figure~\ref{Fig::betaPlots} displays the time-scale $\tau_\kappa$ for epicyclic perturbations (solid black lines) and the time-scale $\tau_\beta$ for shear (dashed black lines), both of which depend {\it only} on $\beta$. We see that these time-scales intersect at around $\beta=0.5$, indicating that for $\beta<0.5$, galactic shear exerts more control over cloud evolution than do epicyclic perturbations, but that for $\beta>0.5$, the opposite is true.

The left-hand panel of Figure~\ref{Fig::betaPlots} compares the dynamic ranges of $\tau_\beta$ and $\tau_\kappa$ to the dynamic range of the time-scale $\tau_{\rm ff,g}$ for gravitational free-fall, which is additionally controlled by the Toomre $Q$ parameter. The variation in $\tau_{\rm ff,g}$ with $Q$ is indicated by the grey filled region, where higher values of $Q$ correspond to higher values of $\tau_{\rm ff,g}$. The variation of $\tau_{\rm ff,g}$ with $\phi_P$ has been neglected in this figure, due to its small effect relative to that of the Toomre $Q$ parameter. For $Q \la 4$, gravitational free-fall is more influential than either $\tau_\beta$ or $\tau_\kappa$ over all values of $\beta$, but its value becomes comparable with these time-scales for $Q \ga 4$. For $Q \sim 15$, gravitational free-fall is the least influential mechanism of the three except at $\beta \rightarrow 1$, where $\tau_\beta$ diverges asymptotically. We can conclude that across the window of $(\beta,Q,\Omega)$ parameter space defined in Section~\ref{Sec::ParamSpace}, there exist points for which each of the three time-scales $\tau_\beta$, $\tau_\kappa$ and $\tau_{\rm ff,g}$ is more influential than the other two. It is therefore necessary to consider all three of these time-scales and their contributions to the cloud lifetime.

In the right-hand panel of Figure~\ref{Fig::betaPlots}, we see that the time-scale $\tau_{\rm cc}$ for cloud-cloud collisions (blue shaded region) is shorter than $\tau_\beta$ and $\tau_\kappa$ at all $\beta$ for $Q \la 2$, but longer than these time-scales for most $\beta$ at $Q \ga 2$. Looking at Figure~\ref{Fig::betaPlots} alone, we conclude that when calculating the cloud lifetime, the contributions of all three time-scales must be considered. However, a comparison between the left-hand and right-hand panels of Figure~\ref{Fig::betaPlots} hints that the influence of cloud-cloud collisions might always be less significant than the influence of gravitational free-fall. Both of the time-scales $\tau_{\rm ff,g}$ and $\tau_{\rm cc}$ have the same functional dependence on the Toomre $Q$ parameter (see Equations~\ref{Eqn::tau_ffg} and~\ref{Eqn::tau_cc}), such that the grey and blue regions in Figure~\ref{Fig::betaPlots} have equal areas in the space of $(\beta,\log{\tau/\Omega^{-1}})$. As $Q$ is decreased, both time-scales decrease monotonically, and since $\tau_{\rm cc}>\tau_{\rm ff,g}$ for both $Q=0.5$ (lower limits of blue and grey shaded regions) and $Q=15$ (upper limits of blue and grey shaded regions), this implies that cloud-cloud collisions can {\it never} be more influential than gravitational collapse. In the following section, we will perform a more detailed comparison between time-scales, taking the $\phi_P$-dependence of $\tau_{\rm ff,g}$ into account, and will show that the influence of cloud-cloud collisions on the cloud lifetime is indeed limited by its inability to compete with gravitational free-fall.

The centre panel of Figure~\ref{Fig::betaPlots} presents the dynamic range of the time-scale $\tau_{\Omega_{\rm P}}$ for spiral arm crossings, for the extended parameter space $(\beta,Q,\Omega,m,\Omega_{\rm P}/\Omega)$. We have used the pattern speed $\Omega_{\rm P}$ for the Milky Way in this figure. Since $\tau_{\Omega_{\rm P}} \rightarrow \infty$ at the radius of co-rotation ($\Omega_{\rm P}/\Omega=1$), the dynamic range for each value of $m$ has no upper limit, so is represented by arrows. Red arrows indicate a dynamic range below the radius of co-rotation ($\Omega_{\rm P}/\Omega<1$), while blue arrows indicate a dynamic range above the radius of co-rotation ($\Omega_{\rm P}/\Omega>1$). In general, the frequency with which a cloud encounters spiral arms is multiplicative in $m$, such that the lower limit on the time-scale $\tau_{\Omega_{\rm P}}$ is brought to lower and lower values as $m$ is increased. There are clearly values of $(\beta,\Omega,m,\Omega_{\rm P}/\Omega)$ for which each of the time-scales $\tau_\beta$, $\tau_\kappa$ and $\tau_{\Omega_{\rm P}}$ has a greater influence on cloud evolution than the other two time-scales. In addition, $\tau_{\Omega_{\rm P}}$ is independent of $Q$, so we can infer from the overlap of the red and blue arrows with the grey and blue shaded regions that there exist values of $(\beta,Q,\Omega,m,\Omega_{\rm P}/\Omega)$ for which $\tau_{\Omega_{\rm P}} = \tau_{\rm ff,g}$ or $\tau_{\Omega_{\rm P}} = \tau_{\rm cc}$. It is therefore clear that, in addition to considering the contributions of $\tau_\kappa$, $\tau_\beta$, $\tau_{\rm ff,g}$ and $\tau_{\rm cc}$ to the cloud lifetime, we should take into account the effect of spiral arm crossings on time-scale $\tau_{\Omega_{\rm P}}$.

In summary, Figure~\ref{Fig::betaPlots} demonstrates that, over the ranges of physical parameters displayed in Table~\ref{Tab::Params}, equality between all pairs of time-scales can be obtained, with the possible exception of equality between $\tau_{\rm ff,g}$ and $\tau_{\rm cc}$. In the next section, we will confirm that there are non-negligible regions of parameter space for which all five cloud evolutionary mechanisms have comparable time-scales, and all contribute to setting the cloud lifetime.

\section{Cloud lifetimes throughout parameter space: the dominance and coexistence of cloud evolutionary mechanisms}
\label{Sec::Coexistence}
Here we establish the procedure by which cloud lifetimes will be characterised in our theory, throughout the parameter space of observable quantities defined in Table~\ref{Tab::Params}. We begin by introducing an equation to quantify the cloud lifetime, using the time-scales for cloud evolution derived in Section~\ref{Sec::time-scales}. We apply this equation first to the simplest case of clouds in galaxies with no spiral arms ($m=0$), in the $(\beta,Q,\Omega)$ fundamental plane. We present figures of the minimum time-scale of cloud evolution throughout this plane, and use these figures to determine the regions of parameter space for which each time-scale is {\it dominant} over all others in setting the cloud lifetime. We then present figures of the cloud lifetime throughout the $(\beta,Q,\Omega)$ fundamental plane, in which we determine the regions of {\it relevance} for each mechanism. This tells us where different cloud evolutionary mechanisms coexist. Having developed the machinery of our theory for the simplest case, we finally extend our formalism to encompass the extended parameter space $(\beta,Q,\Omega,m,\Omega_{\rm P}/\Omega)$ appropriate to galaxies with spiral arms.

\subsection{Calculation of cloud lifetime}
Our theory is expansive, making as few assumptions as possible about the size, structure and gravitational boundedness of molecular clouds. It does not aim to capture the precise way in which GMCs are affected by the five dynamical cloud evolutionary mechanisms presented in Section~\ref{Sec::time-scales}. Instead, we calculate the cloud lifetime by characterising each mechanism according to its rate $\tau^{-1}$.

In combining the five rates of cloud evolution, we make the assumption that galactic shear is primarily a dynamically dispersive process, while the other four mechanisms are dynamically compressive. This categorisation of evolutionary mechanisms is based on the effect of each on gravitational collapse and star formation. An increase in the level of galactic shear is found to lower the efficiency of star formation~\citep{Leroy2008}, the majority of which occurs within dense, collapsing regions of molecular gas~\citep{Hartmann2001,Elmegreen2007,Dobbs2011,Dobbs2013b}. This suggests that galactic shear suppresses star formation by supporting against gravitational free-fall. Dynamically, this is an intuitive result, because gravitational collapse causes molecular gas to increase its density, while galactic shear disrupts the structure. By contrast, the other four cloud evolutionary mechanisms, characterised by the time-scales $\tau_{\rm ff,g}$, $\tau_{\Omega_{\rm P}}$, $\tau_\kappa$ and $\tau_{\rm cc}$, each have the potential to compress molecular gas and thus to promote gravitational collapse and star formation. We therefore add together the four rates $\tau_{\rm ff,g}^{-1}$, $\tau_{\Omega_{\rm P}}^{-1}$, $\tau_\kappa^{-1}$ and $\tau_{\rm cc}^{-1}$, but {\it subtract} the rate of galactic shear, $\tau_\beta$. In doing this, we implicitly assume that the destruction of molecular clouds by stellar feedback occurs on a time-scale that is much shorter than the five dynamical time-scales presented here, such that the processes of compression, star formation and cloud destruction are all quantified within each of the time-scales $\tau_{\rm ff,g}$, $\tau_{\Omega_{\rm P}}$, $\tau_\kappa$ and $\tau_{\rm cc}$. The addition of dynamical rates gives a cloud lifetime of 

\begin{equation}
\tau = |\tau_{\kappa}^{-1}+\tau_{\Omega_{\rm P}}^{-1}+\tau_{\text{ff,g}}^{-1}+\tau_{\text{cc}}^{-1}-\tau_{\beta}^{-1}|^{-1},
\label{Eqn::Lifetime}
\end{equation}
where we take the absolute value of the sum of the rates, so that the cloud lifetime is still positive if shear outpaces all other mechanisms of cloud evolution. We will show below that this only occurs in a very small part of parameter space.

This method of calculating the cloud lifetime makes two major assumptions. Firstly, it ignores non-linearities arising from interactions between the different cloud evolutionary mechanisms. It assumes that the rates of the different mechanisms can simply be added together. Secondly, Equation~(\ref{Eqn::Lifetime}) is statistical in the sense that it combines rates. At each point in parameter space, it represents the ensemble-averaged value of the cloud lifetime for a theoretically-infinite population of clouds. It therefore ignores the {\it discrete} and {\it random} (i.e.~Poissonian) nature of cloud-cloud collisions and spiral arm crossings. In practice, these events may occur at any time between the `birth' of a molecular cloud and the value of the time-scale $\tau_{\rm cc}$ or $\tau_{\Omega_{\rm P}}$.

The major advantage of taking this simplified approach is that it allows us to introduce an analytic expression for the cloud lifetime in terms of observable, physical quantities, and to systematically examine its behaviour throughout the parameter space $(\beta,Q,\Omega,\phi_P,m,\Omega_{\rm P}/\Omega)$. The use of cloud evolutionary rates means that we do not need to define a time of cloud `birth' and therefore do not require an arbitrary theoretical threshold between states of cloud existence and non-existence. By comparing the predictions of our simple dynamical theory to observations of molecular cloud lifetimes, we will be able to assess the importance of complex effects such as non-linearity, cloud chemistry, and stellar feedback. Under the assumption that deviations from Equation~(\ref{Eqn::Lifetime}) are driven by these effects, we will be able to separate the influence of these effects from the dynamics of the ISM itself.

It could be argued that Equation~(\ref{Eqn::Lifetime}) underestimates the cloud lifetime if a large number of clouds undergo global collapse, because the rate of collapse of a cloud cannot be augmented or accelerated by other mechanisms once it becomes decoupled from the large-scale turbulent flow. However, we will see below that as $Q$ decreases and global collapse becomes more prevalent among molecular clouds, the gravitational free-fall time-scale $\tau_{\rm ff,g}$ becomes strongly dominant over the other time-scales, which therefore become irrelevant. Therefore, our model of the cloud lifetime naturally produces a situation in which the time-scales $\tau_\kappa$, $\tau_\beta$, $\tau_{\Omega_{\rm P}}$ and $\tau_{\rm cc}$ make a negligible contribution to the cloud lifetime when many clouds are undergoing global gravitational collapse.

As pointed out by~\cite{Elmegreen2007}, molecular clouds can be destroyed in a number of different ways. A cloud can undergo {\it consumption}, by which it is eventually eroded away by star formation and stellar feedback, {\it dispersal} such that it is torn apart, possibly into smaller entities, or {\it phase change} such that it is converted back into atomic form. Within our formalism, a {\it phase change} may be caused by stellar feedback, in which case it is analogous to cloud {\it consumption}.

By comparing the rate of galactic shear $\tau_\beta^{-1}$ to the combined rates of the dynamically-compressive mechanisms $\tau_\kappa^{-1}+\tau_{\rm ff,g}^{-1}+\tau_{\Omega_{\rm P}}^{-1}+\tau_{\rm cc}^{-1}$, our theory can discern the regimes in which clouds will typically undergo {\it dispersal} rather than {\it consumption}. We define the regimes ({\it s}) as those regions for which

\begin{equation}
\label{Eqn::dispersive_rate}
\tau_\beta^{-1}>\tau_\kappa^{-1}+\tau_{\Omega_{\rm P}}^{-1}+\tau_{\rm ff,g}^{-1}+\tau_{\rm cc}^{-1},
\end{equation}
and we expect the primary mechanism of cloud destruction in these regions to be related to cloud {\it dispersal}, with minimal opportunity for star formation. We define the regimes  ({\it c}) as those regions of parameter space for which

\begin{equation}
\label{Eqn::compressive_rate}
\tau_\beta^{-1}<\tau_\kappa^{-1}+\tau_{\Omega_{\rm P}}^{-1}+\tau_{\rm ff,g}^{-1}+\tau_{\rm cc}^{-1},
\end{equation}
and in these regions, we expect clouds to be destroyed mainly by {\it consumption}, due to gravitational collapse and the ensuing stellar feedback.

\subsection{Regions of dominance and relevance}
\label{Sec::dom_and_rel}
The evolution of clouds in different parts of parameter space, and thus in different galactic environments, is governed by different dynamical processes. We quantify this division of parameter space by defining {\it regions of dominance}. The {\it region of dominance} for a mechanism of cloud evolution is the region of parameter space for which its time-scale is shorter than all other time-scales.

While the mechanism with the shortest time-scale has the greatest influence on cloud evolution, other processes with comparable rates may also contribute. For this reason, we introduce the concept of {\it relevance} in addition to {\it dominance}. The threshold for {\it relevance} should be lenient enough to allow several mechanisms to be relevant at some points in parameter space, but strict enough that only one mechanism is relevant in others. In order to quantify {\it relevance}, we begin by dividing the parameter space into two regimes, ({\it i}) and ({\it ii}).
\begin{enumerate}
\item Lifetime $\tau <$ minimum time-scale $\tau_{\text{min}}$: The minimum time-scale is augmented by other compressive time-scales, resulting in a cloud lifetime that is shorter than the minimum time-scale.
\item Lifetime $\tau >$ minimum time-scale $\tau_{\text{min}}$: Competition between shear support and the compressive time-scales extends the lifetime beyond the minimum time-scale.
\end{enumerate}

In regime ({\it i}), the relevance of a given time-scale should be determined by comparison to $\tau_{\text{min}}$, as $\tau_{\text{min}}$ sets the upper bound on the cloud lifetime. Any mechanism that is not competitive with $\tau_{\text{min}}$ will not have time to appreciably influence the cloud's evolution before the end of its life. Conversely, in regime ({\it ii}) the relevance of a given time-scale should be determined by comparison to the lifetime itself, $\tau$. The extension of cloud lifetime via shear support means that even if a given mechanism occurs at a substantially slower rate than $\tau_{\text{min}}$, it can have an influence on the evolution of the cloud, although this influence will be small relative to $\tau_{\text{min}}$.

Within these two regimes, the {\it regions of relevance} are defined by computing the ratio of each time-scale with either the minimum time-scale in regime ({\it i}) or the cloud lifetime in regime ({\it ii}). Where this ratio is smaller than $2$, the time-scale is deemed to be relevant. That is, the regions of relevance for a cloud evolutionary mechanism in regime ({\it i}) are those regions for which its rate is no less than half the rate of the dominant cloud evolutionary mechanism. In regime ({\it ii}), the regions of relevance are those regions for which a given cloud evolutionary mechanism occurs at a rate no less than half the sum of all cloud evolutionary rates, as on the right-hand side of Equation~(\ref{Eqn::Lifetime}).

Since several different cloud evolutionary mechanisms may be {\it relevant} in each point in parameter space, the {\it regions of relevance} reveal the environmental conditions for which cloud evolution is either controlled by a single dynamical process, or is controlled by a combination of dynamical processes that coexist.

\subsection{Galaxies without spiral arms, $m=0$}
\label{Sec::m0}
\begin{figure*}
 \includegraphics[width=\linewidth]{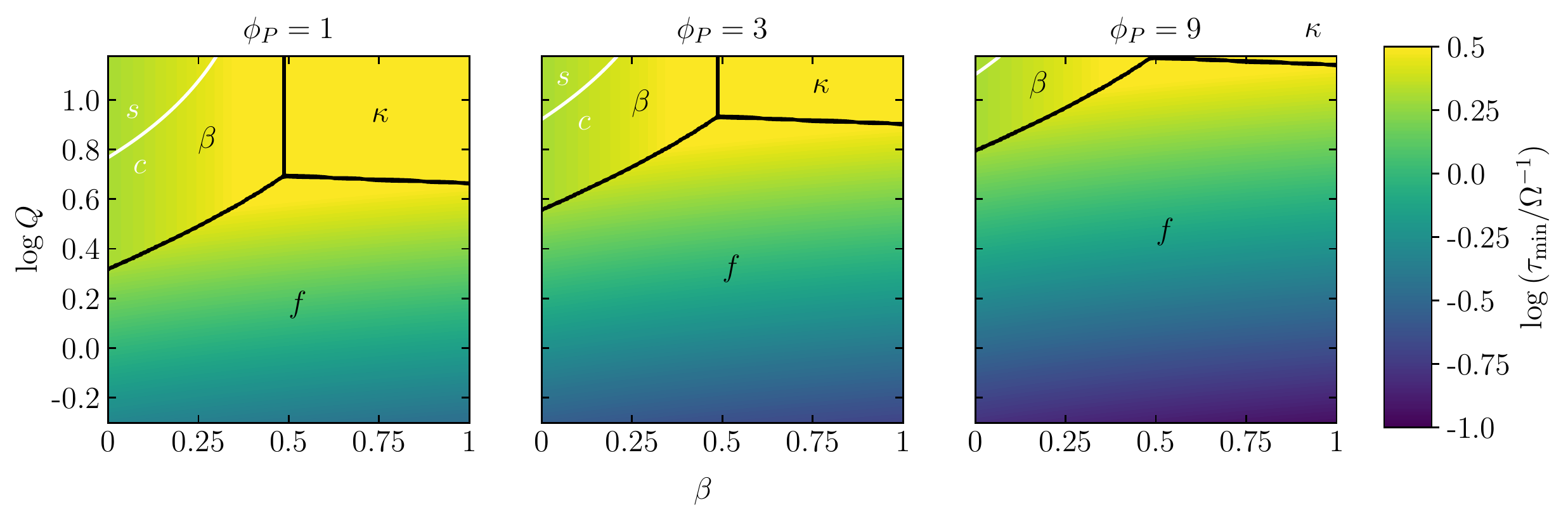}
 \caption{The minimum time-scale and its value for three cross-sections of the parameter space $(\beta,\log{Q},\Omega,\phi_P)$, where $\phi_P$ is given by Equation~(\ref{Eqn::phi_rho_defn}) and no spiral arms are present ($m=0$). The dependence on $\Omega$ is included as a normalisation of the time-scale, as all evolutionary time-scales depend on $\Omega$ in the same way. The dominant time-scales here are $\tau_{\kappa}$, $\tau_{\beta}$ and $\tau_{\text{ff,g}}$, denoted by $\kappa$, $\beta$ and $f$ respectively. The solid black lines delineate boundaries along which two time-scales are equal; the \textit{regions of dominance} are separated by these lines. The solid white lines divide the regions for which the rate of galactic shear is higher than the combined rates of all other mechanisms ({\it s}) from the regions in which it is lower ({\it c}).}
 \label{Fig::QvsBeta_absmin_m0}
\end{figure*}

We begin with the case of clouds in galaxies without spiral arms, such that we may neglect the arm-crossing time-scale $\tau_{\Omega_{\rm P}}$ and its parameters $m$ and $\Omega_{\rm P}/\Omega$. The parameter space is then composed of the fundamental plane $(\beta,Q,\Omega)$ with a (weak) secondary dependence on $\phi_P$.

\subsubsection{Regions of dominance, $m=0$}
\label{Sec::m0_absmin}
We first divide the parameter space into {\it regions of dominance}, by distinguishing the mechanism with the greatest influence on cloud evolution for each region of parameter space. We determine which process occurs at the fastest rate by computing the normalised minimum cloud evolutionary time-scale $\tau_{\rm min}/\Omega^{-1}$ across all values of $\beta$, $Q$ and $\phi_P$. In Figure~\ref{Fig::QvsBeta_absmin_m0}, the value of the minimum time-scale is indicated by the colours, while the solid black lines represent divisions between the regions of dominance, along which two time-scales are equal in value.

The most gravitationally-unstable regions of parameter space at the bottom of each panel ($Q \la 4$, i.e.~gas-rich, star-forming galaxies) are dominated comprehensively by gravitational free-fall `$f$'. Conversely, regions of higher gravitational stability are dominated either by galactic shear `$\beta$' for flatter rotation curves ($Q \ga 4$ and $\beta \la 0.5$, i.e.~in early-type galaxies or ETGs and outer galactic bulges) or by epicyclic perturbations `$\kappa$' for approximately solid-body rotation ($Q \ga 4$ and $\beta \ga 0.5$, i.e.~near galactic centres). For gas that is highly gravitationally-stable with a flat rotation curve (very top left corner of each panel with $Q \sim 15$ and $\beta \sim 0$), the dynamically-dispersive mechanism of galactic shear may dominate over the combined influence of all dynamically-compressive mechanisms, such that $\tau_\beta^{-1}>\tau_{\rm ff,g}^{-1}+\tau_\kappa^{-1}+\tau_{\rm cc}^{-1}$. These regions are labelled $({\it s})$ in Figure~\ref{Fig::QvsBeta_absmin_m0}, enclosed by a solid white line. In region $({\it s})$, many clouds will be pulled apart by galactic shear before they have the chance to collapse and form stars. In region ({\it c}), dynamically-compressive evolutionary mechanisms dominate, so clouds are more likely to be destroyed by gravitational collapse and the subsequent stellar feedback.

As the stellar contribution to the midplane surface density of the ISM is increased from $\phi_P=1$ (i.e.~pure gas discs, left-hand panel of Figure~\ref{Fig::QvsBeta_absmin_m0}) to $\phi_P=9$ (i.e.~galaxies with large stellar contributions, right-hand panel of Figure~\ref{Fig::QvsBeta_absmin_m0}), the time-scale for gravitational collapse decreases in value. This causes the gravity-dominated region of parameter space `$f$' to move upwards into regions of higher gravitational stability. For a single value of the Toomre $Q$ stability parameter, an increase in the proportion of stars destabilises the gas in the galactic midplane and thus reduces the gravitational free-fall time-scale. However, it should be noted that this effect is mitigated by, and may even be reversed by, the direct correlation between $\phi_P$ and $Q$. As the stellar contribution is increased, the velocity dispersion of the midplane gas and the epicyclic frequency of the galaxy disc are also increased, stabilising the gas and thus increasing the value of $Q$. As the gas fraction is decreased, $Q$ therefore increases in proportion to $\phi_P$. As shown in Equation (\ref{Eqn::tau_ffg}), the free-fall time-scale is inversely proportional to $\sqrt{\phi_P}$, but directly proportional to $Q$. Therefore, it may actually be the case that $\tau_{\rm ff,g}$ {\it increases} with an increase in $\phi_P$ (a decrease in the gas fraction), in contrast to the implication of Figure~\ref{Fig::QvsBeta_absmin_m0}. For the remainder of the paper, we set $\phi_P = 3$, the value most appropriate to the Milky Way~\citep[e.g.][]{KrumholzMcKee2005}.

\subsubsection{Regions of relevance, $m=0$}
\label{Sec::m0_lifetimes}
\begin{figure*}
 \includegraphics[width=\linewidth]{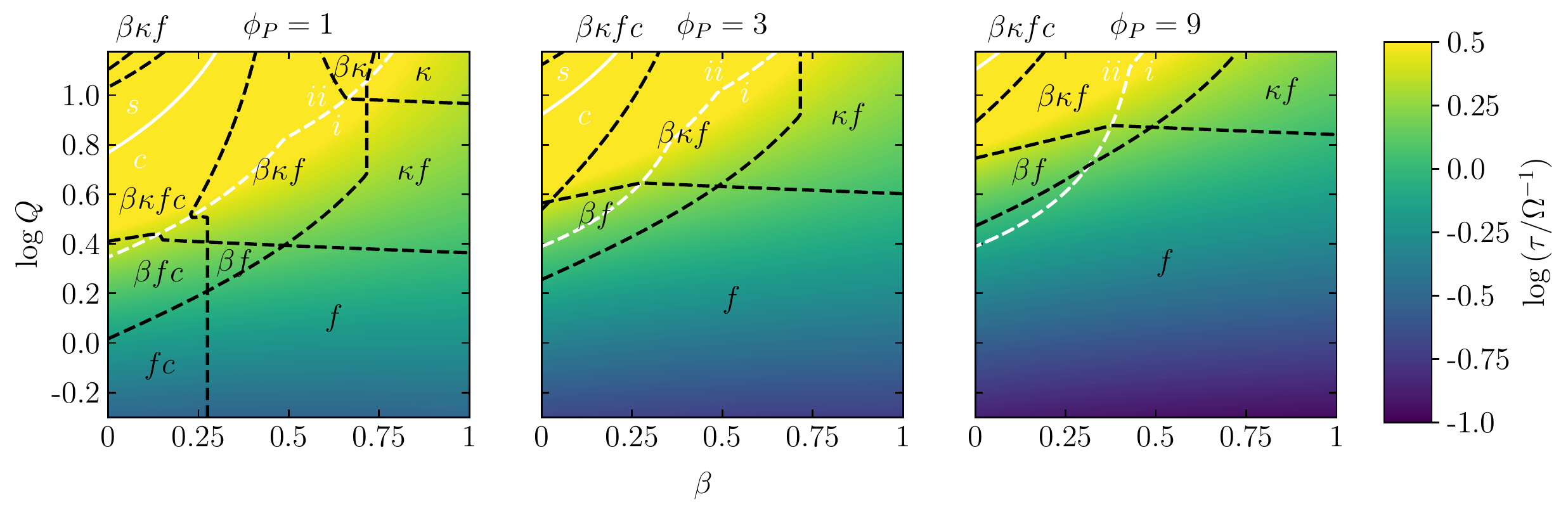}
 \caption{The predicted cloud lifetime for three cross-sections of the parameter space $(\beta,\log{Q},\Omega,\phi_P)$, without spiral arms $(m=0)$. The dependence on $\Omega$ is included as a normalisation of the time-scale, as all evolutionary time-scales depend on $\Omega$ in the same way. The relevant time-scales are $\tau_{\kappa}$, $\tau_{\beta}$, $\tau_{\text{ff,g}}$ and $\tau_{\text{cc}}$, denoted by $\kappa$, $\beta$, $f$ and $c$ respectively. The dashed black lines enclose the {\it regions of relevance} for each time-scale (see Section~\ref{Sec::dom_and_rel}). The dashed white lines divide the regions for which the cloud lifetime is longer than the minimum evolution time-scale ({\it ii}) from the regions in which it is shorter ({\it i}). The solid white lines divide the regions for which the rate of galactic shear is higher than the combined rates of all other mechanisms ({\it s}) from the regions in which it is lower ({\it c}).}
 \label{Fig::QvsBeta_lifetimes_m0}
\end{figure*}

We now examine the coexistence of cloud evolutionary mechanisms by dividing the parameter space into {\it regions of relevance}. We use Equation~(\ref{Eqn::Lifetime}) to compute the value of the normalised cloud lifetime $\tau/\Omega^{-1}$, represented by the coloured contours in Figure~\ref{Fig::QvsBeta_lifetimes_m0}, across all values of $\beta$, $Q$ and $\phi_P$. The regions of relevance throughout parameter space are enclosed by black dashed lines.

In Figure~\ref{Fig::QvsBeta_lifetimes_m0}, the bottom right side of the white dashed line in each panel corresponds to regime ({\it i}), the region of parameter space for which the cloud lifetime is shorter than the minimum time-scale of cloud evolution, $\tau < \tau_{\rm min}$. Regime ({\it i}) spans all but the top left corner of each panel, where galactic shear `$\beta$' is most relevant. This is because galactic shear is the only dynamically-dispersive cloud evolutionary process, so it is required for $\tau > \tau_{\rm min}$. Without galactic shear, the cloud lifetime results from a combination of several dynamically-compressive mechanisms, hence it has a shorter value than any one of the individual time-scales. As mentioned in Section~\ref{Sec::dom_and_rel}, we determine the relevance of each time-scale in regime ({\it i}) by comparison to $2 \times \tau_{\rm min}$.

In the top left corner of each panel of Figure~\ref{Fig::QvsBeta_lifetimes_m0}, the ISM is highly gravitationally-stable with a flat rotation curve and can therefore be effectively supported by galactic shear `$\beta$'. Such an environment would be found, for example, in outer galaxy bulges. The competition between cloud compression and shear support `$\beta f$' elongates the cloud lifetime such that it becomes longer than the minimum time-scale of cloud evolution, $\tau>\tau_{\rm min}$. This condition defines regime ({\it ii}), which is separated from regime ({\it i}) by the white dashed line. In regime ({\it ii}), the relevance of each time-scale is determined by comparison to $2 \times \tau$, as the extended cloud lifetime allows slower-acting processes to play a small role in cloud evolution before the cloud is destroyed. In region `$\beta \kappa f c$' at the very top left corner of each panel, the cloud survives so long that all mechanisms of cloud evolution have time to influence its evolution.

For gravitationally-stable regions of the ISM with $Q \ga 4$ (top half of each panel in Figure~\ref{Fig::QvsBeta_lifetimes_m0}), the coexistence of different cloud evolutionary mechanisms is of crucial importance in setting the cloud lifetime. For this region of parameter space, there is a large difference in value between the normalised cloud lifetime $\tau/\Omega^{-1}$ (colours in Figure~\ref{Fig::QvsBeta_lifetimes_m0}) and the normalised minimum time-scale of evolution (colours in Figure~\ref{Fig::QvsBeta_absmin_m0}). In the top right-hand corner of each panel (i.e.~near galactic centres), epicyclic perturbations are augmented by gravitational collapse in the region `$\kappa f$' of Figure~\ref{Fig::QvsBeta_lifetimes_m0}, reducing the cloud lifetime relative to the dominant time-scale of epicyclic perturbations (region $\kappa$ of Figure~\ref{Fig::QvsBeta_absmin_m0}). Conversely, in the top left-hand corner of each panel (i.e.~in ETGs and outer galaxy bulges) the cloud lifetime is elongated by the competition between galactic shear and all the compressive mechansisms of cloud evolution, including gravity, epicyclic perturbations and cloud-cloud collisions (e.g.~in region `$\beta \kappa f c$').

For regions of the ISM that are gravitationally-unstable ($Q \la 4$, i.e.~gas-rich, star-forming galaxies), the overwhelming domination of gravitational free-fall means that the cloud lifetime is almost equal to the minimum time-scale of cloud evolution. This can be seen by comparison of the coloured contours in Figures~\ref{Fig::QvsBeta_absmin_m0} and~\ref{Fig::QvsBeta_lifetimes_m0}.

The rarity of regions `$c$' in Figure~\ref{Fig::QvsBeta_lifetimes_m0} reveals that cloud-cloud collisions are the least competitive mechanism of cloud evolution. They are only relevant in the case of very gas-rich, highly-shearing environments ($\phi_P = 1$ and $\beta \la 0.25$, i.e.~the left side of the left-hand panel). Such environments may be found in high-redshift galaxies for low values of Toomre $Q$~\citep[e.g.][]{Genzel2014} or in the outskirts of low-redshift galaxies for high values of $Q$~\citep[e.g.][]{Leroy2008}. For higher stellar contributions, $\phi_P = 3$ and $\phi_P = 9$, cloud-cloud collisions are relevant only in the very top left corner of each panel, and only in conjunction with {\it all} other mechanisms of cloud evolution. As noted in Section~\ref{Sec::tau_cc}, the influence of cloud-cloud collisions is likely to be overestimated in this region of parameter space, so their effect is likely to be even smaller than indicated by Figure~\ref{Fig::QvsBeta_lifetimes_m0}. In general, cloud-cloud collisions are only relevant in very few circumstances due to their inability to compete with the time-scale $\tau_{\rm ff,g}$ for gravitational collapse.

\subsection{Galaxies with spiral arms, $m \neq 0$}
The introduction of spiral arms requires the introduction of parameters $m$ and $\Omega_{\rm P}/\Omega$. These refer to the number of spiral arms and the ratio of the pattern speed to the orbital speed, respectively. We show the same values of $\tau_{\rm min}/\Omega^{-1}$ and $\tau/\Omega^{-1}$ as in Section~\ref{Sec::m0}, but in the new extended parameter space $(\beta, Q, \Omega, m, \Omega_{\rm P}/\Omega)$. We examine the $(\beta, Q, \Omega)$ plane in Figures~\ref{Fig::QvsBeta_absmin} and~\ref{Fig::QvsBeta_lifetimes}, and we examine the $(\Omega_{\rm P}/\Omega, Q,\Omega)$ plane in Figures~\ref{Fig::OmegavsBeta_absmin} and~\ref{Fig::OmegavsBeta_lifetimes}. For each cross-section, we examine the cases of $m=1,2$ and $4$ spiral arms. As previously discussed, $\phi_P$ is set to its fiducial Milky Way value of $\phi_P = 3$.

\subsubsection{Regions of dominance, $m \neq 0$}
\label{Sec::spiral_absmin}

\begin{figure*}
 \includegraphics[width=\linewidth]{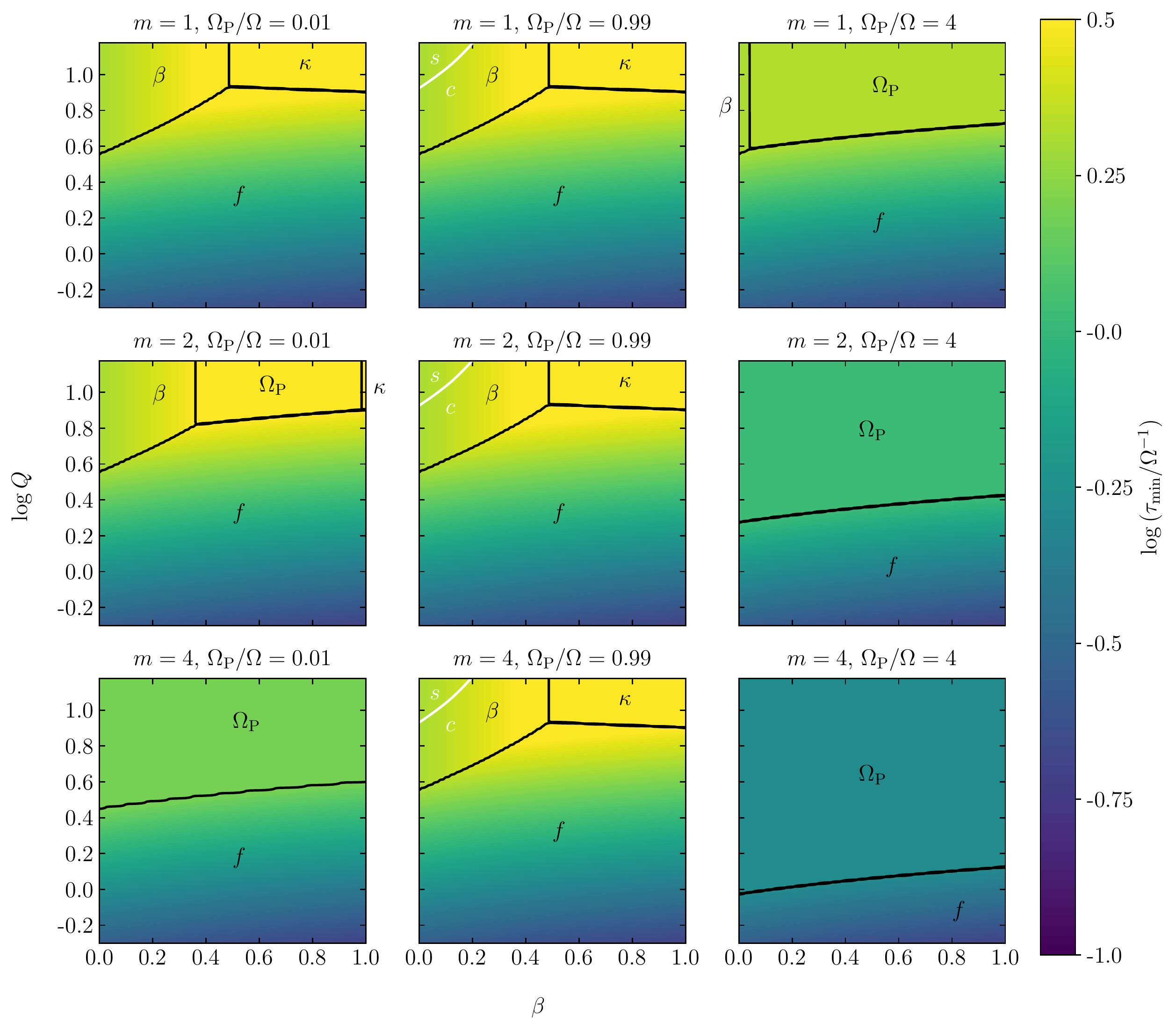}
 \caption{The minimum time-scale and its value for three cross-sections of the parameter space $(\beta,\log{Q},\Omega,\Omega_{\rm P}/\Omega)$ in the $(\beta,Q,\Omega)$ fundamental plane, for three numbers of spiral arms $m=1,2,4$ and for $\phi_P=3$. The dependence on $\Omega$ is included as a normalisation of the time-scale, as all evolutionary time-scales depend on $\Omega$ in the same way. The relevant time-scales here are $\tau_{\kappa}$, $\tau_{\beta}$, $\tau_{\Omega_{\rm P}}$ and $\tau_{\text{ff,g}}$, denoted by $\kappa$, $\beta$, $f$ and $\Omega_{\rm P}$ respectively. The solid black lines delineate boundaries along which two time-scales are equal; the \textit{regions of dominance} are separated by these lines. The solid white lines delineate the values of $Q$ and $\beta$ above which the rate of galactic shear is higher than the combined rates of all other mechanisms, as discussed in Section~\ref{Sec::beta}.}
 \label{Fig::QvsBeta_absmin}
\end{figure*}

\begin{figure*}
 \includegraphics[width=\linewidth]{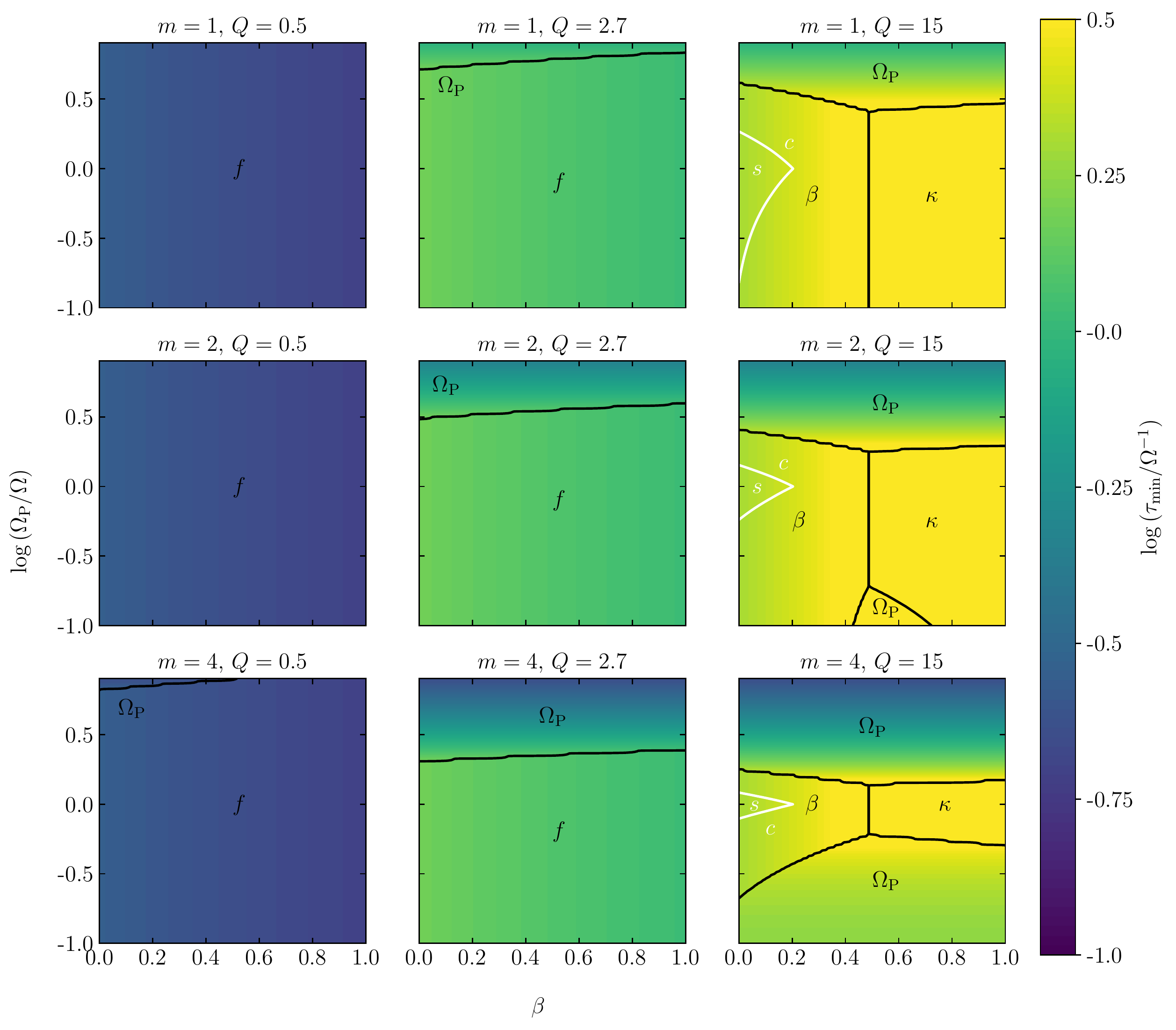}
 \caption{The minimum time-scale and its value for three cross-sections of the parameter space $(\beta,\log{Q},\Omega,\Omega_{\rm P}/\Omega)$ in the $(\beta,\Omega,\Omega_{\rm P}/\Omega)$ fundamental plane, for three numbers of spiral arms $m=1,2,4$ and for $\phi_P=3$. The dependence on $\Omega$ is included as a normalisation of the time-scale, as all evolutionary time-scales depend on $\Omega$ in the same way. The relevant time-scales here are $\tau_{\kappa}$, $\tau_{\beta}$, $\tau_{\Omega_{\rm P}}$ and $\tau_{\text{ff,g}}$, denoted by $\kappa$, $\beta$, $f$ and $\Omega_{\rm P}$ respectively. The solid black lines delineate boundaries along which two time-scales are equal; the \textit{regions of dominance} are separated by these lines. The solid white lines delineate the values of $Q$ and $\beta$ above which the rate of galactic shear is higher than the combined rates of all other mechanisms, as discussed in Section~\ref{Sec::beta}.}
 \label{Fig::OmegavsBeta_absmin}
\end{figure*}

\begin{figure*}
 \includegraphics[width=\textwidth]{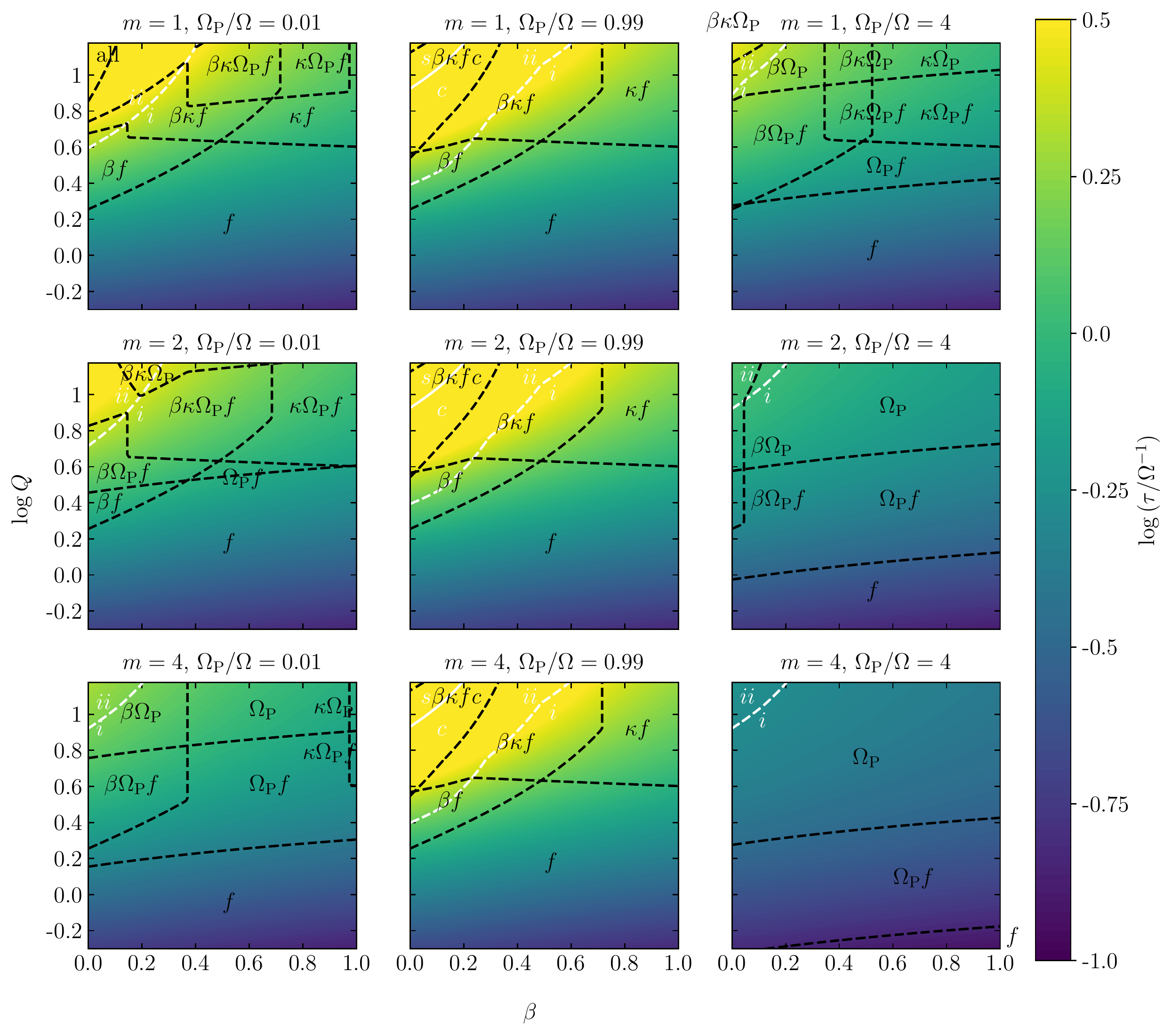}
 \caption{The predicted cloud lifetime for three cross-sections of the parameter space $(\beta,\log{Q},\Omega,\Omega_{\rm P}/\Omega)$ in the $(\beta,Q,\Omega)$ fundamental plane, for three numbers of spiral arms $m=1,2,4$ and $\phi_P=3$. The dependence on $\Omega$ is included as a normalisation of the time-scale, as all evolutionary time-scales depend on $\Omega$ in the same way. The relevant time-scales are $\tau_{\kappa}$, $\tau_{\beta}$, $\tau_{\Omega_{\rm P}}$, $\tau_{\text{ff,g}}$ and $\tau_{\text{cc}}$, denoted by $\kappa$, $\beta$, $\Omega_{\rm P}$, $f$ and $c$ respectively. The dashed black lines divide the regions within which each time-scale is relevant from the regions in which it is irrelevant, as described in Section~\ref{Sec::dom_and_rel}. The dashed white lines divide the regions for which cloud lifetime is longer than the minimum evolution time-scale (above the line) from the regions in which it is shorter (below the line). These are labelled (ii) and (i) respectively. The solid white lines delineate the values of $Q$ and $\beta$ above which the rate of galactic shear is higher than the combined rates of all other mechanisms.}
 \label{Fig::QvsBeta_lifetimes}
\end{figure*}

In Figures~\ref{Fig::QvsBeta_absmin} and~\ref{Fig::OmegavsBeta_absmin} we display the contours of the minimum normalised time-scale $\tau_{\rm min}/\Omega^{-1}$ and solid black lines delineating the {\it regions of dominance} for each time-scale. The setup is identical to that in Figure~\ref{Fig::QvsBeta_absmin_m0}, but extended over the new variables $m$ and $\Omega_{\rm P}/\Omega$.

The panels in the central column of Figure~\ref{Fig::QvsBeta_absmin} describe cloud evolution at the radius of co-rotation ($\Omega_{\rm P}/\Omega=0.99$) in spiral galaxies with $m=1, 2$ or $4$ spiral arms. Each panel is an exact copy of the central panel in Figure~\ref{Fig::QvsBeta_absmin_m0}, describing cloud evolution in flocculent or elliptical galaxies, where $m=0$. This is because the midplane gas at the radius of co-rotation moves in sychronisation with the spiral arms, such that they never interact with molecular clouds, and play no role in cloud evolution.

The left-hand column of Figure~\ref{Fig::QvsBeta_absmin}, with pattern speed $\Omega_{\rm P}/\Omega=0.01$ far within the radius of co-rotation, is very similar to the central, co-rotating column, but with a region of dominance for spiral arm interactions `$\Omega_{\rm P}$' encroaching from the top of each panel as the number of spiral arms is increased. The first noticable change for spiral galaxies, relative to elliptical and flocculent galaxies, occurs for grand design spirals ($m=2$), where highly-stable, low-shear gas ($Q \ga 6$, $\beta \ga 0.4$, e.g.~gas in the presence of a nuclear spiral) in the top right-hand corner of the panel switches from epicycle-dominated `$\kappa$' to spiral arm-dominated `$\Omega_{\rm P}$' evolution. This spiral arm-dominated region extends to even lower values of gravitational stability in the case of four spiral arms ($m=4$, bottom panel in the left-hand column), where it also takes over from the dominance of shear (`$\beta$' in the top and middle panels of the column) for highly-stable, highly-sheared gas ($Q \ga 4$, for all $\beta$). That is, given a sufficiently large number of spiral arms, spiral arm perturbations become the dominant mechanism for cloud evolution in the outer regions of galactic bulges as well as near galactic centres.

The influence of spiral arm crossings is further increased for pattern speeds far outside the radius of co-rotation (right-hand column of Figure~\ref{Fig::QvsBeta_absmin}), because the ratio $\Omega_{\rm P}/\Omega$ can become very large if the midplane angular velocity $\Omega$ is very small. In fact, our model imposes no limit on the increase of $\Omega_{\rm P}/\Omega$ as $\Omega \rightarrow 0$. There is no mathematical reason why the pattern speed cannot be made so high that spiral arms dominate throughout the entire space of $(\beta, Q, \Omega)$. Physically, however, this behaviour is limited by the weakening of the spiral shock at large galactocentric radii, such that the perturbation dies off as $\Omega_{\rm P}/\Omega \rightarrow \infty$. Our model does not explicitly account for this effect, but for this reason we only examine values of the pattern speed ratio between $\Omega_{\rm P}/\Omega=0.01$ and $\Omega_{\rm P}/\Omega=4$.

The right-hand column of Figure~\ref{Fig::QvsBeta_absmin} displays a more extreme version of the pattern shown for clouds inside the radius of co-rotation (left-hand column). Again, the region of spiral arm dominance `$\Omega_{\rm P}$' encroaches from the top right corner of each panel as the number of spiral arms is increased from $m=1$ through $m=4$, but unlike the panels in the left-hand column, this effect is already significant in galaxies with $m=1$ (i.e.~a single spiral arm). When two spiral arms are introduced (i.e.~for grand design spirals), spiral arm crossings dominate cloud evolution down to gravitational stability values of $Q \sim 2$, and with four spiral arms, they are dominant down to $Q \sim 1$. This is well into the realm of clumpy, star-forming galaxies, either in the local Universe~\citep[e.g.][]{Fisher2017} or at high redshift~\citep[e.g.][]{Genzel2014}.

Figure~\ref{Fig::OmegavsBeta_absmin} displays the same information as Figure~\ref{Fig::QvsBeta_absmin}, but takes a different cross-section through $(\beta, Q, \Omega, m, \Omega_{\rm P}/\Omega)$ parameter space---through the $(\beta, \Omega_{\rm P}/\Omega, \Omega)$ plane. In the left-hand column, the panels display the overwhelming dominance of gravitational free-fall `$f$' for highly gravitationally-unstable gas with $Q \sim 0.5$, in galaxies with one, two and four spiral arms (i.e.~gas-rich, clumpy, star-forming spirals). Moderately-stable gas ($Q = 2.7$, central column) in spiral galaxies is also dominated by gravitational free-fall for most galactocentric radii (traced by $\Omega_{\rm P}/\Omega$), with spiral arm perturbations becoming dominant only far outside the radius of co-rotation (large $\Omega_{\rm P}/\Omega$). At these radii, high absolute differences between the angular speed of the spiral arms and the angular speed of the midplane gas can be obtained, particularly as the rotation curve flattens. As in Figure~\ref{Fig::QvsBeta_absmin}, the dominance of spiral arm crossings becomes more prominent as the number of spiral arms is increased.

The right-hand column of Figure~\ref{Fig::OmegavsBeta_absmin} shows that, for very high-stability gas with $Q \sim 15$, galactic shear `$\beta$' and epicyclic perturbations `$\kappa$' govern cloud evolution under specific environmental conditions. Although spiral arm perturbations `$\Omega_{\rm P}$' dominate at galactocentric radii far from the radius of co-rotation (i.e.~the outer and inner regions of spiral galaxies), galactic shear dominates near the radius of co-rotation for $\beta \la 0.5$, (i.e.~when the rotation curve is approximately flat at the radius of co-rotation) and epicyclic perturbations dominate at co-rotation for $\beta \ga 0.5$ (i.e.~when the rotation curve is approximately solid-body at this radius). While the regions of dominance for spiral arm crossings become very large as $m$ increases, epicyclic perturbations and galactic shear retain their dominance for a significant span of galactocentric radii. Even in the case of four spiral arms, using the pattern speed $\sim 0.026 \pm 0.002$ Myr$^{-1}$ of the Milky Way~\citep{Gerhard2011}, and assuming that the radius of co-rotation is at $\sim 8$ kpc with an average rotational velocity of $\sim 200$ km s$^{-1}$ for nearby galactocentric radii, a span of radii from $\sim 5$ kpc to $\sim 13$~kpc are not dominated by spiral arm crossings, according to the bottom right-hand panel of Figure~\ref{Fig::QvsBeta_absmin}.

Note that the solid white lines in the right-hand column of Figure~\ref{Fig::OmegavsBeta_absmin} are analogous to those in the central column of Figure~\ref{Fig::QvsBeta_absmin}. For highly-stable gas at the radius of co-rotation (regime {\it s}), galactic shear outpaces the combination of all compressive evolutionary mechanisms. Clouds in this region of parameter space are therefore likely to be torn apart by galactic shear, while clouds in region ({\it c}) are more likely to be destroyed by gravitational collapse and stellar feedback.

\subsubsection{Regions of relevance, $m \neq 0$}
\label{Sec::spiral_lifetimes}

\begin{figure*}
 \includegraphics[width=\textwidth]{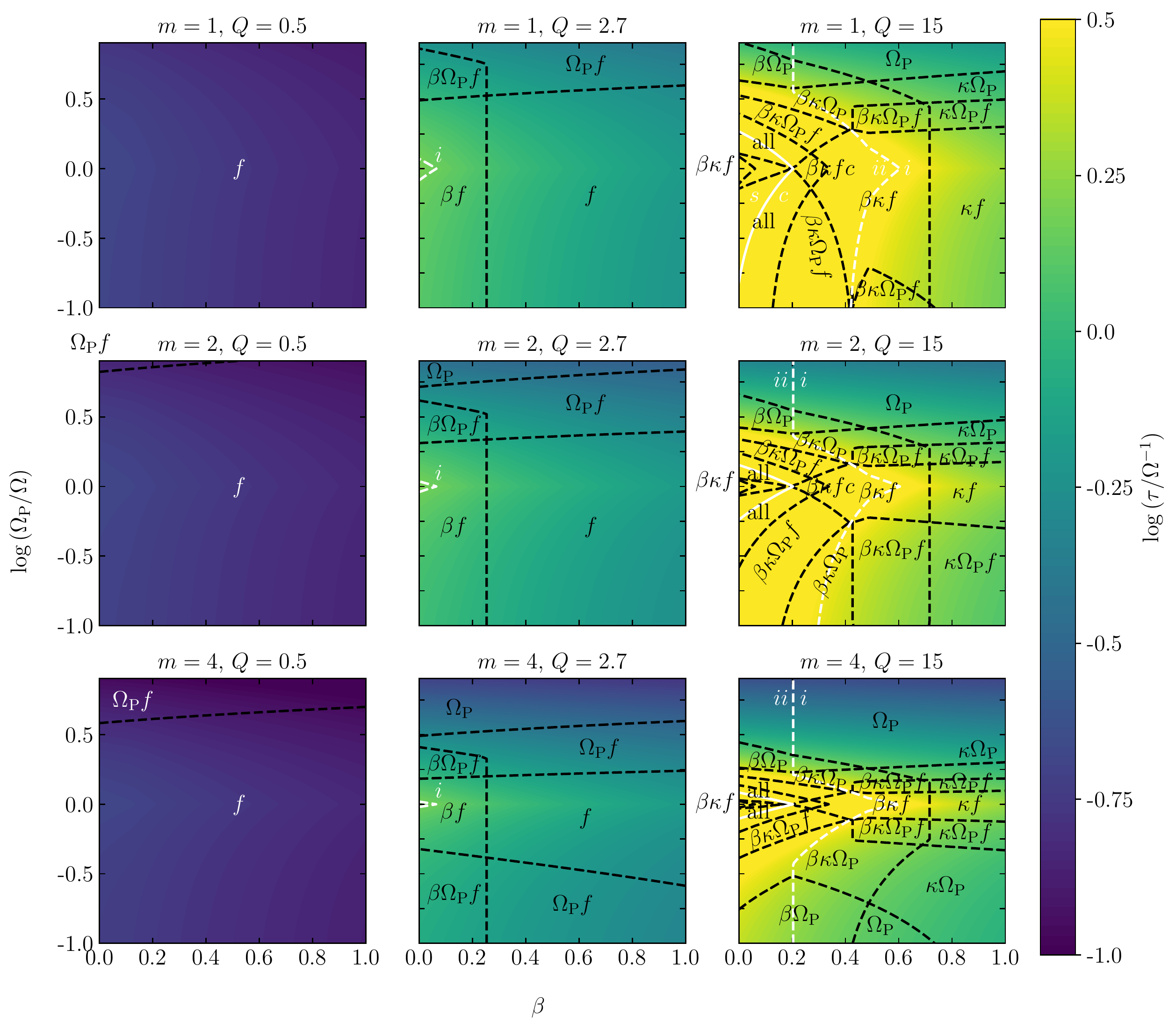}
 \caption{The predicted cloud lifetime for three cross-sections of the parameter space $(\beta,\log{Q},\Omega,\Omega_{\rm P}/\Omega)$ in the $(\beta,\Omega,\Omega_{\rm P}/\Omega)$ fundamental plane, for three numbers of spiral arms $m=1,2,4$ and $\phi_P=3$. The dependence on $\Omega$ is included as a normalisation of the time-scale, as all evolutionary time-scales depend on $\Omega$ in the same way. The relevant time-scales are $\tau_{\kappa}$, $\tau_{\beta}$, $\tau_{\Omega_{\rm P}}$, $\tau_{\text{ff,g}}$ and $\tau_{\text{cc}}$, denoted by $\kappa$, $\beta$, $\Omega_{\rm P}$, $f$ and $c$ respectively. The dashed black lines divide the regions within which each time-scale is relevant from the regions in which it is irrelevant, as described in Section~\ref{Sec::dom_and_rel}. The dashed white lines divide the regions for which cloud lifetime is longer than the minimum evolution time-scale (above the line) from the regions in which it is shorter (below the line). These are labelled (ii) and (i) respectively. The solid white lines delineate the values of $Q$ and $\beta$ above which the rate of galactic shear is higher than the combined rates of all other mechanisms.}
 \label{Fig::OmegavsBeta_lifetimes}
\end{figure*}

In Figures~\ref{Fig::QvsBeta_lifetimes} and~\ref{Fig::OmegavsBeta_lifetimes} we display the contours of the normalised cloud lifetime, $\tau/\Omega^{-1}$, with dashed black lines delineating the {\it regions of relevance}. The setup is identical to that in Figure~\ref{Fig::QvsBeta_lifetimes_m0}, but extended over the $(\beta,Q,\Omega,m,\Omega_{\rm P}/\Omega)$ parameter space. As in Figure~\ref{Fig::QvsBeta_lifetimes_m0}, the parameter space is divided into two regimes labelled ({\it i}) and ({\it ii}), separated by a white dashed line. In regime ({\it i}) the the {\it regions of relevance} are determined by comparison to the minimum cloud evolutionary time-scale, while in regime ({\it ii}) the {\it regions of relevance} are determined by comparison to the cloud lifetime. The threshold for relevance is less than twice the magnitude of $\tau_{\rm min}$ in regime ({\it i}), or twice the magnitude of $\tau$ in regime ({\it ii}).

As for the regions of dominance displayed in Figure~\ref{Fig::QvsBeta_absmin}, the panels in the central column of Figure~\ref{Fig::QvsBeta_lifetimes} display the case of spiral arm co-rotation ($\Omega_{\rm P}/\Omega = 0.99)$ and so have the same division of parameter space as for flocculent gas reservoirs ($m=0$, central panel of Figure~\ref{Fig::QvsBeta_lifetimes_m0}). The left-hand column of Figure~\ref{Fig::QvsBeta_lifetimes}, with pattern speed $\Omega_{\rm P}/\Omega=0.01$ within the radius of co-rotation, is similar to the central column, but with a region of relevance for spiral arm interactions emerging from the top of each panel as the number of spiral arms is increased. Although spiral arm crossings do not dominate cloud evolution at any point in $(\beta, Q, \Omega)$ parameter space for a single spiral arm inside the radius of co-rotation ($m=1$ and $\Omega_{\rm P}/\Omega=0.01$ in the top left panel of Figure~\ref{Fig::QvsBeta_absmin}), they do still play a non-trivial role in cloud evolution (`$\Omega_{\rm P}$' regions of relevance for $Q \ga 6$ in the top left panel of Figure~\ref{Fig::QvsBeta_lifetimes}). This has the effect of reducing the cloud lifetime relative to the case of spiral arm co-rotation, which can be seen by comparing the colours of the regions `$\beta \kappa \Omega_{\rm P} f$', `$\kappa \Omega_{\rm P} f$' and `all' in the top left panel of Figure~\ref{Fig::QvsBeta_lifetimes} to those of the regions `$\beta \kappa f$', `$\kappa f$' and `$\beta \kappa f c$' in one of the panels of the central column. Due to the compressive effect of spiral arm crossings on molecular clouds, `$\Omega_{\rm P}$' augments the epicyclic perturbations `$\kappa$', the gravitational free-fall `$f$' and the cloud-cloud collisions `$c$', and competes against the galactic shear `$\beta$', so that the elongation of the cloud lifetime by shear support is reduced. This effect becomes more and more pronounced as the number of spiral arms is increased through $m=2$ and $m=4$ (middle and bottom panels of the left-hand column). The regime ({\it ii}) of cloud lifetimes that are longer than the dominant evolutionary time-scale (enclosed by a white dashed line) is therefore progressively eroded by the introduction of more spiral arms. In fact, the regime ({\it s}), or which galactic shear has a stronger influence than the combined rates of all compressive evolutionary mechanisms (enclosed by a solid white line in the panels of the central column), is completely removed by the introduction of even a single-arm spiral pattern ($m=1$) at the pattern speeds considered. By causing additional compression of gravitationally-stable gas, particularly in the high-shear regime ($Q \ga 4$, $\beta < 0.5$, i.e.~for outer galaxy bulges), spiral arms encourage the collapse of molecular clouds and so shorten their lifetimes. Conversely, the evolution of clouds formed in gas that is highly gravitationally-unstable (i.e.~in gas-rich, star-forming galaxies) is overwhelmingly governed by gravity, to the extent that no other process is significant in determining the cloud lifetime.

Above the radius of co-rotation, the effect of spiral arm crossings is further enhanced by the larger absolute difference between the spiral arm pattern speed and the angular velocity of the midplane gas for $\Omega_{\rm P}/\Omega=4$ (right-hand column of Figure~\ref{Fig::QvsBeta_lifetimes}). The three panels of the left-hand column and the bottom two panels of the right-hand column therefore form a sequence of increasing influence for spiral arm crossings. In the most extreme case of four spiral arms above the radius of co-rotation (bottom right-hand panel), the region of relevance `$\Omega_{\rm P}$' for spiral arm crossings extends into the highly gravitational-unstable region of parameter space for $Q \la 1$. In this case, spiral arm perturbations heavily influence the cloud lifetime in gas-rich, star-forming spiral galaxies.

Figure~\ref{Fig::OmegavsBeta_lifetimes} displays the same information as Figure~\ref{Fig::QvsBeta_lifetimes}, but shows this information as a cross-section through the $(\beta, \Omega_{\rm P}/\Omega, \Omega)$ plane, rather than as a cross-section through the $(\beta, Q, \Omega)$ plane. Note that the dashed white lines are analogous to those in Figure~\ref{Fig::QvsBeta_lifetimes}, separating the regime ({\it i}), in which $\tau < \tau_{\rm min}$, from the regime ({\it ii}), in which $\tau > \tau_{\rm min}$. Clouds that form from regions of highly-unstable molecular gas at $Q \sim 0.5$ (left-hand column)  in star-forming, gas-rich galaxies are governed almost exclusively by gravitational collapse `$f$', except at galactocentric radii far outside the radius of co-rotation, when four spiral arms are involved (bottom panel of the left-hand column). Clouds that form outside the radius of co-rotation in moderately-stable gas ($Q=2.7$, central column) may be influenced by spiral arm perturbations before they collapse under gravity, as depicted by the regions of relevance `$\Omega_{\rm P}$' and `$\Omega_{\rm P} f$' for spiral arms, encroaching from the top of each panel. In the case of four strong spiral arms, such perturbations may also affect moderately-stable clouds inside the radius of co-rotation (bottom side of the central bottom panel). Galactic shear also plays a significant role in cloud evolution for regions of moderately-stable gas with flat rotation curves ($\beta \la 0.2$, i.e.~the main discs of spiral galaxies), where its coexistence `$\beta \Omega_{\rm P} f$' with spiral arm perturbations and gravitational free-fall slightly extends the cloud lifetime relative to the case of near solid-body rotation ($\beta \ga 0.5$). Finally, the right-hand column of Figure~\ref{Fig::OmegavsBeta_lifetimes} demonstrates that, for clouds formed in highly gravitationally-stable gas that hosts a spiral pattern, the relevant mechanisms of cloud evolution depends very delicately on galactocentric radius (parameterised by $\Omega_{\rm P}/\Omega$) and the slope of the rotation curve (parameterised by $\beta$).

\section{Cloud properties throughout parameter space}
\label{Sec::Discussion}
In this section, we use the {\it regions of dominance} and {\it regions of relevance} identified in Section~\ref{Sec::Coexistence} to systematically predict the observational properties of molecular clouds in different parts of parameter space, and thus in different galactic environments. We begin by characterising the predicted properties of clouds in those regions of parameter space that are controlled by gravitational collapse on a time-scale $\tau_{\rm ff,g}$, and by galactic shear on a time-scale $\tau_\beta$. These two mechanisms of cloud evolution form the basis of our analysis, as the majority of star formation is found to occur in dense, gravitationally-bound regions within molecular clouds~\citep{Hartmann2001,Elmegreen2007,Dobbs2011,Dobbs2013b}. The rate of collapse and the subsequent levels of stellar feedback therefore exert a large influence over the galactic SFR, and play an important role in setting the cloud lifetime. Galactic shear is the only mechanism of cloud evolution that manifestly competes against gravitational collapse, by stretching radially-correlated gas in the azimuthal direction. Shear is able to destroy clouds in the opposite sense to free-fall, by dispersing the molecular gas.

Both gravitational collapse and galactic shear dominate large parts of our parameter space and form large regions of coexistence with each of the other cloud evolutionary mechanisms. Ultimately, the star-forming properties of molecular clouds will depend on their tendency towards collapse rather than dispersion, and so it is the relationship of each mechanism to gravitational collapse, or its prevention by shear support, that is most interesting observationally. We therefore characterise the predicted properties of molecular clouds for coexisting pairs of cloud evolutionary mechanisms, including either gravitational free-fall or galactic shear. Although a large number of regions are characterised by the overlap of more than two time-scales, the cloud properties in such regions can be inferred to a great extent from these pairings. The only exception arises at very high levels of gravitational stability in the absence of shear, where epicyclic perturbations are most likely to be relevant (top right-hand corners of Figures~\ref{Fig::QvsBeta_lifetimes_m0},~\ref{Fig::QvsBeta_lifetimes} and~\ref{Fig::OmegavsBeta_lifetimes}).

\subsection{Dominance of gravitational collapse ($f$)}
\label{Sec::ffg_dominant}
In flocculent galaxies with high gas fractions ($m=0$ and $Q \la 4$), our theory predicts gravity `$f$' to dominate the evolution of molecular clouds, without exception (see Figure~\ref{Fig::QvsBeta_absmin_m0}). For strong spiral arm patterns with $m=2$ or $m=4$ (i.e.~in grand design spiral galaxies), dominance at $Q \la 4$ may be shared between gravitational free-fall and spiral arm perturbations `$\Omega_{\rm P}$', but only above the radius of co-rotation (see Figures~\ref{Fig::QvsBeta_absmin} and~\ref{Fig::OmegavsBeta_absmin}). Within these gravity-dominated regions of parameter space, it is often the case that gravitational collapse is the {\it only} relevant mechanism of cloud evolution, up to values of gravitational stability as high as $Q \approx 3$ (see e.g.~Figure~\ref{Fig::QvsBeta_lifetimes_m0}, for example). The relatively short cloud lifetimes in such environments are consistent with clouds having a short quiescent phase, followed by hierarchical or global gravitational collapse. Given that star formation is mainly limited to dense, gravitationally-bound regions within molecular clouds~\citep{Hartmann2001,Elmegreen2007,Dobbs2011,Dobbs2013b}, a large fraction of these clouds should host star-forming regions, such that the average SFE per unit mass for clouds in these regions of parameter space should be significantly higher than the average SFE per unit mass of all observable clouds.

\subsection{Dominance of galactic shear ($\beta$)}
\label{Sec::beta}
In elliptical galaxies, outer galactic bulges and galaxy outskirts, large regions of gas exist that are both gravitationally stable and have approximately flat rotation curves ($Q \ga 4$ and $\beta \la 0.5$, e.g.~Figures 9 and 10 of \citealt{Leroy2008}). In such environments, cloud evolution is dominated by galactic shear (see the top left-hand corners of each panel in Figure~\ref{Fig::QvsBeta_absmin_m0}). The only exception arises in the presence of spiral arms, where spiral arm perturbations dominate cloud evolution in galaxy outskirts (corresponding to $\Omega_{\rm P}/\Omega \ga 2$ in the central and right-hand columns of Figure~\ref{Fig::OmegavsBeta_absmin}). In particular, the solid white lines in these figures enclose the regions of parameter space ({\it s}) for which galactic shear dominates over the combination of all other cloud evolution mechanisms, such that
\begin{equation}
\tau_{\beta}^{-1} > \tau_\kappa^{-1}+\tau_{\Omega_{\rm P}}^{-1}+\tau_{\text{ff,g}}^{-1}+\tau_{\text{cc}}^{-1}.
\end{equation}
In these regions of parameter space, we predict that clouds of molecular gas are dispersed by shear before they can be encouraged to collapse via any other mechanism. We therefore expect the GMCs in such regions to contain few gravitationally-bound, star forming regions, and to contribute correspondingly little to the galactic SFR. In Figures~\ref{Fig::QvsBeta_lifetimes} and~\ref{Fig::OmegavsBeta_lifetimes} we see that the time-scales on which gas is dispersed are up to several orbital times $\Omega^{-1}$. In the outer regions of galactic discs, where orbital times are long and the rotation curve is flat, diffuse envelopes of quiescent gas with low SFRs may survive for up to hundreds of Myr. In the outer regions of the Central Molecular Zone (CMZ) of the Milky Way, which presents regions of highly gravitationally-stable, highly-shearing gas~\citep{Krumholz2015}, shear support could greatly extend cloud lifetimes~\citep{Jeffreson2018} and explain the very low SFE observed by~\cite{Longmore2013}.

\subsection{Galactic shear/free-fall coexistence ($\beta f$)}
\label{Sec::beta_f}
In the case of flocculent galaxies without a strong spiral arm pattern ($m=0$, i.e.~the central panel of Figure~\ref{Fig::QvsBeta_lifetimes_m0}), coexistence between galactic shear and gravitational collapse is the most influential pairing of cloud evolutionary mechanisms, occupying a large fraction of parameter space `$\beta f$' for $0<\beta<0.7$ and $Q > 1.5$. These are the areas of parameter space where {\it both} $\tau_{\rm ff,g}$ and $\tau_\beta$ are shorter than twice the minimum time-scale $t_{\rm min}$, or shorter than twice the cloud lifetime $\tau$, if $\tau>\tau_{\rm min}$. The shear parameter is typically low (indicative of an approximately flat rotation curve) and the Toomre $Q$ stability parameter is typically high (indicative of low gas fractions). In these regions of parameter space, the dispersive effect of galactic shear will elongate the cloud lifetime and slow down the formation of gravitationally-bound regions within GMCs, where the efficiency of star formation per unit mass is highest, or will prevent their formation altogether~\cite[see also][]{Meidt2018,Meidt2018b}.

In the case that galactic shear is relevant, but GMC evolution is still governed by dynamically-compressive mechanisms, i.e.~in regime ({\it c}) of the region `$\beta f$' (top left corner of the central panel in Figure~\ref{Fig::QvsBeta_lifetimes_m0}), shear will slow the formation of bound regions within GMCs, but will not necessarily prevent their formation over a long period of time. That is, it will lower the SFE per unit time, but not the SFE per unit mass. We therefore expect a significant fraction of the mass in such clouds to be converted to stars, but much more slowly and over a much longer lifetime than the star formation in clouds with lower levels of shear.

In the case that galactic shear is both relevant {\it and} dominant over all dynamically-compressive cloud evolutionary mechanisms, i.e.~in regime ({\it s}) of the region `$\beta f$' (top left corner of the central panel in Figure~\ref{Fig::QvsBeta_lifetimes_m0}), shear may gradually tear molecular clouds apart, and therefore prevent the formation of bound, star-forming regions, even over a long period of time. In this case, both the SFE per unit time and the SFE per unit mass will be significantly reduced. For clouds in regime ({\it s}), we therefore expect very long lifetimes combined with very low integrated levels of star formation, such that only a small fraction of the cloud mass is converted to stars.

Overall, for high levels of gravitational stability ($Q \ga 4$), we expect a longer cloud lifetime and a lower SFE per unit time for a flat rotation curve ($\beta \sim 0$) than for approximately solid-body rotation ($\beta \sim 1$). Indeed,~\cite{Leroy2008} observe that in spiral galaxies with low levels of galactic shear ($\beta \approx 1$), the average SFE per unit time is almost three times higher than the average SFE per unit time at high levels of galactic shear ($\beta \approx 0$), with a much smaller spread. The larger spread of SFEs down to lower values at $\beta \approx 0$ corresponds with our parameter space diagrams (e.g.~Figure~\ref{Fig::QvsBeta_lifetimes_m0}), which show a much larger range of cloud lifetimes for $\beta=0$ than they do for $\beta=1$. The influence of galactic shear is also expected to manifest itself through elevation of the virial parameter, and observational studies are indeed beginning to find an inverse correlation between the virial parameter and the SFE per unit time~\citep{Leroy2017b}.

The elongation of the cloud lifetime, due to the competition between galactic shear and gravitational collapse, should become particularly noticeable about the division between regime ({\it s}), where the rate of shear outpaces the sum of all the dynamically-compressive evolutionary rates, and regime ({\it c}), where the dynamically-compressive mechanisms outpace the shear. These divisions are indicated by solid white lines in the Figures of Section~\ref{Sec::Coexistence}, on which the cloud lifetime is theoretically infinite, according to Equation~(\ref{Eqn::Lifetime}). In practice, the balance between shear support and the other cloud evolutionary mechanisms can never be sufficiently finely-tuned to give an infinite cloud lifetime, due firstly to the influence of small-scale, non-dynamical influences on cloud evolution, and secondly because cloud-cloud collisions, spiral arm crossings and gravitational collapse are discrete stochastic events that may occur at any time in a cloud's lifecycle. However, the key point remains that, in the vicinity of these lines, we predict cloud lifetimes to be longer than in any other region of parameter space.

Previous work has suggested that shear support is ineffective by looking at $\sim 30$~pc (regions of) clouds, much smaller than the Toomre scale, in the specific environment of the solar neighbourhood~\citep{Dib2012}. Indeed, we do not expect shear support to be effective under these conditions. Firstly, we will show in Section~\ref{Sec::MilkyWay} that the solar neighbourhood occupies a part of parameter space where shear may be relevant, but never dominates over gravitational collapse. Secondly, zooming in on scales much smaller than the Toomre scale implies looking at locally-collapsing regions that are decoupled from the galactic-scale flow, such that shear is already marginalised. This is immediately obvious in Figure~\ref{Fig::DynamicRanges}, where the dynamic ranges of $\tau_\beta$ and $\tau_{\rm ff,cl}$ do not overlap. Once a gravitationally-bound and locally-collapsing region has formed, we expect it to collapse on a time-scale $\tau_{\text{ff,cl}}$, independent of its environment. Our theory considers a wider range of objects than those that are gravitationally-bound and collapsing, and we find that for gravitationally-stable regions of the ISM with $Q \ga 4$, galactic shear has a significant or even dominant influence on cloud evolution.

\subsection{Dominance of spiral arm interactions ($\Omega_{\rm P}$)}
\label{Sec::OmegaP_dominant}
In spiral galaxies, we find large areas of parameter space that are dominated by spiral arm crossings. These regions of spiral arm dominance are preferentially located outside the radius of co-rotation ($\Omega_{\rm P}/\Omega > 1$) and at higher levels of gravitational stability, as can be seen by comparing the sizes of the `$\Omega_{\rm P}$' regions in the centre ($Q \sim 2.7$) and right-hand ($Q \sim 15$) columns of Figure~\ref{Fig::OmegavsBeta_absmin} (see also Figure~\ref{Fig::QvsBeta_absmin}). Spiral arm crossings may also be dominant below the radius of co-rotation in grand-design spirals ($m=2$ and $m=4$ panels in the left-hand column of Figure~\ref{Fig::QvsBeta_absmin}), provided that the inner disc region is sufficiently gravitationally-stable ($Q \ga 3$).

In the following two sections, we will describe our predictions regarding molecular cloud evolution in these spiral arm-dominated regions of parameter space. Spiral arms are found to have a strong influence on the organisation of star-forming molecular gas, both in observations~\citep[e.g.][]{Elmegreen1983,Meidt2013} and in hydrodynamical simulations~\citep[e.g.][]{Dobbs2008b,Dobbs2011b}, sometimes even leading to a `beads-on-a-string' morphology in which the majority of massive GMCs live in spiral arms. However, the effect of spiral density waves on GMC evolution (and consequently on the galactic SFR), relative to that in non-spiral galaxies, is highly contested. Many observational comparisons between the galactic SFR in grand design and flocculent galaxies ~\citep[e.g.][]{Elmegreen1986,Schinnerer2017} have found that spiral arms make no significant difference to the rate of star formation on kiloparsec scales. On the other hand,~\cite{Hart2017} find that two-armed spiral galaxies form stars more efficiently than flocculent galaxies, even though the absolute SFR is unaffected. Similarly, hydrodynamical simulations by~\cite{Dobbs2011b} find that it is possible for spiral arms to enhance the SFE per unit mass by promoting the formation of more massive, longer-lived and gravitationally-collapsing GMCs.

\subsubsection{Spiral arm crossing/free-fall coexistence ($\Omega_{\rm P} f$)}
At low and moderate levels of gravitational stability ($Q \la 6$, i.e.~in star-forming regions of galaxies), there exist large regions of parameter space `$\Omega_{\rm P} f$' in which gravitational free-fall and spiral arm crossings coexist (see e.g.~left-hand and right-hand columns of Figure~\ref{Fig::QvsBeta_lifetimes}). Due to gravitational instability, we expect that clouds in such environments will contain dense, gravitationally-bound, star-forming regions~\citep[e.g.][]{Hartmann2001,Elmegreen2007,Dobbs2011,Dobbs2013b} that are decoupled from the large-scale galactic dynamics and that eventually destroy the cloud via stellar feedback. We expect that spiral arm perturbations will sweep up and and collect these star-forming clouds, creating a larger number of massive GMCs than would be expected due to gravity alone. As the degree of gravitational instability is increased, the effect of spiral arm crossings on cloud evolution and the cloud lifetime is reduced, because GMCs are more likely to be destroyed by stellar feedback before encountering a spiral arm. Accordingly, the most gravitationally-unstable gas (bottom of each panel, $Q \la 3$) displays a much smaller difference in colour than does the moderately-stable gas (middle of each panel, $3 \la Q \la 6$).

\subsubsection{Spiral arm crossing/shear coexistence ($\beta \Omega_{\rm P}$)}
\label{Sec::beta_OmegaP}
In the outskirts of spiral galaxies, where levels of gravitational stability may be high and the rotation curve is relatively flat (e.g.~the top left-hand corner of each panel in the left-hand column of Figure~\ref{Fig::QvsBeta_lifetimes}), galactic shear and spiral arm perturbations coexist in regions `$\beta \Omega_{\rm P}$' of parameter space. Due to the high levels of gravitational stability and shear support, we expect that these regions of parameter space correspond to the most diffuse molecular gas, with the lowest levels of gravitational collapse and star formation. We therefore expect that GMCs in such environments are unlikely to be destroyed by stellar feedback before encountering a spiral arm. They are more likely to be affected by spiral arm crossings than GMCs in more gravitationally-unstable regions of parameter space. We expect that the spiral arm density wave will shock and compress diffuse molecular gas into a state of higher density, and therefore induce gravitational collapse in some clouds that were previously super-virial and shear-supported. This effect becomes stronger as the number of spiral arms is increased, causing the cloud lifetime to be reduced from $>3$ orbital times in the case of a single spiral arm (top left-hand corner of the $m=1$ panel) to $\sim 1$ orbital time in the case of four spiral arms (top left-hand corner of the $m=4$ panel).

\subsection{Dominance of epicyclic perturbations}
\label{Sec::kappa}
Epicyclic perturbations are dominant only in regions of high gravitational stability and approximately solid-body rotation (e.g.~the top right-hand corner of each panel in the central column of Figure~\ref{Fig::QvsBeta_absmin}). Furthermore, in spiral galaxies, they are only dominant in the vicinity of the radius of co-rotation (e.g.~Figure~\ref{Fig::OmegavsBeta_absmin}). Due to the high minimum value of $\tau_\kappa$ (see e.g.~Figure~\ref{Fig::DynamicRanges}), epicyclic perturbations are often relevant only in conjunction with two or three other mechanisms of cloud evolution; most often with free-fall or shear.

\subsubsection{Epicyclic perturbations/free-fall coexistence ($\kappa f$)}
\label{Sec::kappa_f}
In correspondence with the regions of dominance for epicyclic perturbations, coexistence between epicyclic perturbations and gravitational free-fall occurs only for high levels of gravitational stability ($Q \ga 4$). This indicates that the relevance of gravitational free-fall has little to do with gravitational instability and more to do with the weak competition provided by epicyclic perturbations, which sets the longest value of the minimum time-scales $\tau_{\rm min}/\Omega^{-1}$ throughout all of parameter space (see e.g.~Figure~\ref{Fig::QvsBeta_absmin_m0}). Molecular clouds in these regions of parameter space will host few gravitationally-bound, star-forming regions, and will contribute little to the galactic SFR, in comparison to molecular clouds that are dominated by gravitational free-fall. In regions of very high shear parameter $\beta \rightarrow 1$, these clouds will experience little dispersion by galactic shear, so that the small compressions introduced by orbital eccentricity on a time-scale $\tau_\kappa$ may induce isolated instances of gravitational collapse and star formation. However given the high values of Toomre $Q$ and the length of the time-scale $\tau_\kappa$, we do not necessarily expect these instances to be common.

\subsubsection{Shear/epicyclic perturbation coexistence ($\beta \kappa$)}
\label{Sec::beta_kappa}
Regions of coexistence `$\beta \kappa$' for epicyclic perturbations and shear generally occur at high values of gravitational stability and non-negligible levels of galactic shear ($Q \ga 4$ and $\beta \la 0.8$, e.g.~top side of each panel in Figure~\ref{Fig::QvsBeta_lifetimes_m0}). Furthermore, the influence of spiral arms tends to overpower the effect of epicyclic perturbations, such that they are only relevant for galaxies with one or no spiral arms ($m=1$ or $m=0$), or close to the radius of co-rotation in galaxies with strong spiral patterns (e.g.~the $\Omega_{\rm P}/\Omega=0.99$ panels in Figure~\ref{Fig::QvsBeta_lifetimes}). Although the regions `$\beta \kappa$' are small, the interplay between galactic shear and epicyclic perturbations may be of critical importance in key galactic environments. For instance, the Central Molecular Zone (CMZ) of the Milky Way hosts a dense gas stream at a galactocentric radius of $R\sim100$~pc \citep[e.g.][]{Molinari2011}, with a very high gas fraction, such that $\phi_P \sim 1$~\citep{Henshaw2016b}. Depending on the distance a cloud has travelled along the stream, the degree of gravitational stability may be as low as $Q \sim 1.7$~\citep{Kruijssen2014b} or as high as $Q \sim 5$~\citep[c.f.][]{Henshaw2016b}. Additionally, the degree of shear is weak but non-trivial, such that $\beta \sim 0.7$~\citep{Krumholz2015}. Due to this combination of physical parameters, a molecular cloud in the $100$~pc stream may belong to any one of the three regions `$f$', `$\kappa f$' or `$\beta \kappa f$' in the left-hand panel of Figure~\ref{Fig::QvsBeta_lifetimes_m0}, depending on its gravitational stability and thus its position on the stream. That is, clouds in one section of the stream may be supported by galactic shear (region `$\beta \kappa f$'), while clouds in another section may simply collapse (regions `$\kappa f$' and `$f$'). The relative importance of epicyclic perturbations in these regions of parameter space is consistent with the theory that they trigger the collapse of shear-supported molecular clouds, as these clouds pass through the pericentre of an eccentric orbit along the $100$~pc stream~\citep{Longmore2013b,Kruijssen2015,Henshaw2016b,Kruijssen2018}. We investigate the mechanisms governing the CMZ cloud lifecycle in more detail in~\cite{Jeffreson2018}.

\subsection{Dominance of cloud-cloud collisions}
\label{Sec::cc}
In Figures~\ref{Fig::QvsBeta_absmin_m0},~\ref{Fig::QvsBeta_absmin} and~\ref{Fig::OmegavsBeta_absmin}, we see that the cloud-cloud collision time-scale is never dominant over the time-scale for gravitational collapse, as predicted in Section~\ref{Sec::beta_Q_variation}. Furthermore, Figures~\ref{Fig::QvsBeta_lifetimes_m0},~\ref{Fig::QvsBeta_lifetimes} and~\ref{Fig::OmegavsBeta_lifetimes} demonstrate that with the exception of extremely high midplane gas fractions ($\phi_P \approx 1$), $\tau_{\text{cc}}$ is only relevant for high values of gravitational stability and flat rotation curves ($Q \sim 15$ and $\beta \sim 0$), where competition between gravitational free-fall and galactic shear extends the cloud lifetime. Since all mechanisms of cloud evolution are relevant in this region of parameter space, and the importance of cloud-cloud collisions at high $Q$ is likely to be overestimated in our theory (see Section~\ref{Sec::tau_cc}), we conclude that cloud-cloud collisions only make a meaningful contribution to the cloud lifetime in shearing, gas-right environments ($\beta \sim 0$ and $\phi_P \sim 1$), such as extended high-redshift galaxies (at low $Q$) and low-redshift galaxy outskirts (at high $Q$). In these discs, we expect that cloud-cloud collisions may shorten the cloud lifetime and slightly enhance the cloud-scale SFR by inducing gravitational collapse in high-density regions of molecular clouds which are not already bound and collapsing. This effect will be negligible for the galactic-scale SFR, because at low to middle values of the Toomre $Q$ parameter, most star formation will occur in bound, collapsing regions that have already decoupled from the galactic-scale dynamics.

\section{Cloud lifetimes in a sample of galaxies}
\label{Sec::Galaxies}
Here we examine the cloud lifetimes predicted by our theory for a sample of four galaxies (the Milky Way, M31, M51, and M83). We use surface density profiles, rotation curves, and velocity dispersion profiles from the literature to derive the quantities $\beta$, $Q$ and $\phi_P$ that define our parameter space. We also use information from the literature to justify our choice of the number of spiral arms $m$ for each galaxy, as well as our choice of pattern speed $\Omega_{\rm P}$. We set the stellar contribution to the surface density of the ISM to the value of $\phi_P=3$ appropriate to Milky Way-like disc galaxies~\citep{KrumholzMcKee2005}.

We calculate the cloud lifetime using Equation~(\ref{Eqn::Lifetime}), with each of the constituent time-scales as derived in Section~\ref{Sec::time-scales}, and consider our predictions as a function of galactocentric radius. For comparison, we also display the divisions between the {\it regions of relevance} discussed in Sections~\ref{Sec::Coexistence} and~\ref{Sec::Discussion}. In the case of M51, we compare our predictions for the cloud lifetime to the predictions of~\cite{Meidt2015}.

Before proceeding, it is important to note that our theory predicts the cloud lifetime according to the entire reservoir of gas in the galactic midplane, quantified by the midplane gas density $\rho_g$. That is, we do not consider the emissivity of molecular gas tracers such as CO and their variation with the local gas density, as described in~\cite{Leroy2017}. This approach is required to consider and compare the influences of large-scale dynamical processes, which affect molecular gas at all density scales, not only those density scales that are CO-bright. Our approach allows us to determine the processes that are {\it dominant} in setting the cloud lifetime, in the sense that they influence molecular clouds on the shortest time-scales. On the other hand, care must be taken when our predictions of the cloud lifetime are compared to those derived from CO observations. In regions of gas that are purely molecular and have average densities above $\sim 10 M_{\odot}{\rm pc}^{-3}$, the vast majority of gas has a high CO emissivity, such that our predictions should correspond well with CO observations. However, as the average midplane ${\rm H}_2$ density decreases, an increasing fraction of the cloud lifetime predicted by our theory becomes `invisible' in the sense that it is not strongly CO-emitting. Our theory therefore over-predicts the CO-traced cloud lifetime by an increasing amount as the average ${\rm H}_2$ density of the galactic midplane decreases. We expect this effect to be particularly significant in the lower-density outskirts of galactic discs.

\subsection{Milky Way}
\label{Sec::MilkyWay}
\begin{figure}
 \includegraphics[width=0.45\textwidth]{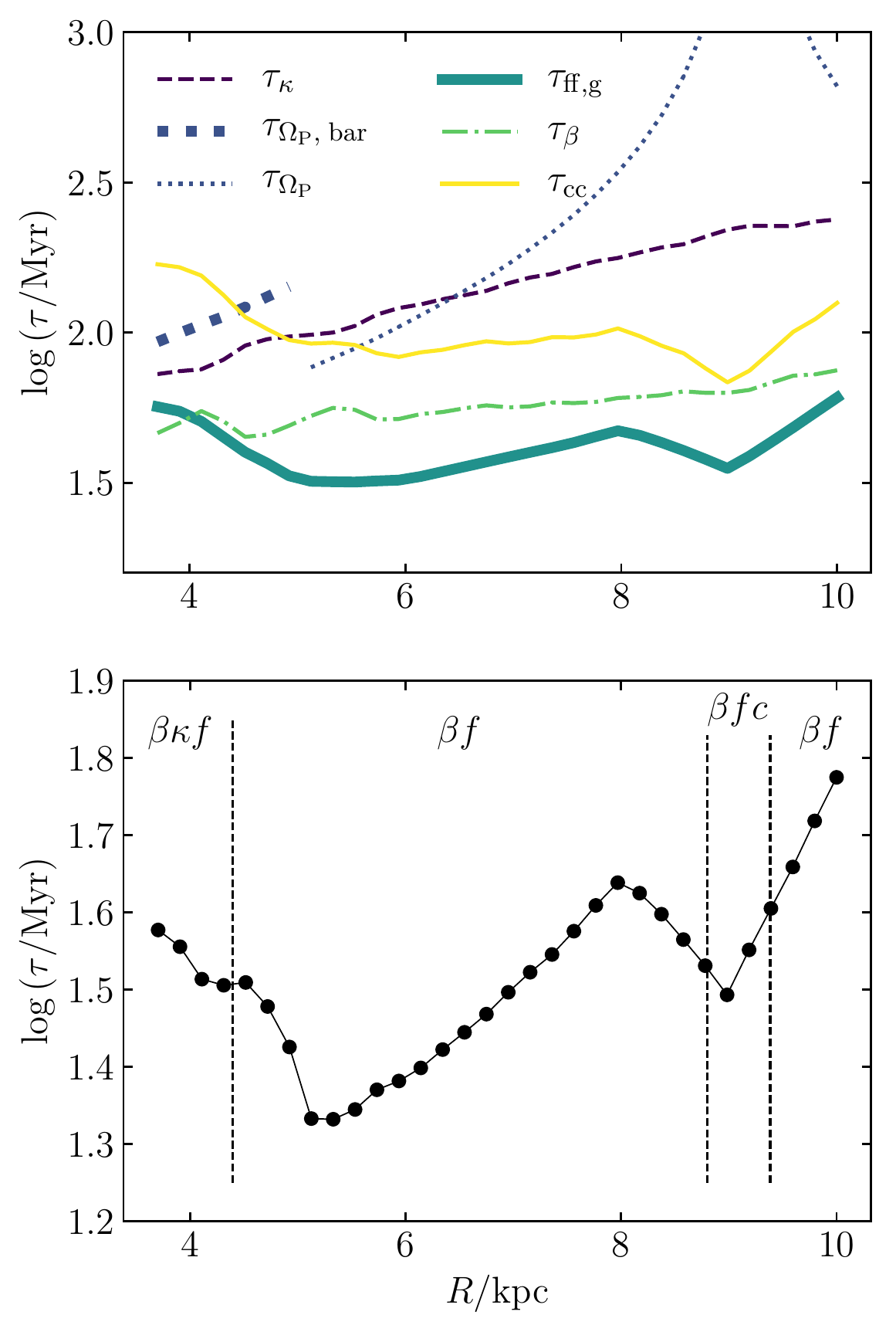}
 \caption{Variation in the predicted cloud lifetimes for the Milky Way with galactocentric radius. The upper panel gives each of the time-scales of dynamical evolution, including the overall minimum time-scale $\tau_{\text{min}}$. The time-scale for spiral arm crossings ($m=4$) is represented by the thin dotted line, while the time-scale for interactions with the Galactic bar ($m=2$) is represented by the thick dotted line. The lower panel gives the lifetime calculated using Equation~(\ref{Eqn::Lifetime}). The regions of relevance are labelled as in Figures~\ref{Fig::QvsBeta_lifetimes_m0},~\ref{Fig::QvsBeta_lifetimes} and~\ref{Fig::OmegavsBeta_lifetimes}, where $f$ corresponds to $\tau_{\rm ff,g}$, $\beta$ corresponds to $\tau_\beta$, $\kappa$ corresponds to $\tau_\kappa$ and $c$ corresponds to $\tau_{\rm cc}$.}
 \label{Fig::MilkyWay}
\end{figure}

We predict the cloud lifetimes for the Milky Way using the surface densities from~\cite{Kennicutt2012}, the rotation curve from~\cite{Bland-Hawthorn2016}, and the velocity dispersions from~\cite{Heiles2003}, each of which vary with galactocentric radius. These observational data determine the parameters $\beta$ and $Q$. We use a pattern speed of $\Omega_{\rm P} \approx 0.026 \pm 0.002$ Myr$^{-1}$ from~\cite{Gerhard2011} along with the rotation curve to set the value of $\Omega_{\rm P}/\Omega$. We represent the galactic bar inside $\sim 5$~kpc~\citep{Wegg2015} using $m=2$.

The number of spiral arms $m$ that should be used for $>5$~kpc is not clear, as a large degree of uncertainty remains as to the strength of the spiral pattern for the Milky Way~\citep{Antoja2016}. It is not obvious whether $m=4$ for strong spiral arms, or whether $m=0$ due to flocculence. However we find that this uncertainty does not have a large effect on our predictions, because the low observed pattern speed ensures that the time-scale for spiral arm crossings is long outside $5$~kpc. In the top panel of Figure~\ref{Fig::MilkyWay}, we display the variation in each time-scale with galactocentric radius, and the time-scale for spiral arm collisions $\tau_{\Omega_{\rm P}}$ is displayed as a thick dotted line within the region of the Galactic bar ($<5$~kpc), and as a thin dotted line outside this region ($>5$~kpc). The discontinuity between these two sections corresponds to the tip of the Galactic bar. We have used $m=4$ for strong spiral arms, and even in this extreme case, the time-scale for spiral arm crossings is always longer than twice the minimum time-scale (the thick green line representing $\tau_{\rm ff,g}$ in the top panel) and between five and six times longer than the cloud lifetime (the black filled markers in the bottom panel). The regions of relevance displayed in the bottom panel of Figure~\ref{Fig::MilkyWay}, which compare each time-scale to twice the minimum time-scale $\tau_{\rm min}$ if $\tau_{\rm min}$ is longer than the cloud lifetime, and compare each time-scale to twice the minimum cloud lifetime $\tau$ if $\tau>\tau_{\rm min}$, confirm the small contribution made by spiral arm crossings---$\tau_{\Omega_{\rm P}}$ is never a {\it relevant} time-scale.

Comparing the top and bottom panels of Figure~\ref{Fig::MilkyWay} indicates that the cloud lifetime is primarily controlled by gravitational collapse and shear support, over the radial interval considered. The dip in $\tau_{\text{ff,g}}$ in the vicinity of the solar neighbourhood is due to the dip in the time-scale for gravitational collapse (and to a lesser extent in the time-scale for cloud-cloud collisions), which in turn is caused by a dip in the value of the Toomre $Q$ parameter at this radius. This behaviour is not mirrored by the shear time-scale, reducing the degree of support provided to the cloud. The slight global increase in cloud lifetime with galactocentric radius is due simply to the $\Omega^{-1}$-dependence of all the time-scales.

In the vicinity of the solar neighbourhood ($R \sim 8$~kpc), the cloud lifetimes we predict for the Milky Way are between $30$ and $40$~Myr. This contrasts with the lifetimes of gravitationally-bound, star-forming regions described by~\cite{Elmegreen2000} and~\cite{Hartmann2001}, which have lifetimes of $<5$ Myr within the solar neighbourhood. As described in Section~\ref{Sec::tau_ffcl}, these regions are gravitationally-decoupled from the galactic-scale dynamics and therefore have lifetimes independent of dynamical time-scales. Our model is more general and accounts not only for globally-collapsing regions but for clouds of many different sizes and structures, which may be globally or hierarchically collapsing, bound or unbound. We predict that such a sample of clouds has a median lifetime of $33$~Myr, or $0.8$ free-fall times $\tau_{\rm ff,g}$ in the midplane of the galaxy.

\subsection{M31}
\begin{figure}
 \includegraphics[width=.45\textwidth]{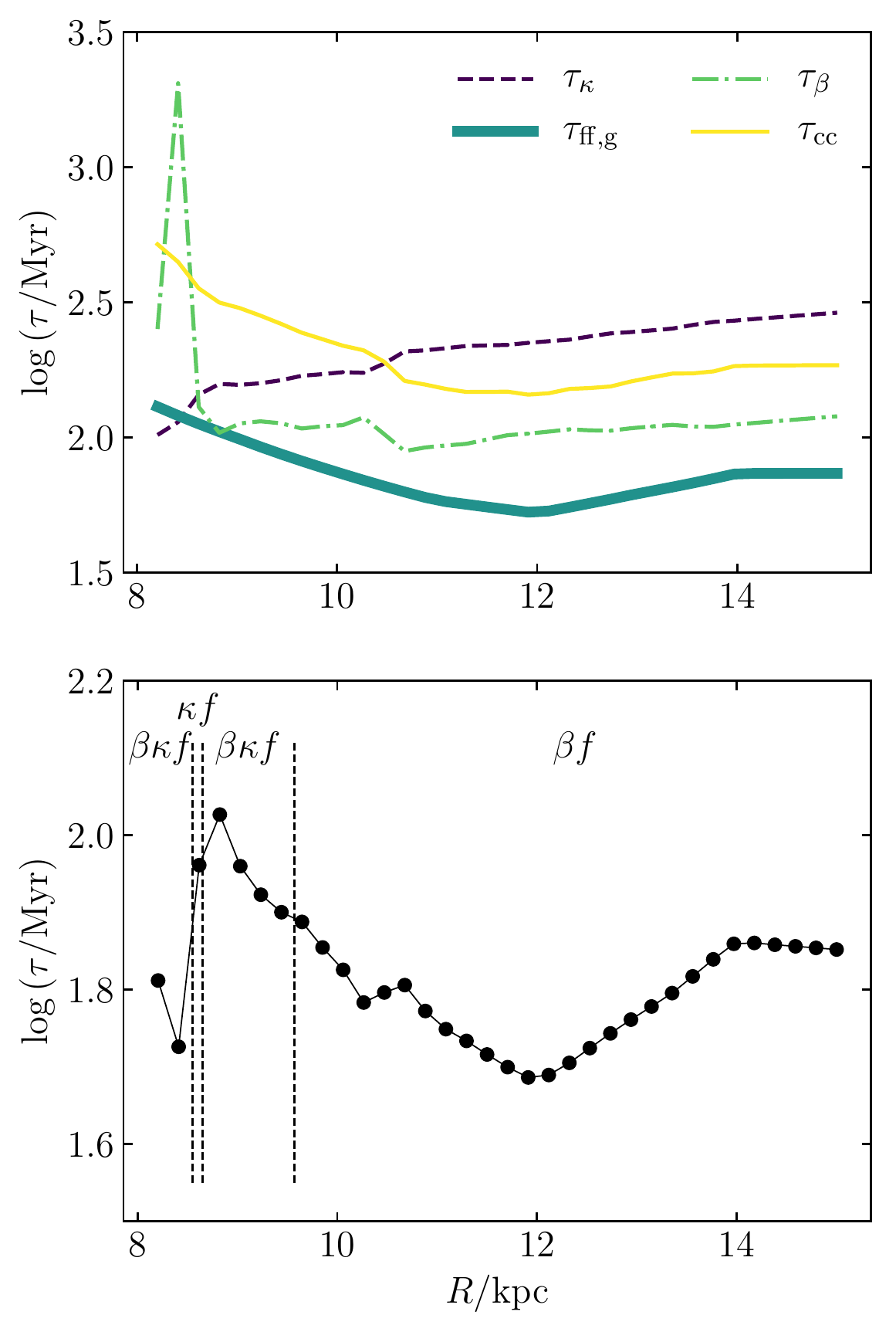}
 \caption{Variation in cloud lifetimes predicted for the galaxy M31 by our model, with radius from its centre in kpc. The upper panel gives each of the time-scales of dynamical evolution, including the overall minimum time-scale $\tau_{\text{min}}$. The lower panel gives the lifetime calculated using Equation~(\ref{Eqn::Lifetime}) (black filled circles). The regions of relevance are labelled as in Figures~\ref{Fig::QvsBeta_lifetimes_m0},~\ref{Fig::QvsBeta_lifetimes} and~\ref{Fig::OmegavsBeta_lifetimes}, where $f$ corresponds to $\tau_{\rm ff,g}$, $\beta$ corresponds to $\tau_\beta$, and $\kappa$ corresponds to $\tau_\kappa$.}
 \label{Fig::M31}
\end{figure}

To predict the cloud lifetimes for M31, we use the surface densities provided by Schruba et al.~(in prep.), the rotation curve from~\cite{Corbelli2010}, and the velocity dispersions from~\cite{Braun2009}, to calculate the values of the parameters $\beta$ and $Q$. Given that there is no evidence for the presence of spiral arms in M31, we set $m=0$ in our model, to produce the cloud lifetimes shown in the bottom panel of Figure~\ref{Fig::M31}, against galactocentric radius. The top panel shows the variation in individual time-scales, while the bottom panel displays the resulting cloud lifetime calculated using Equation~(\ref{Eqn::Lifetime}). The dashed lines delineate the regions of relevance for each cloud evolutionary mechanism.

In the case of M31, the general trend of increasing cloud lifetime due to the $\Omega^{-1}$ dependence of all time-scales is obscured by long cloud lifetimes between galactocentric radii of around $8$-$10$~kpc and a large dip at around $12$~kpc. This radius corresponds to a ring of gas that characterises the morphology of M31~\citep[e.g.][]{Braun2009}. The dip at $\sim 12$~kpc is due to a dip in the Toomre $Q$ parameter at this radius, causing the time-scales for gravitational free-fall and cloud-cloud collisions to decrease in magnitude. The long cloud lifetimes from $8$ to $10$~kpc are manifestly due to the proximity of the shear and gravitational free-fall time-scales at this radius. The shear time-scale $\tau_\beta$ in the top panel of Figure~\ref{Fig::M31} comes very close to the gravitational free-fall time-scale $\tau_{\rm ff,g}$, such that the rates are nearly balanced. Galactic shear competes against gravitational collapse to disperse the cloud rather than to collapse it, the balance between these two time-scales leads to longer cloud lifetimes via Equation~(\ref{Eqn::Lifetime}).

In general, M31 displays longer values of all cloud evolutionary time-scales than the Milky Way, by a consistent factor of $\sim 3$. This leads to a higher median lifetime of $\sim 64$~Myr, corresponding to $\sim 0.9$ free-fall times $\tau_{\rm ff,g}$ in the plane of the galaxy. The orbital time $\Omega^{-1}$ is the only dynamical parameter that affects all time-scales in the same way, indicating that the slower angular speed of midplane gas rotation at these larger radii is responsible for the elongation of cloud lifetimes in M31, relative to the Milky Way.

\subsection{M51}
\label{Sec::M51}
\begin{figure}
 \includegraphics[width=.45\textwidth]{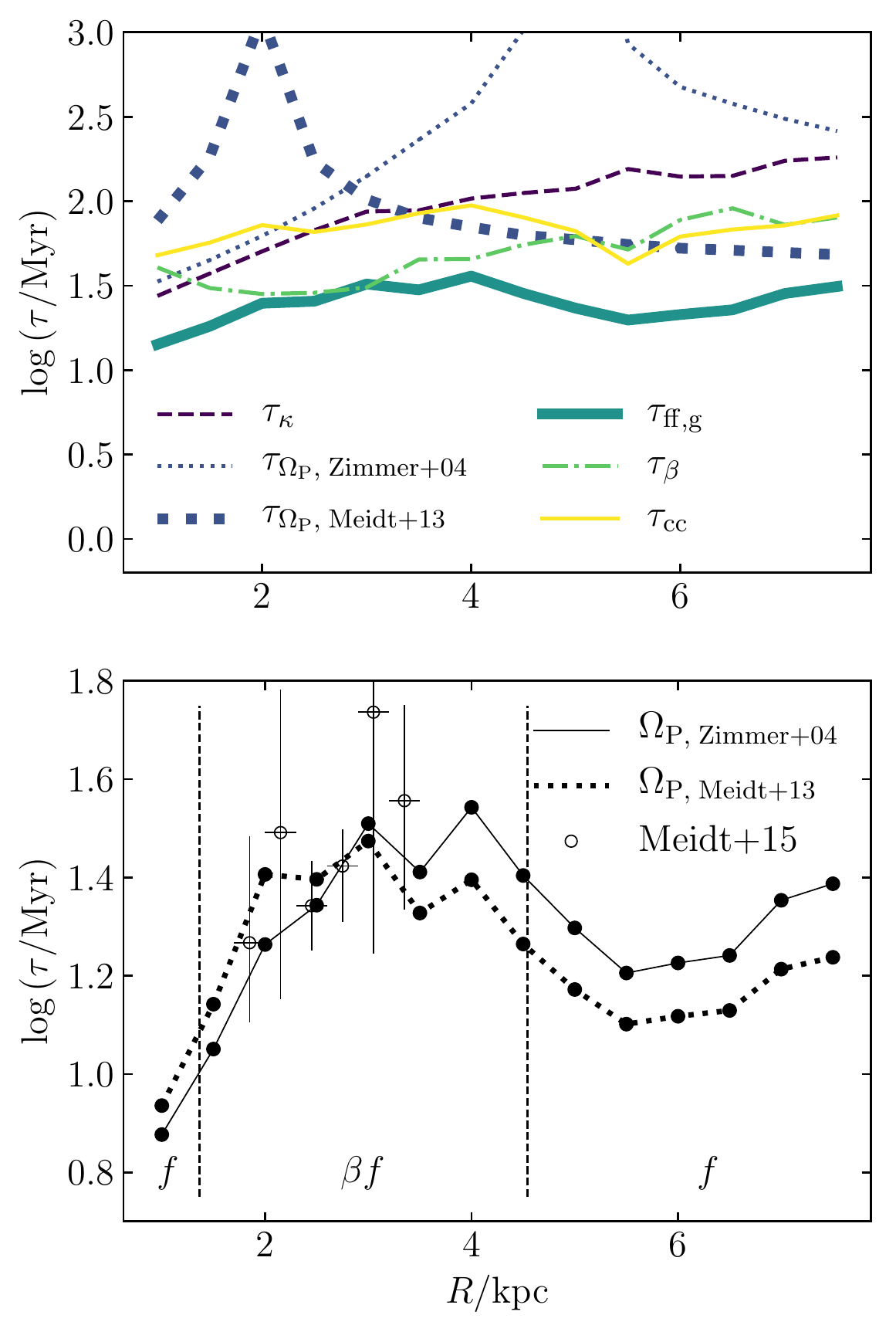}
 \caption{Variation in cloud lifetimes predicted for the galaxy M51 by our model, with radius from its centre in kpc. The upper panel gives each of the time-scales of dynamical evolution, including the overall minimum time-scale $\tau_{\text{min}}$. The lower panel gives the lifetime calculated using Equation~(\ref{Eqn::Lifetime}), with two different values of the pattern speed (see Section~\ref{Sec::M51}). Black filled circles represent lifetimes calculated using the value from~\protect\cite{Zimmer2004} (Zimmer+04), while red filled circles represent lifetimes calculated using the higher value from~\protect\cite{Meidt2013} (Meidt+13). The black open circles represent the cloud lifetimes calculated in~\protect\cite{Meidt2015} (Meidt+15). We see that there is satisfactory agreement between their results and our predictions for both pattern speeds. The regions of relevance are labelled only for the~\protect\cite{Zimmer2004} pattern speed, as in Figures~\ref{Fig::QvsBeta_lifetimes_m0},~\ref{Fig::QvsBeta_lifetimes} and~\ref{Fig::OmegavsBeta_lifetimes}, where $f$ corresponds to $\tau_{\rm ff,g}$ and $\beta$ corresponds to $\tau_\beta$.}
 \label{Fig::M51}
\end{figure}

Our predictions for the cloud lifetimes in M51 are calculated using the surface densities, velocity dispersions and rotation curve from~\cite{Schuster2007}, from which we calculate the values of the parameters $\beta$ and $Q$. The striking grand design structure of M51 consists of two main spiral arms~\citep{Henry2003}, such that we take $m=2$ in our models at all galactocentric radii. The pattern speed of these spiral arms has been extensively studied at varying radii, and while~\cite{Zimmer2004} uses the Tremaine-Weinberg method to derive a global pattern speed of $\Omega_{\rm P} \approx 0.039$ Myr$^{-1}$ for these spiral arms,~\cite{Meidt2008,Meidt2013} find evidence for at least two different pattern speeds inside a radius of $4$ kpc, both of which are significantly higher than this value, around $0.09$ Myr$^{-1}$. In the top panel of Figure~\ref{Fig::M51}, it can be seen that the time-scale $\tau_{\Omega_{\rm P}}$ for spiral arm crossings is significantly altered by the choice of pattern speed. The thin dotted line corresponds to the value from~\cite{Zimmer2004}, while the thick dotted line corresponds to the value from~\cite{Meidt2013}. In the bottom panel, we see that the predicted cloud lifetimes are also altered by choosing one of these pattern speeds over the other. The~\cite{Zimmer2004} value is represented by a solid line, while the~\cite{Meidt2013} value is represented by a dotted line. Both sets of cloud lifetimes agree well with the lifetimes estimated from observations by~\cite{Meidt2015}, who use the higher pattern speed from~\cite{Meidt2013} (black open circles).

For simplicity, we show the regions of relevance in Figure~\ref{Fig::M51} (dashed lines) only for the~\cite{Zimmer2004} value of $\Omega_{\rm P}$, as discussion about the exact radial dependence of the multiple pattern speeds in~\cite{Meidt2013} is ongoing. The authors suggest that the higher pattern speed may even apply to a bar which terminates at its radius of co-rotation ($\approx 2.3$ kpc) and give way to a much lower value of $0.056$ Myr$^{-1}$. We note that while spiral arm crossings are never relevant for the~\cite{Zimmer2004} pattern speed (thin dotted line in the top panel of Figure~\ref{Fig::M51}), they are relevant at $R \ga 6$~kpc for the~\cite{Meidt2013} value (thick dotted line).

Our broad brush strokes theory captures the overall dominance of gravitational free-fall and galactic shear, as well as the transition between shear-dominated and gravity-dominated behaviour proposed in~\cite{Meidt2015}. Shear is much more important at smaller galactocentric radii, while at larger radii ($R>4.5$~kpc), only gravity is significant. Accordingly, these are the galactocentric radii at which~\cite{Meidt2015} find strong signatures of high-mass star formation. In the context of our models, we would also expect the mechanism of cloud destruction to depend on the degree of shear support provided against the four dynamically-compressive time-scales. We would expect more clouds to be destroyed by dispersion for galactocentric radii $R<4.5$~kpc, and for more clouds to be destroyed by collapse and feedback for $R>4.5$~kpc. This prediction also agrees with the discussion in~\cite{Meidt2015}.

Overall, the smaller galactocentric radii at which the observations for M51 are taken, relative to the observations for the Milky Way and M31, mean that the orbital time $\Omega^{-1}$ is shorter for M51. This is reflected in the values of all the cloud evolutionary time-scales, which are $\sim 2$ shorter for M51 than for the Milky Way, giving a cloud lifetime which is also a factor of $\sim 2$ shorter. The median cloud lifetime is $\sim 21$~Myr, equivalent to $\sim 0.8$~free-fall times $\tau_{\rm ff,g}$ in the midplane of the galaxy. The longest lifetimes occur where the competition between galactic shear and gravitational collapse is closest.

\subsection{M83}
\label{Sec::M83}
\begin{figure}
 \includegraphics[width=.45\textwidth]{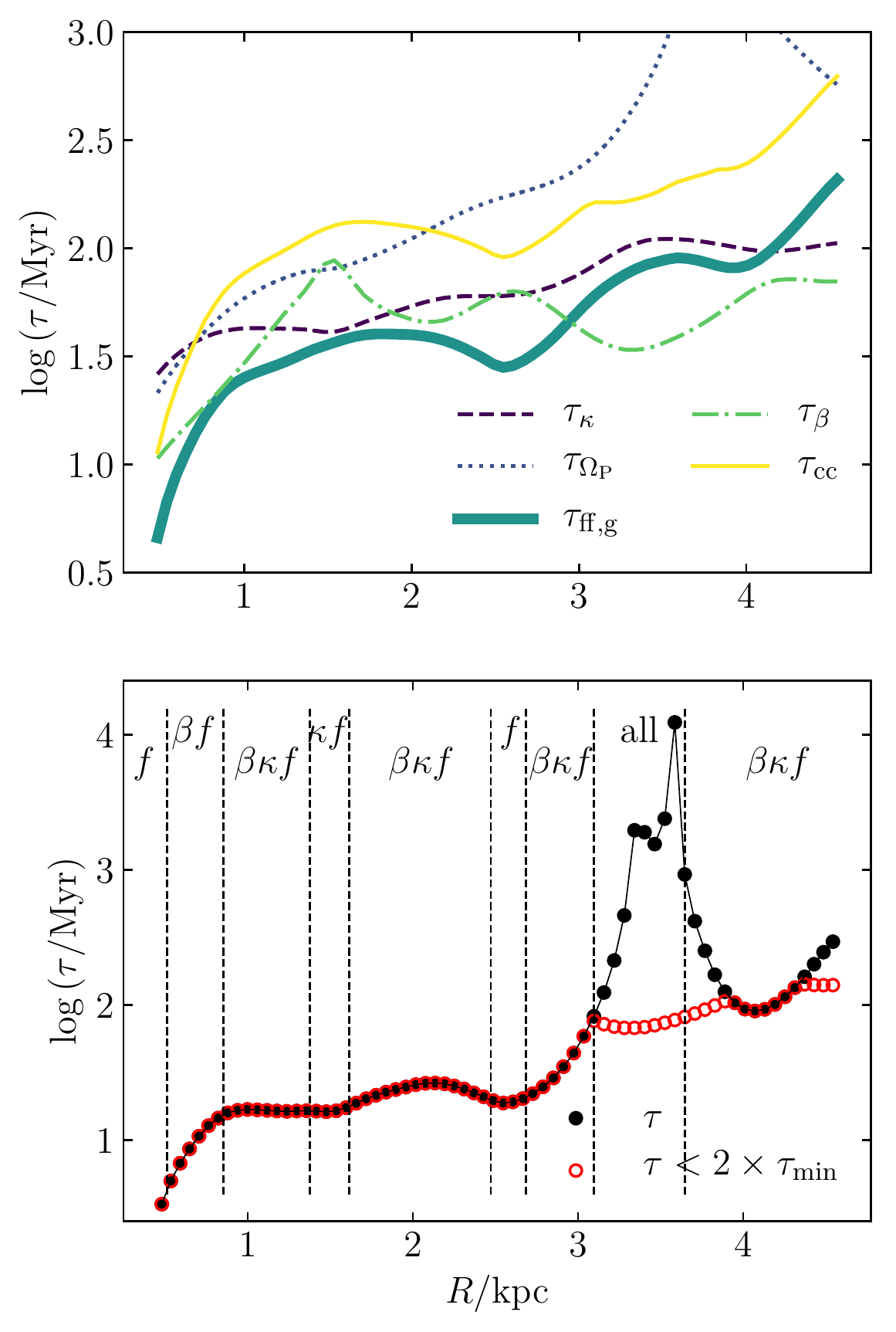}
 \caption{Variation in cloud lifetimes predicted for the galaxy M83 by our model, with radius from its centre in kpc. The upper panel gives each of the time-scales of dynamical evolution, including the overall minimum time-scale $\tau_{\text{min}}$. The black filled circles in the lower panel give the lifetime calculated using Equation~(\ref{Eqn::Lifetime}), while the red open circles give the lifetime below a threshold of twice the minimum cloud evolutionary time-scale. This prevents the cloud lifetime from becoming unreasonably long due to an unphysical degree of precision in the balance between galactic shear and the other cloud evolutionary mechanisms (see Section~\protect\ref{Sec::M83}). The regions of relevance are labelled as in Figures~\ref{Fig::QvsBeta_lifetimes_m0},~\ref{Fig::QvsBeta_lifetimes} and~\ref{Fig::OmegavsBeta_lifetimes}, where $f$ corresponds to $\tau_{\rm ff,g}$, $\beta$ corresponds to $\tau_\beta$, $\kappa$ corresponds to $\tau_\kappa$, and `all' indicates that all time-scales are relevant.}
 \label{Fig::M83}
\end{figure}

We predict the cloud lifetimes in M83 using the observational data given in Figure 6 of~\cite{Freeman2017}. These consist of the total (atomic and molecular) gas surface density curve, the rotation curve from~\cite{Walter2008} and the total velocity dispersion inferred from the Atacama Large Millimeter/submillimenter Array (ALMA) CO and THINGS 21-cm data, weighted by the relative surface densities of the atomic and molecular components. From these data, we obtain the values $\beta$, $Q$ and $\Omega$ in our fundamental plane. Like M51, M83 has two main spiral arms, so we use a value of $m=2$ and a pattern speed of $\Omega_{\rm P}=0.045$~Myr$^{-1}$ from~\cite{Zimmer2004}. The predicted cloud lifetimes are displayed in the bottom panel of Figure~\ref{Fig::M83} as a function of galactocentric radius, while the individual time-scales are shown in the panel above.

To remove noise on short scales from these data, we have used a Savitzky-Golay filter with 4th-order polynomials and a window size of 35 data points~\citep{SavitzkyGolay1964}. With this setup, each smoothed data point is calculated using $(35-1)/2=17$ data points to either side of its own position, weighted by a set of 35 analytic coefficients. These analytic coefficients are constrained by the 4th-order polynomial coefficients, of which there are five. In simple terms, each window of 35 data points is therefore described using five coefficients, giving an effective `smoothing length' of $\sim 0.4$~kpc. This procedure removes noise on scales $\sim 0.1$~kpc without removing the physical oscillations on kpc-scales, which are most visible in the profiles of $\tau_\beta$ and $\tau_\kappa$, and that also persist in the cloud lifetime (bottom panel of Figure~\ref{Fig::M83}).

The large peak in the cloud lifetime at $R \sim 3.5$ in the bottom panel of Figure~\ref{Fig::M83} illustrates an important caveat in our use of Equation (\ref{Eqn::Lifetime}) to combine the time-scales of cloud evolution. The peak occurs when the rate of galactic shear approaches the combined rates of the dynamically-compressive mechanisms of cloud evolution, such that $\tau_\beta^{-1} \rightarrow \tau_\kappa^{-1} + \tau_{\Omega_{\rm P}}^{-1} + \tau_{\rm ff,g}^{-1} + \tau_{\rm cc}^{-1}$. At such closely-balanced values of dispersive and compressive evolutionary rates, the cloud lifetime predicted by Equation (\ref{Eqn::Lifetime}) becomes very sensitive to $\beta$, which in turn is very sensitive to the rotation curve. Any noise in the rotation curve may then produce sharp peaks in the cloud lifetime, with $\tau \rightarrow \infty$. That is, the simple statistical approach we have taken to account for the coexistence of cloud evolutionary mechanisms and the phenomenon of shear support (adding their rates) means that cloud lifetimes can increase asymptotically if the shear and the collective other time-scales are balanced. This asymptotic increase in the cloud lifetime is unphysical, and results from the following characteristics of our theory:
\begin{enumerate}
	\item Our theory is statistical in the sense that it combines rates. However, when stochastically drawing from these rates and their associated probabilities, one cloud evolutionary mechanism will always occur first in practice, with the exception of galactic shear and epicyclic perturbations, which have a continuous effect over time. This stochasticity `smoothes out' the sensitivity of cloud lifetime to the rotation curve, in the sense that it does not allow extremely close balance between cloud evolutionary mechanisms. Even if large-scale dynamics alone are responsible for the evolution of molecular clouds, the cloud lifetime cannot become infinite when shear balances all other time-scales, due to their random nature.
	\item Our theory considers the influence of large-scale dynamics on the cloud lifetime, but not the influences of many other possible non-dynamical evolutionary mechanisms on smaller scales. When the large-scale dynamical mechanisms of cloud evolution approximately negate each other, all other cloud evolutionary mechanisms become comparatively more important. We would therefore expect that in regions of parameter space where the dynamical mechanisms of cloud evolution predict very long cloud lifetimes, GMCs will actually be destroyed on much shorter time-scales by non-dynamical or smaller-scale mechanisms of cloud evolution than those which we have considered, such as ionisation by UV radiation, or stellar feedback from gravitationally-bound regions on sub-cloud scales~\citep[e.g.][]{Elmegreen2007}.
\end{enumerate}

To illustrate the effect of these stated assumptions, we include an additional set of `cloud lifetimes' in the bottom panel of Figure~\ref{Fig::M83}, represented by the open circles. These values are identical to the lifetimes from Equation (\ref{Eqn::Lifetime}) (indicated by the filled circles), but with an imposed upper limit of twice the minimum evolutionary time-scale. At $R \sim 3.5$~kpc, the minimum time-scale is the shear time-scale $\tau_\beta$, so the cloud lifetime is limited by the assumption that the influence of galactic shear can only be reduced partially, by a factor of $1/2$, due to competition with dynamically-compressive mechanisms of cloud evolution. This causes the maximum predicted lifetime to drop to $\tau \sim 100$~Myr at $R \sim 3.5$~kpc. Given that galactic shear has a continuous effect on an object that is extended along the galactocentric radial direction, while spiral arm crossings and cloud-cloud collisions are discrete, stochastic events, such partial balance is more likely to occur than the infinite lifetime produced by exact balance between compressive and dispersive dynamical time-scales. By this reasoning, the $\sim 10^4$~Myr height of the sharp peak in the cloud lifetime between $3$ and $4$~kpc should not be viewed as a physically accurate prediction. However, we do predict that the close balance between dynamically-compressive and dynamically-dispersive mechanisms of cloud evolution results in elongated cloud lifetimes between $3$ and $4$~kpc in M83.

Overall, we find that cloud lifetimes in M83 are controlled primarily by the interplay between gravitational collapse and galactic shear, with gravity dominating below $R\approx 3$ kpc ($\tau_{\rm ff,g}$ is manifestly the shortest time-scale for $R \la 3$ in the top panel of Figure~\ref{Fig::M83}) and shear dominating above $R \approx 3$ kpc ($\tau_\beta$ is manifestly the shortest time-scale for $R \ga 3$ in the top panel of Figure~\ref{Fig::M83}). Correspondingly, we expect a decrease of the SFE at $R>3$~kpc, which in turn may explain the drop of the cluster formation efficiency at these radii \citep{Adamo2015}. Epicyclic perturbations are also relevant at most galactocentric radii, as indicated by the regions of relevance labelled with $\kappa$ in the bottom panel of Figure~\ref{Fig::M83}, however $\tau_\kappa$ is never dominant over $\tau_{\rm ff,g}$ or $\tau_\beta$. The predicted importance of gravity and shear is consistent with the work of~\cite{Freeman2017}, who have found that significant gravitational contraction and regulation by galactic shear is implied by the cloud mass distribution. These authors also show that both the maximum cloud mass and the average mass of the five most massive GMCs drop systematically from the central regions of the galaxy out to $4.5$~kpc. Our predictions demonstrate a corresponding systematic {\it increase} in the cloud lifetime between these galactocentric radii (ignoring the sharp peak between $3$ and $4$~kpc), due primarily to the $\Omega^{-1}$-dependence of all cloud evolutionary time-scales. Combining our theoretical predictions with the observational analysis of~\cite{Freeman2017} therefore suggests that the cloud mass may be anti-correlated with the cloud lifetime. Such a result would also be consistent with the view that the majority of star formation occurs in the most massive clouds~\citep{Murray2011,Reina-Campos2017}, which collapse faster, have higher SFRs, and are shorter-lived.

\section{Conclusions}
\label{Sec::Conclusions}
We have developed an analytic theory for the molecular cloud lifetime that depends on the large-scale dynamics of the ISM, independent of theoretical assumptions about the size, structure, mode of gravitational collapse, and gravitational boundedness of GMCs. This theory includes the mechanisms of gravitational free-fall, galactic shear, epicyclic perturbations, spiral arm interactions and cloud-cloud collisions. We have made as few assumptions as possible about the exact way in which clouds interact with each of the proposed environmental mechanisms, instead characterising the influence of each by its time-scale and by describing its dynamical relationship to the other evolutionary mechanisms in the context of a hydrostatic equilibrium galaxy disc. Our approach represents an advancement both in its systematic approach and in its expansiveness, by accounting simultaneously for multiple influences on cloud evolution. Here we briefly summarise the main results and conclusions of our work.

\subsection{Theoretical approach}
To our knowledge, no theory prior to this work has produced a formalism for considering the interplay between environmental mechanisms of cloud evolution across parameter space. Our systematic approach enables an overview of the influences on GMC evolution to be obtained without having to make assumptions that may only be relevant to a subset of GMCs. By comparing our simple analytic expression for the cloud lifetime to observations that will soon be available through currently-ongoing efforts (see Section~\ref{Sec::FutureWork}), the accuracy of our predictions will be testable. The comparison between observation and theory will shed light on the influence of ISM dynamics relative to the effects not considered in our theory, such as the degree of non-linear interplay between the evolutionary mechanisms and the details of stellar feedback. Our theory also provides a broad brush strokes platform upon which physical complexity and detail can be systematically built. We briefly discuss our ongoing work in this direction in Section~\ref{Sec::FutureWork}.

\subsection{Significance of each cloud evolutionary mechanism}
We find that the large-scale dynamical mechanisms affecting cloud evolution can be described within a fundamental parameter space $(\beta,Q,\Omega)$, with a secondary dependence on $\phi_P$, a parameter that reflects the (inverse of the) local gas fraction. These are all observable properties of the ISM that can be derived using the galactic rotation curve, along with the gas and stellar surface densities and velocity dispersions. In galaxies with spiral arms, the fundamental parameter space is extended to include the number of spiral arms $m$ and the pattern speed as a fraction of the ISM angular velocity, $\Omega_{\rm P}/\Omega$. We have predicted the environmentally- and dynamically-dependent cloud lifetimes throughout the region of parameter space that is currently observable, considering the possibility of augmentation and competition between cloud evolutionary mechanisms. For each cloud evolutionary mechanism, we have systematically determined the {\it regions of dominance} (i.e.~the regions of parameter space for which each mechanism occurs on a shorter time-scale than all other mechanisms) and the {\it regions of relevance} (i.e.~the regions of parameter space for which each mechanism occurs on a time-scale no longer than twice the minimum time-scale, or no longer than twice the cloud lifetime, if the cloud lifetime is elongated by competition between different mechanisms). Via examination of the resulting parameter space maps, we have reached the following main conclusions.
\begin{enumerate}
\item Throughout the currently-observable region of parameter space, chosen according to the ranges of physical parameters presented in~\cite{Leroy2008}, gravitational collapse on a time-scale $\tau_{\rm ff,g}$ is by far the most prevalent and influential mechanism in setting the cloud lifetime. Gravitational free-fall is {\it relevant} throughout most of our parameter space, excluding the most gravitationally-stable (high Toomre $Q$) regions of spiral galaxies. It is {\it dominant} for levels of gravitational stability up to $Q \sim 4$ in most spiral galaxies and up to $Q \sim 8$ in elliptical and flocculent galaxies.
\item In galactic environments for which the rotation curve is approximately flat and gravitational stability is high ($\beta \la 0.5$ and $Q \ga 4$), galactic shear provides a {\it relevant} degree of support against the dynamically-compressive mechanisms of gravitational free-fall, epicyclic perturbations and cloud-cloud collisions. This causes the cloud lifetime to be elongated. At the highest levels of shear ($\beta \sim 0$), shear support {\it dominates} cloud evolution down to Toomre $Q$ values as low as $Q \sim 4$. With the introduction of spiral arms, the regions of {\it relevance} and {\it dominance} for galactic shear are reduced, but still occupy a significant region of parameter space in the cases of low pattern speeds $\Omega_{\rm P}/\Omega$ and few spiral arms $m$.
\item Spiral arm crossings on a time-scale $\tau_{\Omega_{\rm P}}$ are most {\it relevant} at high values of gravitational stability (high Toomre $Q$), where interactions between clouds and spiral arms may occur before clouds are destroyed by gravitational collapse and the resulting stellar feedback. At high pattern speeds $\Omega_{\rm P}/\Omega$ and high numbers of spiral arms $m$, spiral arm crossings may be {\it relevant} at all values of $Q$, with regions of {\it dominance} extending as low as $Q \sim 1$. At lower numbers of spiral arms and lower pattern speeds, the dominance of spiral arm crossings is limited to values of $Q \ga 6$.
\item In non-spiral galaxies, the regions of {\it relevance} and {\it dominance} for epicyclic perturbations on a time-scale $\tau_\kappa$ are restricted to high values of Toomre $Q$, where the resulting orbital eccentricity of clouds and the ISM can influence the evolution of GMCs before gravitational collapse sets in. They are also restricted to high values of the shear parameter $\beta$, where clouds are not significantly dispersed before an epicycle is completed. Such conditions exist near the centres of disc galaxies like the Milky Way. In galaxies with spiral arms, epicyclic perturbations are only relevant close to the radius of co-rotation, where spiral arms play no role in cloud evolution, or in the case of a weak spiral pattern below the radius of co-rotation. When present, the influence of spiral arms easily dominates over the effect of epicycles.
\item Cloud-cloud collisions on a time-scale $\tau_{\rm cc}$ are never {\it dominant}. They are {\it relevant} only in the case of a pure gas disc $\phi_P = 1$ or at very high levels of stability and shear ($\beta \sim 0$ and $Q \sim 15$) in galaxies without spiral arms, where all cloud evolutionary mechanisms become relevant. This negligible influence of cloud-cloud collisions throughout most of parameter space is primarily due to the similar environmental scaling of the free-fall time and the cloud collision time, such that the rate of gravitational collapse always outpaces the rate of collisions, especially at low values of Toomre $Q$. The gravitationally-bound clouds needed to sustain regular cloud-cloud collisions are typically destroyed by collapse and stellar feedback before a collision can occur.
\end{enumerate}

Combining the above considerations for individual cloud evolutionary mechanisms, we can describe which mechanisms set the cloud lifetime in the specific parts of the parameter space that describes galaxies. Our conclusions in this area are as follows.
\begin{enumerate}
	\item In highly star-forming, gas-rich galaxies such as high-redshift galaxies, with large reservoirs of gravitationally-unstable gas ($Q \la 4$), the cloud lifetime is controlled almost exclusively by gravitational free-fall.
	\item Far from the radius of co-rotation in grand design spiral galaxies ($|\Omega_{\rm P}/\Omega-1| \gg 1$ with $m=2$ or $m=4$), spiral arm perturbations have a {\it dominant} influence on the cloud lifetime for all $Q \ga 1$. Gravitational collapse is still {\it dominant} up to the case of marginal stability ($Q \la 1$).
	\item Close to galactic centres, where molecular gas may be highly gravitationally-stable with low levels of shear ($\beta \ga 0.5$ and $Q \sim 15$), the cloud lifetime is controlled by epicyclic perturbations (i.e.~orbital eccentricity). If strong spiral arms are also present (e.g.~in the presence of a nuclear spiral), then the cloud lifetime is set by a combination of epicycles  and spiral arm perturbations.
	\item For outer galaxy bulges, containing regions of highly-stable, highly-shearing gas ($Q \ga 6$ and $\beta \la 0.5$), cloud lifetimes are governed by the competition between gravitational collapse and galactic shear, along with a lesser contribution from epicyclic perturbations. Clouds in this region of parameter space tend to have longer lifetimes due to the high degree of shear support, allowing dynamical mechanisms with longer time-scales to significantly influence cloud evolution. This region of parameter space is therefore characterised by the coexistence of many different cloud evolutionary mechanisms. For very high levels of stability and shear ($Q \sim 15$ and $\beta \sim 0$), the influence of galactic shear dominates over the combined influence of all the dynamically-compressive mechanisms of cloud evolution, such that clouds are more likely to be destroyed by dispersal than by collapse.
\end{enumerate}

The combination of all evolutionary mechanisms across parameter space gives rise to a number of implications for star formation and cloud evolution as a function of the galactic environment. Our conclusions in this area are as follows.
\begin{enumerate}
\item Our results imply that typically, the cloud lifetime increases with the degree of gravitational stability $Q$.
\item Many gravitationally-stable environments with $Q \ga 4$ are dominated by spiral arm crossings on a time-scale $\tau_{\Omega_{\rm P}}$. Due to the high degree of gravitational stability in these clouds, we expect that they contain few gravitationally-bound, star-forming regions that are decoupled from the galactic-scale dynamics. By inducing compression of previously unbound regions of the ISM, spiral arm interactions may therefore increase the galactic-scale SFE per unit mass, and decrease the cloud lifetime.
\item We expect a transition from star-forming clouds to quiescent clouds as the rotation curve flattens and the degree of gravitational stability increases. As $\beta$ decreases and $Q$ increases, the relevance of shear support increases, leading to a suppression of gravitational collapse, and consequently star formation. This may contribute to the low SFR observed across much of the Galactic Centre environment, which displays high values of gravitational stability, $Q \sim 15$, and a non-trivial degree of galactic shearing, $\beta=0$--$0.7$~\citep[e.g.][]{Krumholz2015}.
\item Across our parameter space, comparison of cloud evolutionary time-scales suggests that cloud-cloud collisions are rarely an important factor in setting the course of cloud evolution or the cloud lifetime. Gravitational collapse of the ISM occurs on a much shorter time-scale, suggesting that collisions between clouds do not play an important role in setting either the cloud-scale SFR or the galactic-scale SFR.
\end{enumerate}

\subsection{Cloud lifetimes}
We have applied our theory to four galaxies and find that the predicted cloud lifetimes typically range between $10$ and $100$ Myr, with the exception of a few extreme regimes in M83 and one extreme regime in M31, where shear balances almost exactly with gravitational collapse. In practice, this delicate balance of rates cannot ever be maintained, due to the stochastic, quantised nature of several of the considered mechanisms.
\begin{enumerate}
\item Milky Way ($R=4$--$10$~kpc): Cloud lifetimes fall between $21$~Myr and $60$~Myr with a median of $33$~Myr. The dominant mechanisms of cloud evolution are galactic shear and gravitational collapse of the ISM, with minor contributions from the spiral arms and cloud-cloud collisions.
\item M31 ($R=8$--$15$~kpc): Cloud lifetimes fall between $49$~Myr and $106$~Myr with a median of $64$~Myr, mainly due to relatively high values of the orbital time-scale $\Omega^{-1}$ at these galactocentric radii. The lifetimes are dominated by galactic shear and gravitational collapse of the ISM.
\item M51 ($R=1$--$8$~kpc): Cloud lifetimes are the shortest of the four galaxies, falling between $8$~Myr and $35$ Myr with a median of $21$~Myr. This is due mainly to the short values of the orbital time $\Omega^{-1}$ at the radii considered. These predictions are in good agreement with the lifetimes estimated by~\cite{Meidt2015}. Again, the dominant mechanisms of evolution are galactic shear and gravitational collapse.
\item M83 ($R=0.5$--$5$~kpc): M83 hosts a transition between short and long cloud lifetimes. Cloud lifetimes are short (10-30 Myr) for $R<3$~kpc and long ($\sim100$~Myr) for $R>3$~kpc, with a median of $\sim 25$~Myr.
\end{enumerate}

\subsection{Future work}
\label{Sec::FutureWork}
Our theory for molecular cloud lifetimes represents a first step towards a more detailed understanding of cloud evolution. While this theory provides a reasonably accurate description of the highly limited number of observed cloud lifetimes presently available and is straightforward to interpret by virtue of its analytic approach, we have also identified areas where its idealised nature may obstruct further insights. Our predicted cloud lifetimes do not consider the constraints on observable molecular gas densities imposed by the use of tracers such as CO, and will therefore overestimate the cloud lifetime in regions of galaxies with low average ${\rm H}_2$ densities. Additionally, in galactic environments where the time-scales of different mechanisms are expected to be very similar, our predicted cloud lifetimes are highly sensitive to the assumption that individual evolutionary mechanisms are not correlated and occur ergodically in time, in the sense that the the behaviour of a typical cloud corresponds approximately to the average behaviour of an infinite succession of clouds observed over an infinitely-long time period. We are currently extending this work by exploring the presented parameter space with hydrodynamical simulations of disc galaxies, allowing us to consider the (plausibly non-linear) interaction between the mechanisms introduced here. Such numerical simulations enable a closer study of the detailed physics that have been absorbed into the dynamical time-scales in the present work, such as stellar feedback. They will allow us to examine the relationship between cloud lifetimes observable via CO emission, as well as the cloud lifetimes predicted by our analytic theory. Finally, they will provide insight into how galaxy evolution drives changes of the cloud lifetime, and possibly the SFE, as the host galaxy evolves through the parameter space identified here.

In addition to these theoretical and numerical efforts, systematic observational measurements of the cloud lifecycle have now become accessible with the Atacama Large Millimeter/submillimeter Array (ALMA). By applying the statistical method of \citet{Kruijssen2014} and and~\citet{Kruijssen2018b} to high-resolution ALMA maps of galaxies in the local Universe and potentially out to high redshift (e.g.~Kruijssen et al.~in prep.; Hygate et al.~in prep.; Schruba et al.~in prep.; Chevance et al.~in prep.; Ward et al.~in prep.), it will be possible to systematically test our theory and (inevitably) revise, extend, or rule out its key physical ingredients. This will provide important insight into the cloud life-cycle and will contribute to the ongoing progress towards a bottom-up, cloud-scale synthesis of galactic star formation.”

\section*{Acknowledgements}
SMRJ and JMDK gratefully acknowledge funding from the German Research Foundation (DFG) in the form of an Emmy Noether Research Group (grant number KR4801/1-1, PI Kruijssen). JMDK acknowledges support from the European Research Council (ERC) under the European Union's Horizon 2020 research and innovation programme via the ERC Starting Grant MUSTANG (grant agreement number 714907, PI Kruijssen) and from Sonderforschungsbereich SFB 881 `The Milky Way System' (subproject P1) of the DFG. We are grateful to M\'{e}lanie Chevance, Bruce Elmegreen, Eve Ostriker, Marta Reina-Campos, and Andreas Schruba for helpful discussions. We thank an anonymous referee for their careful reading of the manuscript.

\bibliographystyle{mnras}
\bibliography{bibliography}

\appendix
\section{An upper bound on the epicyclic amplitude}
\label{Sec::AppendixA}
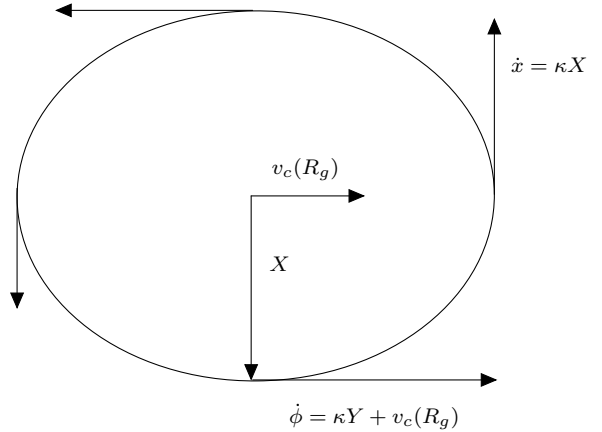
\begin{figure}
\label{Fig::Epicycles}
\centering
\begin{scaletikzpicturetowidth}{.9\linewidth}
\begin{tikzpicture}[line cap=round,line join=round,>=triangle 45,x=1.0cm,y=1.0cm]
\clip(5.,1.) rectangle (15.,7.5);
\draw [rotate around={0.5846305207051659:(9.14,4.5)}] (9.14,4.5) ellipse (3.1400784301664126cm and 2.453180088700454cm);
\draw [->] (9.08,4.5) -- (10.58,4.5);
\draw [->] (9.100040638097349,2.06) -- (12.32,2.06);
\draw [->] (12.28,4.52) -- (12.28,6.86);
\draw (12.4,6.44) node[anchor=north west] {$\dot{x} = \kappa X$};
\draw (9.48,1.85) node[anchor=north west] {$\dot{\phi} = \kappa Y + v_c(R_g)$};
\draw (9.25,5.1) node[anchor=north west] {$v_c(R_g)$};
\draw [->] (9.08,4.5) -- (9.08,2.05);
\draw (9.22,3.8) node[anchor=north west] {$X$};
\draw [->] (9.1,6.96) -- (6.5,6.96);
\draw [->] (6.,4.6) -- (6.,3.);
\end{tikzpicture}
\end{scaletikzpicturetowidth}
\caption{Schematic representation of a particle undergoing epicyclic motion, in the reference frame of the host galaxy. Equations~\protect\ref{Eqn::epicycles_radial_soln} and~\protect\ref{Eqn::epicycles_tangential_soln} dictate that the maximum tangential velocity occurs at the pericentre of the orbit, where the trajectory of the particle is parallel to the velocity of the guiding centre.}
\end{figure}

We use the conservation of angular momentum to put an upper bound on the relative magnitude of epicyclic oscillations $X/R_g$, following the derivation of~\cite{BinneyTremaine1987}. Within a locally harmonic minimum of the gravitational potential $\Phi(R)$, the radial motion of a cloud is governed by the equation of motion

\begin{equation}
\label{Eqn::epicycles_radial}
\ddot{x} = -\kappa^2 x,
\end{equation}
for epicyclic frequency $\kappa$. This is solved by the sinusoidal form

\begin{equation}
\label{Eqn::epicycles_radial_soln}
x(t) = X \cos{(\kappa t + \alpha)},
\end{equation}
where $X$ is the epicyclic amplitude and $\alpha$ is its phase, both of which are set by the initial conditions that perturb the cloud from a circular orbit. Using the conservation of angular momentum, we then obtain the instantaneous angular velocity $\dot{\phi}$ of the cloud in the reference frame of the host galaxy, given by

\begin{equation}
\label{Eqn::phidot}
\dot{\phi}(R) = \frac{L_z}{R^2} = \frac{L_z}{(R_g+x)^2} = \Omega_g \Big(1-\frac{2x}{R_g}\Big),
\end{equation}
where $L_z$ is the azimuthal angular velocity, $\Omega_g$ is the angular velocity of the guiding centre $\Omega_g=L_z/R_g^2$, and $x$ is defined by Equation~(\ref{Eqn::epicycles_radial_soln}). Using the restriction $X \ll R$ within the epicyclic approximation, We have expanded to first order in $X/R_g$. Equation~(\ref{Eqn::phidot}) can be integrated to obtain

\begin{equation}
\phi = \Omega_g t + \phi_0 - \gamma \frac{X}{R_g} \sin{(\kappa t + \alpha)},
\end{equation}
where $\gamma = 2\Omega / \kappa$. Therefore, the tangential motion in the reference frame of the guiding centre is given by

\begin{equation}
\label{Eqn::epicycles_tangential_soln}
y(t) = -Y \sin{(\kappa t + \alpha)},
\end{equation}
with

\begin{equation}
\label{Eqn::gamma} 
\gamma = \frac{2\Omega}{\kappa} \equiv \frac{Y}{X}.
\end{equation}
Together, Equations~\ref{Eqn::epicycles_radial_soln} and~\ref{Eqn::epicycles_tangential_soln} define a set of axes with origin at the guiding centre, $x$-axis pointing in the radial direction and $y$-axis pointing in the direction of motion of the guiding centre. By the conservation of angular momentum in the reference frame of the host galaxy, it is obvious that the cloud's trajectory at the orbital pericentre must be parallel to the velocity of the guiding centre, as depicted in Figure~\ref{Fig::Epicycles}. At the apocentre, the cloud's trajectory is antiparallel to the guiding centre velocity, and Equation~(\ref{Eqn::phidot}) gives

\begin{equation}
\dot{\phi} = \Omega_g\Big(1-\frac{2X}{R_g}\Big).
\end{equation}
In the reference frame of the host galaxy, the cloud must be moving in the same direction as the guiding centre at its point of closest approach, thus the conservation of angular momentum requires that it always be travelling in this direction relative to the galactic centre, such that

\begin{equation}
\dot{\phi} > 0.
\end{equation}
This gives an upper bound on the amplitude of epicyclic oscillations,

\begin{equation}
\label{Eqn::upper_bound}
\frac{X}{R_g} < \frac{1}{2}.
\end{equation}
A given perturbation from a circular orbit might therefore cause a cloud to undergo epicyclic motion with amplitude $X$ in the range $0 < X/R_g < 1/2$, and since we make no assumption about these initial conditions in our theory, we will assume for now a uniform distribution between the two extremes, such that the typical epicyclic amplitude is given by

\begin{equation}
\frac{X}{R_g} \approx \frac{1}{4}.
\end{equation}

\bsp
\label{lastpage}

\end{document}